\newcommand{\etaSS}{\eta_{\scriptsize\textnormal{SS}}}
\newcommand{\etaC}{\eta_{\scriptsize\textnormal{C}}}
\newcommand{\etaCA}{\eta_{\scriptsize\textnormal{CA}}}
\newcommand{\teta}{\eta^*}
\newcommand{\calL}{\mathcal{L}}
\newcommand{\calC}{\mathcal{C}}

\newcommand{\ini}{\scriptsize\textnormal{i}}
\newcommand{\fin}{\scriptsize\textnormal{f}}
\newcommand{\inter}{\scriptsize\textnormal{int}}
\newcommand{\irr}{\scriptsize\textnormal{irr}}
\newcommand{\lin}{\scriptsize\textnormal{lin}}

\newcommand{\STA}{SST}
\newcommand{\conf}{\scriptsize\textnormal{conf}}
\newcommand{\kin}{\scriptsize\textnormal{kin}}

\newcommand{\be}{\begin{equation}}
\newcommand{\ee}{\end{equation}}
\newcommand{\ba}{\begin{eqnarray}}
\newcommand{\ea}{\end {eqnarray}}

\newcommand{\mean}[1]{\left \langle #1 \right \rangle}

\documentclass[12pt]{iopart}
\usepackage{amsbsy}
\usepackage{color}
\usepackage{graphicx}
\usepackage{cite}
\usepackage{hyperref}
\usepackage{braket}

\usepackage[utf8]{inputenc}

\begin{document}

\title[Driving rapidly while remaining in control]{Driving rapidly while remaining in control: classical shortcuts from Hamiltonian to stochastic dynamics}

\author{David Gu\'ery-Odelin$^1$, Christopher Jarzynski$^{2,3,4}$, 
Carlos A. Plata$^{5}$, Antonio Prados$^5$ and Emmanuel Trizac$^6$}

\address{$^1 $ Laboratoire Collisions, Agr\'egats, R\'eactivit\'e, IRSAMC, Universit\'e de Toulouse, CNRS, UPS, France  \\
$^2$ Department of Chemistry and Biochemistry, University of Maryland, College Park, MD, USA \\
$^3$ Institute for Physical Science and Technology, University of Maryland, College Park, MD, USA \\
$^4$ Department of Physics, University of Maryland, College Park, MD, USA \\
$^5$ F\'isica Te\'orica, Universidad de Sevilla, Apartado de Correos 1065, E-41080 Sevilla, Spain \\
$^6$ Universit\'e Paris-Saclay, CNRS, LPTMS, 91405, Orsay, France}
\vspace{10pt}
\begin{indented}
\item[]November 2022
\end{indented}

\begin{abstract}
   Stochastic thermodynamics lays down a broad framework to revisit the venerable concepts of heat, work and entropy production for individual stochastic trajectories of mesoscopic systems. Remarkably, this approach, relying on stochastic equations of motion, introduces time into the description of thermodynamic processes---which opens the way to fine control them. As a result, the field of finite-time thermodynamics of mesoscopic systems has blossomed. In this article, after introducing a few concepts of control for isolated mechanical systems evolving according to deterministic equations of motion, we review the different strategies that have been developed to realize finite-time state-to-state transformations in both over and underdamped regimes, by the proper design of time-dependent control parameters/driving. The systems under study are stochastic, epitomized by a Brownian object immersed in a fluid; they are thus strongly coupled to their environment playing the role of a reservoir. Interestingly, a few of those methods (inverse engineering, counterdiabatic driving, fast-forward) are directly inspired by their counterpart in quantum control. The review also analyzes the control through reservoir engineering. Besides the reachability of a given target state from a known initial state, the question of the optimal path is discussed. Optimality is here defined with respect to a cost function, a subject intimately related to the field of information thermodynamics and the question of speed limit. Another natural extension discussed deals with the connection between arbitrary states or non-equilibrium steady states. This field of control in stochastic thermodynamics enjoys a wealth of applications, ranging from optimal mesoscopic heat engines to population control in biological systems.\end{abstract}

%
%
%
%
%

\section{Introduction}\label{sec:intro}

Thermodynamics originated in efforts to tame the motive power of fire~\cite{1824Carnot}.
Originally concerned with notions of heat and temperature, the field was formalized during the 19th century into a set of universal principles that 
govern the properties of macroscopic systems in thermal equilibrium, as well as transformations between equilibrium states~\cite{Callen}.
Beginning in the 1860s, statistical physics revealed the microscopic roots of thermodynamics, greatly enhancing its power to predict and explain systems' material properties.
Reversible transformations have traditionally played a central role in the foundations of thermodynamics.
Such transformations occur in an idealized adiabatic (infinitely slow~\cite{Adiabatic}) limit, in which a system's dynamical behavior is essentially irrelevant.
More recent developments, however, have focused on nonequilibrium, finite-time processes, where dynamics become important.
In particular the growing field of {\it stochastic thermodynamics}~\cite{Sekimoto,2012Seifert,2021Peliti} 
extends the concepts of heat, work and entropy production to individual trajectories of microscopic systems, evolving under stochastic equations of motion.

\begin{figure}
\begin{center}       
 \includegraphics[width=\textwidth]{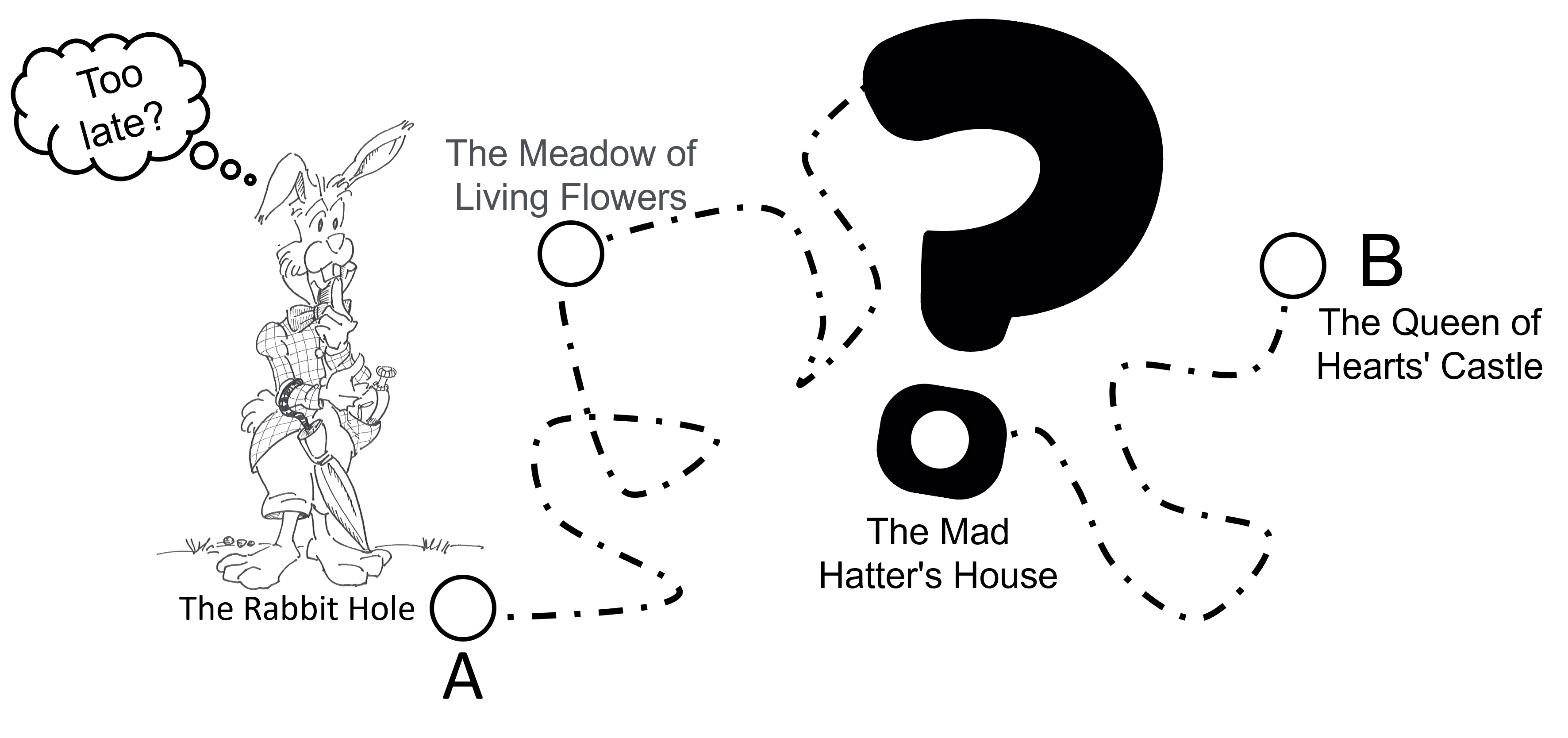}
         \end{center}
   \caption{The White Rabbit is doing his best to overcome his unpunctuality, 
   which is not an easy task in Wonderland. In this review, we study a
   related type of question, and first address the possibility of finding a path from A to B (A and B are fixed).
   When several such paths may exist, a second level of question
   deals with searching for the optimal one; for the White Rabbit, this amounts to finding the fastest.
\label{fig:rabbit}
}
\end{figure}

By introducing time into the description of thermodynamic processes, it becomes possible to formulate new questions, and in particular to investigate the control and optimization of finite-time thermodynamic transformations, see Fig.~\ref{fig:rabbit}.
Recent years have seen a surge of activity in this area.
In this review, we survey finite-time control methods that have been developed within the framework of stochastic thermodynamics.
Many of these methods (though not all!) have been inspired by developments in the control of quantum systems, particularly in the field of {\it shortcuts to adiabaticity}~\cite{2019Guery-Odelin}.
For this reason, we begin with a brief overview of quantum frameworks and features that are especially relevant for control ideas in stochastic thermodynamics. 

The goal of quantum shortcuts to adiabaticity is to steer a system to evolve from an eigenstate $\vert n(0)\rangle$ of an initial Hamiltonian $\widehat H(0)$ to the corresponding eigenstate $\vert n(t_{\rm f})\rangle$ of a final Hamiltonian $\widehat H(t_{\rm f})$. For infinitely slow driving, this is achieved automatically by virtue of the quantum adiabatic theorem~\cite{2004Griffiths}, which guarantees that the system remains in the instantaneous eigenstate $\vert n(t)\rangle$ of $\widehat H(t)$ at all times. For rapid driving, three broad approaches for achieving the above-mentioned goal have emerged: {\it inverse engineering}~\cite{2011Chen,2014Torrontegui}, {\it transitionless} or {\it counterdiabatic driving}~\cite{2003Demirplak,2005Demirplak,2009Berry}, and {\it fast-forward} methods~\cite{2010Masuda,2012Torrontegui}. 


In inverse engineering methods, instead of deducing a system's evolution under a given driving protocol (as is the customary approach in physics), one seeks to engineer a driving protocol that produces the desired evolution.  This is accomplished by exploiting the equations of motion that govern the system's dynamics. In quantum mechanics, inverse engineering methods have been applied to a variety of dynamical frameworks, including the Schr\"odinger equation, the evolution operator, the dynamical invariants, and the density matrix formalism \cite{2019Guery-Odelin}.

In the counterdiabatic approach, for a given time-dependent reference Hamiltonian $\widehat H(t)$, one seeks a Hamiltonian $\widehat H_{CD}(t)$ with the following property: if the system evolves unitarily under $\widehat H + \widehat H_{CD}$ from an initial eigenstate $\vert n(0)\rangle$, then throughout the process the system remains in the $n$'th eigenstate of $\widehat H(t)$.
In other words the system follows the adiabatic trajectory $\vert n(t)\rangle$, even when the driving is rapid.
Explicit expressions for $\widehat H_{CD}$ are given by Eqs.~\eref{eq:HCD-q} and Eq.~\eref{eq:HLCD-q} below~\cite{2003Demirplak,2005Demirplak,2009Berry,2017Patra}.
Both results steer the system exactly along the adiabatic trajectory $\vert n(t)\rangle$, hence the solution to the counterdiabatic problem is not always unique.

In the fast-forward approach~\cite{2010Masuda,2012Torrontegui}, a potential $\widehat U_{FF}(t) = U_{FF}(\widehat{\mathbf{x}},t)$ is designed such that, if the system evolves under $\widehat H + \widehat U_{FF}$ from an initial eigenstate $\vert n(0)\rangle$, then at $t=t_{\rm f}$ the system arrives at the desired final state $\vert n(t_{\rm f})\rangle$. At intermediate times the state of the system takes the form $\psi(\mathbf{x},t) = e^{iS(\mathbf{x},t)/\hbar} \langle \mathbf{x}\vert n(t) \rangle$, where $S(\mathbf{x},t)$ is real. Note that while $\psi(\mathbf{x},t)$ itself is not an eigenstate of $\widehat H(t)$, its coordinate-space probability distribution coincides with the eigenstate probability distribution: $\vert\psi(\mathbf{x},t)\vert^2 = \vert\langle \mathbf{x}\vert n(t)\rangle\vert^2$.

In the shortcuts described above, the goal is to make the system arrive rapidly at a destination it would have reached naturally had the process been carried out quasistatically. In the quantum case, the desired destination is an energy eigenstate, but as we shall see the same goal can be reformulated for classical systems governed by Hamilton's equations (Sec.~\ref{sec:isolated-systems}), and for stochastic systems evolving under overdamped and underdamped Brownian dynamics (Secs.~\ref{sec:inverse-engineering} - \ref{sec:fast-forward}).
In fact, not only is the goal the same---rapid evolution to a quasistatic outcome---but there are close similarities between the various quantum, classical and stochastic shortcuts designed to achieve this goal. 
A number of strategies for constructing shortcuts can be unified within a framework organized around the continuity equation~\cite{2017Patra}. In each case the strategy involves identifying a velocity field $v(x,t)$, or else the corresponding acceleration field $a(x,t)$, then using this field to construct the counterdiabatic or fast-forward Hamiltonian or potential---see Eqs.~\eref{eq:HLCD}, \eref{eq:uffdef}, \eref{eq:HLCD-q} and \eref{eq:UCD-stoch}.


Although the term shortcuts to adiabaticity (STA) is widely used in the context of quantum and Hamiltonian classical dynamics, for stochastic systems other expressions such as {\it engineered swift equilibration}~\cite{2016Martinez} and {\it shortcuts to isothermality}~\cite{2017Li} have been introduced.
While these terms are descriptive within particular contexts, they do not fully capture the broad scope of methods that have been developed.
In this review we will use the general terminology, \textit{swift state-to-state transformations}, and the acronym {\it \STA}, to embrace the entire catalogue of shortcutting---see \ref{app:acronyms}
for a summary of acronyms used throughout.

This review focuses primarily on systems evolving under stochastic dynamics, but we begin in Sec.~\ref{sec:isolated-systems} by describing shortcuts developed for classical mechanical systems governed by Hamiltonian dynamics. Mainly, we describe the generalisation to classical mechanics of counterdiabatic and fast-forward driving.
These methods, both in their quantum and classical versions, are based on the time manipulation of the system Hamiltonian.  Also, we briefly describe other shortcuts---for instance in the framework of the Boltzmann equation. Section~\ref{sec:systems-with-bath} is devoted to shortcuts for systems in contact with a thermal bath and thus described by  stochastic dynamics, codified for instance by the Fokker-Planck equation---focusing on 
rapid transitions between equilibrium states.  Therein, we further generalize three main STA quantum approaches (inverse engineering,
counterdiabatic method,
and fast-forward) to the stochastic framework, by time manipulating the potential energy, i.e. by applying a suitably chosen external force. In addition, we consider a new possibility of tuning the evolution of these systems: engineering the thermal environment.

Loosely speaking, it may argued that the finite-time driving associated with \STA, which involves the exploration of non-equilibrium states, requires more resources than an infinitely slow adiabatic connection---for which the system is at equilibrium for all times.
 To be concrete, for an isothermal process the work performed on the system is minimized when the process is performed infinitely slowly, and in this case the work is equal to the free energy difference between the final and initial states.
For \STA, there emerges an irreversible contribution to the work that depends on the path swept by the system. It is therefore natural to explore how to design protocols that optimally use the resources available by minimizing some ``cost'' function, e.g. the irreversible work for the isothermal connection. Also, the system parameters that become time-dependent in \STA\ must often verify certain constraints. The question of minimizing a certain cost function that is a functional of the trajectory, with constrained parameters, is the central problem of optimal control theory~\cite{1987Pontryagin,2012Liberzon}---a collection of tools that has been employed in the applied mathematics and engineering literature for a long time, but only recently in physics~\cite{2005Bechhoefer,Bechhoefer-book}.
In Sec. \ref{sec:OCT}, we analyze how these tools translate to the setting of stochastic thermodynamics, including a brief discussion of information geometry ideas and the emergence of the so-called classical speed limits. Then, in Sec. \ref{sec:beyond-equilibrium}, we extend previously discussed techniques to transitions between non-equilibrium states. Section~\ref{sec:heat_engine_and_beyond} deals with applications of stochastic shortcuts, including heat engines. Finally, perspectives and conclusions are drawn in Sec.~\ref{sec:conclusions}.

The stochastic dynamics considered in Secs.~\ref{sec:systems-with-bath}-\ref{sec:heat_engine_and_beyond} account for the effects of thermal noise in the surrounding bath.  However, the resulting randomness generally becomes negligible (that is, of relative order $N^{-1/2}$ where $N$ is the number of degrees of freedom of the system) in the macroscopic limit $N\rightarrow\infty$.  In this limit the effects of the bath can typically be modeled by deterministic equations of motion.  For this reason our review generally focuses on systems with one or a few degrees of freedom, where the role of noise can not be neglected.






\section{Shortcuts for isolated classical systems}
\label{sec:isolated-systems}

\subsection{Background and Setup}

As indicated above, the general goal of STA is to steer a system to arrive quickly at a destination that it would have reached without external assistance (so to speak) if the process were performed infinitely slowly. For quantum shortcuts, this desired destination is an energy eigenstate. For shortcuts involving isolated classical systems, the goal is most naturally framed in terms of adiabatic invariants, therefore we begin this section with a discussion of classical adiabatic invariants.

The study of these invariants traces back more than a century~\cite{1895LeCornu,1902Rayleigh} to the problem of a simple pendulum whose length $\ell$ varies with time.
If $\ell$ changes slowly, then so too do both the pendulum frequency $\nu$ 
and its energy $E$.
However, in the harmonic regime of small oscillations the ratio $E/\nu$ remains fixed in the limit of infinitely slow variation of the pendulum length.\footnote{
Note the correspondence between the classical adiabatic invariant, $E/\nu$, and the quantum number $n$: since the $n$'th eigenenergy of a harmonic oscillator is given by
$E_n = h \nu \left( n + \frac{1}{2} \right)$,
the quantum adiabatic invariance of $n$ matches the classical adiabatic invariance of $E/\nu$.} 

\begin{figure}[htp]
   \begin{center}
      \includegraphics[width=0.6\textwidth]{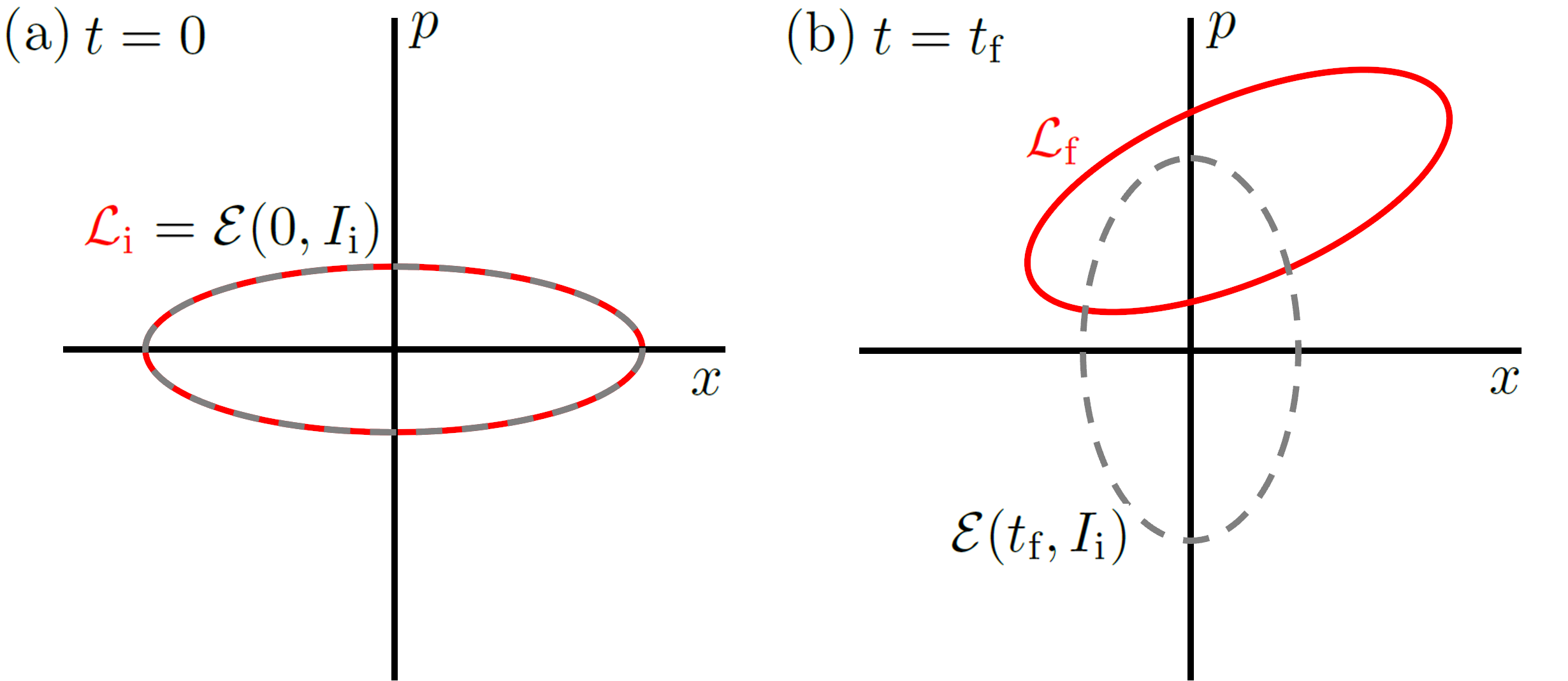}
   \end{center}
   \caption{In (a), the gray loop depicts an energy shell ${\cal E}(0,I_{\ini})$ of $H(z,0)$, and the red loop ${\cal L}_{\ini}={\cal L}(0)$ depicts initial conditions for a set of trajectories that subsequently evolve under $H(z,t)$, where $z=(x,p)$.
   The two loops are identical at $t=0$, as shown.  In (b), the gray loop depicts the energy shell ${\cal E}(t_{\fin},I_{\ini})$ of $H(z,t_{\fin})$, while the red loop ${\cal L}_{\fin} = {\cal L}(t_{\fin})$ depicts the final conditions for the set of trajectories.  In the adiabatic limit the two would coincide, ${\cal L}_{\fin}={\cal E}(t_{\fin},I_{\ini})$, as the action $I$ becomes invariant.  For any choice of $t_{\fin}$, the phase space volume enclosed by both loops in (b) is equal to that enclosed by the loops in (a), by Liouville's theorem, even though the loop ${\cal L}_{\fin}$ is not itself an energy shell.
   }
      \label{fig:shellsAndLoops}
\end{figure}

More generally, consider a classical particle of mass $m$ in one degree of freedom, described by a Hamiltonian
\be
\label{eq:Hzt}
H(z,t) = \frac{p^2}{2m} + U(x,t) ,
\ee
where $z=(x,p)$ denotes a point in phase space.
For any fixed value of $t$, we take $U(x,t)$ to be a confining potential, with closed (periodic) orbits in phase space.
We assume that $U(x,t)$ has a single minimum whose location may depend on $t$;\footnote{For potentials with two or more local minima, the adiabatic invariance of the action breaks down when phase space separatrices are crossed~\cite{1986Tennyson}
To exclude this complicating feature, we assume only a single minimum.};
that $U(x,t)$ varies with time only during the interval $t_{\ini} = 0\le t\le t_{\fin}$;
and that the time-dependence of the potential is turned on and off smoothly at $t=0$ and $t=t_{\fin}$, respectively---more precisely, $U(x,t)$ is twice differentiable with respect to time.
Under these conditions, if we treat $t$ as fixed parameter then every trajectory evolving under $H(z,t)$ is closed (i.e.\ periodic).
If we instead let $t$ denote the running time then the {\it action}
\be
\label{eq:defI}
I(E,t) = \oint_E  dx \, p 
\ee
is an adiabatic invariant~\cite{1980Goldstein}.
The right side indicates a clockwise line integral around the {\it energy shell} $E$, that is the level set $H(z,t) = E$, which forms a closed loop in phase space (see Fig.~\ref{fig:shellsAndLoops}).
We will use the notation ${\cal E}(t,I)$ to denote the energy shell of $H(z,t)$ whose action is $I$.

Imagine a classical trajectory $z(t)$ that evolves under Hamilton's equations of motion as $H(z,t)$ is varied extremely slowly over a long time interval $0\le t\le t_{\fin}$.
Let $E(t) \equiv H(z(t),t)$ denote the slowly evolving energy of this trajectory.
In stating that the action $I$ is an adiabatic invariant, we mean that in the limit of infinitely slow driving its value remains constant along this trajectory:
\be
\label{eq:adiabaticInvarianceAtAllTimes}
I(E(t),t) = I(E(0),0) \equiv I_{\ini} \quad,\quad 0\le t \le t_{\fin}
\ee
even though, in general, $E(t) \ne E(0)$.
For a harmonic oscillator with time-dependent frequency $\nu(t)$, the action defined by Eq.~\eref{eq:defI} is equal to $I(E,t) = E/\nu$, in agreement with the discussion of the simple pendulum, above.

The integral in Eq.~\eref{eq:defI} gives the phase space volume~\footnote{By convention we use the term ``volume'' rather than ``area'', even though phase space is two-dimensional.} enclosed by the energy shell, hence we can equally well write
\be
\label{eq:defI-alt}
I(E,t) = \int dz \, \theta\left[ E - H(z,t) \right]
\ee
where $\theta(\cdot)$ is the unit step function, and $\int dz = \int dx \int dp$ denotes integration over phase space.
Thus the adiabatic invariance of the action can be described as follows:
if a trajectory is initially located on an energy shell ${\cal E}(0,I_{\ini})$ that encloses phase space volume $I_{\ini}$, then for any $t \in [0,t_{\fin}]$ the trajectory will be located on the energy shell ${\cal E}(t,I_{\ini})$, which encloses the same amount of phase space.

Classical shortcuts to adiabaticity (\STA) are concerned with the situation in which the variation of the potential $H(z,t)$ is not slow.
Consider again the Hamiltonian given by Eq.~\eref{eq:Hzt}, only now imagine that the interval $[0,t_{\fin}]$ over which $H$ varies with time is not particularly long---indeed, it can be arbitrarily short, though finite.
In this situation, the action $I$ is generally not invariant:
\be
I(E(t),t) \ne I(E(0),0) \, .
\ee
We will view the non-invariance of the action as a deficiency, to be corrected using tools similar to those devised for quantum \STA / STA.

In what follows, the term {\it slow driving} refers to the adiabatic limit, while {\it fast driving} denotes non-adiabatic time-dependence of $H(z,t)$.

While the discussion above has focused on the evolution of a single trajectory $z(t)$, the invariance of the action for slow driving and the breaking of that invariance for rapid driving are conveniently visualized in terms of closed {\it loops} evolving in phase space; see Fig.~\ref{fig:shellsAndLoops}.
Imagine, at $t=0$, a collection of infinitely many initial conditions distributed over a single energy shell of the initial Hamiltonian, with action $I_{\ini}$.
These initial conditions define a loop ${\cal L}_{\ini}$ that coincides with the energy shell, ${\cal L}_{\ini} = {\cal E}(0,I_{\ini})$, as depicted schematically in Fig.~\ref{fig:shellsAndLoops}(a).
From each of these initial conditions a trajectory $z(t)$ evolves under $H(z,t)$.
At any later time, $t>0$, a snapshot of these trajectories defines a new closed loop ${\cal L}(t)$.
Under slow driving, Eq.~\eref{eq:adiabaticInvarianceAtAllTimes} implies that the loop ${\cal L}(t)$ ``clings'' at all time to the instantaneous shell whose action is $I_{\ini}$, that is, ${\cal L}(t) = {\cal E}(t,I_{\ini})$ for all $t \in [0,t_{\fin}]$.
In particular, the initial loop ${\cal L}_{\ini}={\cal L}(0)$ is mapped onto a final loop ${\cal L}_{\fin} \equiv {\cal L}({t_{\fin}}) = {\cal E}(t_{\fin},I_{\ini})$ that coincides with an energy shell of the final Hamiltonian (see the gray loop in Fig.~\ref{fig:shellsAndLoops}(b)).
In terms of the initial and final energies of the trajectory, we have
\be
\label{eq:I0I1}
I(E_{\ini},0) = I(E_{\fin},t_{\fin}) = I_{\ini} \, ,
\ee
where $E_{\ini}=E(0)$ and $E_{\fin}=E(t_{\fin})$.
Under fast driving, by contrast, the loop ${\cal L}(t)$ strays away from the energy shell ${\cal E}(t,I_{\ini})$ as illustrated for $t=t_{\fin}$ in Fig.~\ref{fig:shellsAndLoops}(b).

The goal of classical \STA\ can be stated as follows: for a rapidly driven Hamiltonian $H(z,t)$ and an action $I_{\ini}$, devise a strategy that evolves the loop ${\cal L}_{\ini} = {\cal E}(0,I_{\ini})$ to the loop ${\cal L}_{\fin} = {\cal E}(t_{\fin},I_{\ini})$, under Hamiltonian evolution.
This is analogous to the situation in quantum \STA, where the goal is to use unitary evolution to evolve a wavefunction from an eigenstate of an initial Hamiltonian to the corresponding eigenstate of the final Hamiltonian.

In the classical case, we attempt to meet this goal by designing an auxiliary term $H_{\rm aux}(z,t)$ such that the desired evolution is generated by the  Hamiltonian
\be
\label{eq:hamsum}
H_{\rm \STA}(z,t) = H(z,t) + H_{\rm aux}(z,t)  \, .
\ee
The auxiliary term effectively steers the evolving loop ${\cal L}(t)$ to the desired target, namely the final energy shell ${\cal E}(t_{\fin},I_{\ini})$.
While the dynamics are generated by $H_{\rm \STA}(z,t)$, we emphasize that the energy shells ${\cal E}(t,I)$ are always defined with respect to the original Hamiltonian $H(z,t)$.

We now identify three different flavors of this problem, which differ from one another in how ambitiously the above-mentioned goal is addressed.
\begin{itemize}
\item  In one version of the problem, $H_{\rm aux}$ is independent of the choice of action $I_{\ini}$, and the action remains invariant throughout the entire process, i.e.\ Eq.~\eref{eq:adiabaticInvarianceAtAllTimes} is satisfied at all times and for any choice of $E_{\ini}=E(0)$.
This represents the strongest version of the problem, and we will refer to it as {\it global counterdiabatic (GCD)} driving.
\item In a somewhat more relaxed version, we continue to insist that the action be invariant throughout the process (Eq.~\eref{eq:adiabaticInvarianceAtAllTimes}), but now we allow $H_{\rm aux}$ to depend on the choice of $I_{\ini}$, equivalently on the choice of $E_{\ini}$; we will call this {\it local counterdiabatic (LCD)} driving.
\item In the most relaxed version of the problem, we allow $H_{\rm aux}$ to depend on the choice of $I_{\ini}$ (as with LCD driving), and we further allow the invariance of the action to be broken at intermediate times $0<t<t_{\fin}$, insisting only that $I(E_{\fin},t_{\fin})=I(E_{\ini},0)$ (Eq.~\eref{eq:I0I1}).
We refer to this version as {\it fast-forward (FF)} driving.
\end{itemize}

The designations global and local reflect the distinction that in one case (GCD) a single $H_{\rm aux}$ must succeed for all energy shells, i.e.\ globally, while in the other (LCD) we are free to design $H_{\rm aux}$ based on the energy shell under consideration, i.e.\ locally.
The terms counterdiabatic and fast-forward are taken from the literature on quantum shortcuts---the former signifies the suppression of non-adiabatic excitations~\cite{2003Demirplak}, while the latter evokes a mechanism for rapidly arriving at a desired final destination, regardless of the intermediate path taken to get there~\cite{2010Masuda}.

All three versions of the classical \STA\ problem---as defined above, for one degree of freedom---have been solved, in the sense that in each version an explicit recipe has been devised for constructing an auxiliary Hamiltonian that achieves the desired steering of trajectories.
As described in greater detail below, for GCD driving the auxiliary Hamiltonian is generally a complicated, non-linear function of both position ($x$) and momentum ($p$).
For LCD driving, the auxiliary Hamiltonian is simpler, taking the form $H_{\rm aux}(z,t) = pv(x,t)$, with $v(x,t)$ given by Eq.~\eref{eq:vdef} below.
Finally, for FF driving, the auxiliary Hamiltonian does not depend on momentum at all, in other words the desired goal can be satisfied by adding a time dependent potential energy function $U_{\rm FF}(x,t)$ to the original Hamiltonian.
Perhaps not surprisingly, as we relax our demands on the performance of the auxiliary Hamiltonian, from global to local to fast-forward, the form of $H_{\rm aux}$ becomes simpler, at least in its dependence on momentum: $H_{\rm aux}(x,p,t) \rightarrow pv(x,t) \rightarrow U_{\rm FF}(x,t)$.

LCD and FF driving are closely related: the function $v(x,t)$ that appears in the LCD auxiliary Hamiltonian (Eq.~\eref{eq:HLCD}) is also used to construct the fast-forward potential $U_{\rm FF}$, as elaborated in Sec.~\ref{subsec:recipes}.
In general, GCD driving is not closely related to the other two.
However, for a particular class of driving protocols that go by the name of {\it scale-invariant driving}~\cite{2014Deffner}, GCD and LCD driving are identical, giving the same $H_{\rm aux}(z,t)$ (see Eq.~\eref{eq:HCD-si} below).
Scale-invariant protocols are those for which the potential $U(x,t)$ in Eq.~\eref{eq:Hzt} has the form~\cite{2013delCampo,2014Deffner}
\be
\label{eq:scaleInvart}
U(x,t) = \frac{1}{\sigma^2} U_0 \left( \frac{x-\mu}{\sigma} \right)
\ee
where $\sigma = \sigma(t)$ and $\mu = \mu(t)$ are functions of time.
By varying $\mu$ we translate the potential, while varying $\sigma$ stretches or squeezes the potential along the $x$ axis, and rescales its magnitude, without otherwise altering its profile.
Potentials of this form give rise to convenient scaling properties of both the classical dynamics and the quantum energy eigenstates~\cite{2014Deffner}. 
Power-law potentials $U(x,t) = \alpha \, [x/L(t)]^c$, with $c=2, 4, \cdots$, offer an illustrative example of scale-invariant driving~\cite{2013Jarzynski}.
The harmonic oscillator with time-dependent stiffness ($c=2$) and the particle-in-a-box, with time-dependent box length ($c\rightarrow\infty$) belong to this class.

We note in passing that if $U(x,t)$ is scale-invariant \eref{eq:scaleInvart} for some $\sigma(t)$ and $\mu(t)$, then so is $U(x,t)+b/(x-\mu)^2$ for any real constant $b$.\footnote{
This follows trivially, since $b/y^2 = (1/\sigma^2) \, b/(y/\sigma)^2$, with $y=x-\mu$.}
Potentials of this form will arise in Sec.~\ref{subsec:boltzmann}.

\subsection{Recipes for classical shortcuts to adiabaticity}
\label{subsec:recipes}

We now describe how to construct $H_{\rm aux}(z,t)$ for the three situations just outlined.
For the derivations of these methods and further details, we refer to the original papers, cited below.

In Secs.~\ref{subsubsec:gcd}-\ref{subsubsec:ff}, we will assume that $U(x,t)$ is twice differentiable with respect to time (see comments after Eq.~\eref{eq:Hzt}), hence both $U$ and $\partial_t U$ are continuous functions of time.
Since $\partial_t U=0$ outside the interval $[0,t_{\fin}]$, this assumption implies the boundary conditions
\begin{equation}
\label{eq:bc}
    \frac{\partial U}{\partial t}(x,0^+) = \frac{\partial U}{\partial t}(x,t_{\fin}^-) = 0 .
\end{equation}
In other words the time-dependence of $U$ is turned on and off smoothly rather than abruptly.
In Sec.~\ref{subsubsec:bc} we will briefly consider the implications of relaxing this assumption.

\subsubsection{Global counterdiabatic driving}
\label{subsubsec:gcd}

Classical GCD mirrors the quantum approach developed by Demirplak and Rice~\cite{2003Demirplak} and Berry~\cite{2009Berry}.
In the quantum case, the auxiliary or {\it counterdiabatic} Hamiltonian is given as a sum over energy eigenstates of the original Hamiltonian $\widehat H(t)$:
\be
\label{eq:HCD-q}
\widehat H_{\rm GCD}(t) = i\hbar \sum_m \left( \vert \dot m\rangle \langle m \vert  -  \langle m \vert \dot m\rangle \vert m\rangle \langle m\vert \right) \, .
\ee
Here $\vert m\rangle = \vert m(t)\rangle$ denotes the $m$'th eigenstate of $\widehat H(t)$, and $\vert \dot m\rangle = \partial_t \vert m(t)\rangle$ is its time derivative.

We are interested in constructing the classical counterpart of $\widehat H_{\rm GCD}(t)$.
While Eq.~\eref{eq:HCD-q} may not seem well-suited to this end, the quantum operator defined (uniquely) by that equation can equivalently be defined by~\cite{2013Jarzynski}:
\numparts
\ba
\label{eq:HCD-q2a}
\left[ \widehat H_{\rm GCD},\widehat H \right] &=& i\hbar \left( \partial_t\widehat H - {\rm diag} \, \partial_t\widehat H \right) \\
\label{eq:HCD-q2b}
\langle n \vert \widehat H_{\rm GCD} \vert n\rangle &=& 0 \qquad \forall \, n
\ea
\endnumparts
where ${\rm diag} \, \widehat A \equiv \sum_m \vert m\rangle \langle m\vert\widehat A\vert m\rangle \langle m\vert$.
Defining $\widehat H_{\rm GCD}$ in this manner is convenient, as the correspondence between the quantum commutator $[ \widehat A,\widehat B ]$ and the classical Poisson bracket $\{ A,B \}$,
together with the correspondence between quantum energy eigenstates and classical energy shells, suggest a natural classical analogue:
\numparts
\ba
\label{eq:HCD-c2a}
\left\{ H_{\rm GCD},H \right\} &=& \partial_t H - \langle \partial_t H\rangle_H \\
\label{eq:HCD-c2b}
\langle H_{\rm GCD} \rangle_E &=& 0 \qquad \forall \, E
\ea
\endnumparts
The notation
\begin{equation}
\langle A\rangle_E = \frac{ \int dz \, \delta\left(E-H\right) A(z) }{ \int dz \, \delta\left(E-H\right) }
\end{equation}
denotes a microcanonical average of an observable $A(z)$ over the energy shell $E$ of $H(z,t)$.
The left side of Eq.~\eref{eq:HCD-c2a} is evaluated at a phase point $z$, and the notation $\langle \partial_t H\rangle_H$ on the right indicates that the average is taken over the energy shell containing $z$;
see Ref.~\cite{2013Jarzynski} for further details, as well as Ref.~\cite{2013Deng} where the same result is derived by means of classical generating functions, and Ref.~\cite{2017Kolodrubetz}, where $\widehat{H}_{\rm GCD}$ and $H_{\rm GCD}$ are framed as {\it adiabatic gauge potentials}.
Yet another interesting approach is taken in Ref.~\cite{2017Okuyama}, where $H_{\rm GCD}$ is constructed from the dispersionless Korteweg-de Vries hierarchy, building on earlier work in the quantum context~\cite{2016Okuyama}.

We have introduced Eqs.~\eref{eq:HCD-c2a} and \eref{eq:HCD-c2b}  as the classical analogue of Eqs.~\eref{eq:HCD-q2a} and \eref{eq:HCD-q2b}.
It can be verified directly from classical analysis~\cite{2013Jarzynski} that the function $H_{\rm GCD}(z,t)$ defined by Eqs.~\eref{eq:HCD-c2a} and \eref{eq:HCD-c2a} has the counterdiabatic property we seek: along a trajectory $z_G(t)$ evolving under $H + H_{\rm GCD}$, the action is preserved exactly:
\be
\label{eq:GCDtraj}
\frac{d}{dt} I\left[H(z_G(t),t),t\right] = 0
\ee
Hence the solution to a quantum problem, Eqs.~\eref{eq:HCD-q2a} and \eref{eq:HCD-q2b}, combined with semiclassical reasoning, has yielded the exact solution to the analogous classical problem, Eqs.~\eref{eq:HCD-c2a} and \eref{eq:HCD-c2b}.

To translate Eqs.~\eref{eq:HCD-c2a} and \eref{eq:HCD-c2b} into an explicit function $H_{\rm GCD}(z,t)$, note that Eq.~\eref{eq:HCD-c2a} implies the following relation, for any points $z_a$ and $z_b$ on an energy shell $E$ of $H(z,t)$:
\be
\label{eq:zazb}
H_{\rm GCD}(z_b,t) - H_{\rm GCD}(z_a,t) = \int_a^b ds \, \left[ \partial_t H(z(s),t) - \langle \partial_t H\rangle_E \right] \, ,
\ee
where $t$ is treated here as a parameter, and $z(s)$ is a trajectory of energy $E$ that evolves with time $s$ under fixed $H(z,t)$, from $z(a)=z_a$ to $z(b)=z_b$.
This relation determines $H_{\rm GCD}(z,t)$ for all points on the energy shell $E$, up to an additive constant whose value is in turn determined by Eq.~\eref{eq:HCD-c2b}.
This additive constant can depend on time, but it has no dynamical relevance.

Equations \eref{eq:HCD-c2a} and \eref{eq:HCD-c2b} thus provides an explicit recipe for constructing $H_{\rm GCD}$.
For scale-invariant driving (see Eq.~\eref{eq:scaleInvart}) this recipe leads to the particularly simple expression~\cite{2014Deffner}
\be
\label{eq:HCD-si}
H_{\rm GCD}(z,t) \, = \,  \frac{\dot\sigma}{\sigma} (x-\mu) p \,+\, 
\dot \mu \, p
\ee
where the dots denote derivatives with respect to time.
In the special cases of the harmonic oscillator and the particle-in-a-box, Eq.~\eref{eq:HCD-si} reproduces results derived previously by other means~\cite{2010Muga,2013diMartino,2013Jarzynski}.

For non-scale-invariant driving, it is difficult to obtain closed-form expressions for $H_{\rm GCD}$, and one must resort to solving Eqs.~\eref{eq:HCD-c2a} and \eref{eq:HCD-c2b} or \eref{eq:zazb} numerically.
An exception is the case of the ``tilted piston'', which involves a particle inside a one-dimensional box with hard walls separated by a distance $L$, with a potential inside the box that is linear in $x$, with slope $\alpha$.
Either $L$ or $\alpha$, or both, can be made time-dependent.
Even for this relatively modest extension of the particle-in-a-box, the exact expression for $H_{\rm GCD}$ is complicated and non-linear in $p$~\cite{2017Patra-a}.

Finally, although Eqs.~\eref{eq:HCD-c2a} \eref{eq:HCD-c2b} uniquely defines $H_{\rm GCD}$, Eq.~\eref{eq:GCDtraj} remains satisfied for a trajectory $z_G(t)$ evolving under $H+H_{\rm GCD}+f(H)$, for any differentiable function $f(\cdot)$.
As a result, to construct a globally counterdiabatic auxiliary Hamiltonian we need only satisfy Eq.~\eref{eq:HCD-c2a} and not necessarily Eq.~\eref{eq:HCD-c2b} -- imposing the latter (i.e.\ setting $f=0$) amounts to a kind of gauge choice.
Analogous comments apply in the quantum case to Eqs.~\eref{eq:HCD-q2a} and \eref{eq:HCD-q2b}.

\subsubsection{Local counterdiabatic driving}
\label{subsubsec:lcd}

In the case of local counterdiabatic driving, we select a value of action, $I_{\ini}$.
For any $t\in[0,t_{\fin}]$, ${\cal E}(t,I_{\ini})$ is the energy shell of $H(z,t)$ with the same action $I_{\ini}$; we refer to this shell as the {\it adiabatic energy shell} for our choice of $I_{\ini}$.
This energy shell represents the desired evolution we wish to generate: if a trajectory begins on the energy shell ${\cal E}(0,I_{\ini})$ at $t=0$, we want to guarantee that it will be found on the shell ${\cal E}(t,I_{\ini})$ at all $t\in [0,t_{\fin}]$.
Unlike with GCD, this goal can be accomplished with an auxiliary Hamiltonian that is linear in momentum, $p$.

The LCD auxiliary Hamiltonian involves a function $v(x,t)$ that is constructed as follows.
At time $t$, the shell ${\cal E}(t,I_{\ini})$ forms a loop in phase space, with left and right turning points, $x_L(t)$ and $x_R(t)$.
Let
\be
\bar p_\pm(x,t) = \pm \left[ 2m ({\cal E}(t,I_{\ini}) - U(x,t)) \right]^{1/2}
\ee
denote the upper and lower branches of this loop, and let
\be
{\cal S}(x,t) = 2 \int_{x_L(t)}^x dx^\prime \, \bar p_+(x^\prime,t)
\ee
denote the volume of phase space enclosed by ${\cal E}(t,I_{\ini})$ between the left turning point and a vertical line located at position $x \in [x_L,x_R]$.
(To avoid confusion, we stress that ${\cal S}$ is {\it not} Hamilton's principal function, which appears in the Hamilton-Jacobi equation~\cite{1980Goldstein}.)
Since ${\cal S}$ increases monotonically with $x$ we can invert it, writing $x=x({\cal S},t)$, with ${\cal S}\in[0,I_{\ini}]$.
We then define a velocity field
\be
\label{eq:vdef}
v(x,t) = \frac{\partial}{\partial t}x({\cal S},t)
= - \frac{\partial{\cal S}/\partial t}{\partial{\cal S}/\partial x}
\ee
using the cyclic identity of partial derivatives.
From Eq.~\eref{eq:bc} it follows that $v(x,0)=v(x,t_{\fin})= 0$.
The LCD auxiliary Hamiltonian is then given by
\be
\label{eq:HLCD}
H_{\rm LCD}(x,p,t) = pv(x,t).
\ee

Adding $H_{\rm LCD}$ to the original Hamiltonian $H$ (as per Eq.~\eref{eq:hamsum}) gives us
\be
\label{eq:hSTAlcd}
H_{\rm \STA}(z,t) = \frac{p^2}{2m} + U(x,t) + pv(x,t)
\ee
which generates the equations of motion
\be
\label{eq:hameqlcd}
\dot x = \frac{p}{m} + v(x,t) \quad,\quad
\dot p = -\frac{\partial U}{\partial x} -p \frac{\partial v}{\partial x}(x,t) .
\ee
Under these equations, the action is conserved exactly for any trajectory $z_L(t)$ (where $L$ is short for $LCD$) launched from the initial adiabatic energy shell~\cite{2017Jarzynski}.
That is,
\be
\label{eq:LCDtraj}
z_L(t) \in {\cal E}(t,I_{\ini})
\ee
for all $t\in[0,t_{\fin}]$.
For such a trajectory, the first terms appearing on the right in Eq.~\eref{eq:hameqlcd} generate motion along the instantaneous energy shell, while the second terms (involving $v$) force the trajectory to remain attached to the evolving energy shell.

For scale-invariant driving, Eq.~\eref{eq:vdef} gives
\be
\label{eq:vsi}
v(x,t) = \frac{\dot\sigma}{\sigma}(x-\mu) + \dot \mu
\ee
As mentioned shortly before Eq.~\eref{eq:scaleInvart}, in this special case the GCD (Eqs.~\eref{eq:HCD-c2a} and \eref{eq:HCD-c2b}) and LCD (Eq.~\eref{eq:HLCD}) prescriptions yield the same auxiliary Hamiltonian, Eq.~\eref{eq:HCD-si}.

\subsubsection{Fast-forward driving}
\label{subsubsec:ff}

Having defined $v(x,t)$ above, it is straightforward to construct the FF auxiliary potential $U_{\rm FF}(x,t)$.
We first introduce an acceleration field
\be
\label{eq:adef}
a(x,t) = \frac{\partial v}{\partial x} v + \frac{\partial v}{\partial t} = \frac{\partial^2}{\partial t^2} x({\cal S},t) .
\ee
$U_{\rm FF}$ is then defined, up to an arbitrary function of time, by
\be
\label{eq:uffdef}
-\frac{\partial U_{\rm FF}}{\partial x} = ma(x,t) .
\ee
Adding $U_{\rm FF}$ to $H$ gives
\be
\label{eq:hSTAff}
H_{\rm \STA}(z,t) = \frac{p^2}{2m} + U(x,t) + U_{\rm FF}(x,t)
\ee
which generates the equations of motion
\be
\label{eq:hameqff}
\dot x = \frac{p}{m} \quad,\quad
\dot p = -\frac{\partial U}{\partial x} +ma(x,t) .
\ee
If two trajectories are launched from identical initial conditions on the energy shell ${\cal E}(0,I_{\ini})$, and one of them, $z_L(t)$, evolves under Eq.~\eref{eq:hameqlcd}, while the other, $z_F(t)$, evolves under Eq.~\eref{eq:hameqff} (where $F$ is short for $FF$), then the two are related by~\cite{2017Jarzynski}
\be
\label{eq:FFboost}
x_F(t) = x_L(t)
\quad,\quad
p_F(t) = p_L(t) + mv(x_L(t),t) \, .
\ee
The two trajectories reunite at $t=t_{\fin}$, as $v(x,t_{\fin})=0$.
Thus the fast-forward trajectory $z_F(t)$ starts on the energy shell ${\cal E}(0,I_{\ini})$ at $t=0$, then strays from ${\cal E}(t,I_{\ini})$ at intermediate times, but ultimately arrives at the final adiabatic energy shell:
\be
\label{eq:FFtraj}
z_F(t_{\fin}) = z_L(t_{\fin}) \in {\cal E}(t_{\fin},I_{\ini})
\ee

In Sec. 2.5 of Ref.~\cite{2017Kolodrubetz}, the close relationship between LCD and FF driving is described in terms of canonical gauge transformations that map Hamiltonians of the form given by Eq.~\eref{eq:hSTAlcd} into those given by Eq.~\eref{eq:hSTAff}.
A similar approach was used in Ref.~\cite{2014Deffner}, and analogous unitary transformations were earlier introduced in the quantum context in Ref.~\cite{2012Ibanez}.

For scale-invariant driving, Eqs.~\eref{eq:vsi} - \eref{eq:uffdef} lead to the following fast-forward potential~\cite{2014Deffner}:
\be
\label{eq:siuff}
U_{\rm FF}(x,t) = -\frac{m}{2}\frac{\ddot\sigma}{\sigma}(x-\mu)^2 - m\ddot \mu x .
\ee
For the special cases of the harmonic oscillator and particle-in-a-box, equivalent results were obtained using an inverse engineering approach, in Refs.~\cite{2010Chen, 2012delCampo}.

Let us focus briefly on the harmonic oscillator, for which the original Hamiltonian is given by
\begin{equation}
\label{eq:ho}
H(z,t) = \frac{p^2}{2m} + \frac{m}{2}\omega^2(t)  x^2  ,
\end{equation}
with $\dot\omega(0)=\ddot\omega(0)=\dot\omega(t_{\fin})=\ddot\omega(t_{\fin})=0$.
The potential $U(x,t)$ in Eq.~\eref{eq:ho} can be cast into scale-invariant form (see Eq.~\eref{eq:scaleInvart}) by setting $U_0(x)=mx^2/2$, $\sigma = \omega^{-1/2}$ and $\mu=0$.
Using Eq.~\eref{eq:siuff}, we can then combine $U$ and $U_{\rm FF}$ into a single quadratic potential (see Eq.~\eref{eq:hSTAff}):
\begin{equation}
H_{\rm \STA}(z,t) = \frac{p^2}{2m} + \frac{m}{2}\Omega^2(t) x^2
\end{equation}
with
\begin{equation}
\label{eq:Omega2}
\Omega^2 = \omega^2 - \frac{3}{4}\frac{\dot\omega^2}{\omega^2} + \frac{1}{2} \frac{\ddot\omega}{\omega} \, .
\end{equation}
Rewriting the right side of Eq.~\eref{eq:Omega2} in terms of $\sigma$ rather than $\omega$, and re-ordering terms, we get
\begin{equation}
\label{eq:ermakov}
\ddot\sigma + \Omega^2\sigma = \sigma^{-3} \, .
\end{equation}
This is the Ermakov equation, which was used in Ref.~\cite{2010Chen} as the starting point for designing a \STA\ protocol for the harmonic oscillator.
When $\Omega$ is constant, this equation admits a simple solution, since $\sigma^2$ follows a harmonic oscillator 
equation, see Eq. \eref{eq:ermakovSolution} below.

\subsubsection{Boundary conditions in time}
\label{subsubsec:bc}

To this point, we have assumed that the time-dependence of $U$ is turned on and off smoothly; see Eq.~\eref{eq:bc}.
For GCD and LCD driving this assumption is unnecessary: as long as $U$ is {\it once}-differentiable with respect to time, the results described in Secs.~\ref{subsubsec:gcd} and \ref{subsubsec:lcd} remain valid.
A discontinuity in $\partial_t U$ at $t=0$ or $t=t_{\fin}$ merely implies that $H_{\rm aux}$ is turned on or off abruptly.

With FF driving the situation is subtler.
If $\partial_t U$ is discontinuous at $t=0$, then so is the velocity field $v(x,t)$, hence the acceleration field $a(x,t)$ (Eq.~\eref{eq:adef}) is undefined.
In this situation, the classical fast-forward method of Sec.~\ref{subsubsec:ff} can be salvaged if we apply the following {\it impulsive} potential at $t=0$:
\begin{equation}
    U_{imp}(x,t) = -m \, \delta(t) \int^x  dx^\prime \, v(x^\prime,0^+) .
\end{equation}
(The choice of the lower limit of integration is arbitrary.)
This potential causes a trajectory with initial phase space conditions $(x,p)$ at $t=0^-$ to jump suddenly to $(x,p+mv(x,0^+))$ at $t=0^+$.
If $\partial_t U$ is discontinuous at $t=t_{\fin}$, then another impulse is required:
\begin{equation}
    U_{imp}(x,t) = +m \, \delta(t-t_{\fin}) \int^x  dx^\prime \, v(x^\prime,t_{\fin}^-) .
\end{equation}
Once these impulses are included, the FF driving works as described in Sec.~\ref{subsubsec:ff}.
In particular all points located on the initial energy shell ${\cal E}(0,I_{\ini})$ at $t=0^-$ evolve to points on the final energy shell ${\cal E}(t_{\fin},I_{\ini})$ at $t=t_{\fin}^+$.\footnote{If $\partial_t U$ is discontinuous at intermediate times $t_1, t_2, \cdots$, then similar impulses at those times can ``rescue'' the fast-forward method.}

For an illustration of the effects of such impulses in the context of the particle-in-a-box under fast-forward driving, see Sec.\ III.A of Ref.~\cite{2014Deffner}.

\subsubsection{Beyond one degree of freedom}
\label{subsubsec:beyond1dof}

It is natural to ask whether the classical GCD, LCD and FF recipes discussed above can be generalized to systems with $n>1$ degrees of freedom.
This question has largely remained unexamined in the literature.

When $n>1$, the definition of the adiabatic invariant itself depends on the dynamics generated by the system Hamiltonian $H(\mathbf{x},\mathbf{p},t)$.
If the dynamics are integrable at all values of $t$, then the phase space coordinates can be written in terms of action-angle variables $(\mathbf{I},\mathbf{w})$, and the action variables $(I_1,\cdots I_n)$ are adiabatic invariants ~\cite{1980Goldstein}.
It is plausible that the methods described above could be applied to construct a separate auxiliary Hamiltonian $H_{\rm aux}^k$ for each action-angle pair $(I_k,w_k)$, such that evolution under $H+\sum_k H_{\rm aux}^k$ would preserve the value of each invariant.

If the dynamics are ergodic over the energy shell at all values of $t$, then the phase space volume
\begin{equation}
    \Omega(E,t) = \int d^n\mathbf{x} \int d^n\mathbf{p} \,\, \theta\left[ E - H(\mathbf{x},\mathbf{p},t) \right]
\end{equation}
is the sole adiabatic invariant~\cite{Hertz1910a,Hertz1910b,Anosov1960,Kasuga1961a,Kasuga1961b,Kasuga1961c,Ott1979,Lochak1988}. Note that the phase space volume is directly related to the Gibbs definition of entropy in the microcanonical ensemble.
In the special case when the ergodic Hamiltonians $H(\mathbf{x},\mathbf{p},t)$, with $t\in[0,t_{\fin}]$, constitute a {\it canonical family},\footnote{$H(\mathbf{x},\mathbf{p},t)$ is a canonical family if $H(t_1)$ can be mapped to $H(t_2)$ by a canonical transformation, for any $t_1, t_2 \in [0,t_{\fin}]$~\cite{1994Robbins}.} the $n$-dimensional analogue of Eqs.~\eref{eq:HCD-c2a} and \eref{eq:HCD-c2b} is known to have a solution~\cite{1995Jarzynski}, suggesting that $H_{\rm GCD}$ can be constructed for canonical families of ergodic Hamiltonians.
However, it is not clear how to extend this approach to the typical situation in which $H(\mathbf{x},\mathbf{p},t)$ is not a canonical family.

Finally, integrable and ergodic Hamiltonians are themselves somewhat special: a generic classical Hamiltonian in $n>1$ degrees of freedom has a mixed phase space, with some trajectories evolving regularly and others irregularly (chaotically)~\cite{1993Ott}.
Adiabatic invariants for such systems are not as unambiguously defined as for integrable or ergodic Hamiltonians, hence the very notion of what constitutes a shortcut to adiabaticity becomes murky.

\subsection{Classical shortcuts and microcanonical ensembles}

The discussion in Sec.~\ref{subsec:recipes} focused on the evolution of individual trajectories (Eqs.~\eref{eq:GCDtraj}, \eref{eq:LCDtraj}, \eref{eq:FFtraj}).
It is also useful to analyze this problem at the statistical level.

In this section, we consider a microcanonical ensemble of initial conditions on the energy shell ${\cal E}(0,I_{\ini})$.
This ensemble can be understood as a probability density that is distributed over the loop ${\cal L}_{\ini}$ shown in Fig.~\ref{fig:shellsAndLoops}(a).
Specifically, if we use the angle $w \in [0,2\pi)$ of action-angle variables $(I,w)$~\cite{1980Goldstein} to label points around the loop, then the microcanonical distribution is uniform in $w$.
We emphasize that the microcanonical distribution represents a particular choice of initial conditions for the ensemble; it does not arise from coupling with a thermal bath.
We now describe what happens to this ensemble, as trajectories evolve from these initial conditions under global counterdiabatic, local counterdiabatic, and fast-forward dynamics.

In the GCD case, the ensemble of trajectories clings to the adiabatic energy shell ${\cal E}(t,I_{\ini})$ during the process, as the action is preserved (Eq.~\eref{eq:GCDtraj}).
Moreover the ensemble remains microcanonical at all times: for any $t\in[0,t_{\fin}]$, a snapshot of the ensemble of trajectories would show them to be distributed uniformly, with respect to the angle variable $w$, on the adiabatic energy shell.

For LCD driving, the ensemble of trajectories also clings to the adiabatic energy shell (Eq.~\eref{eq:LCDtraj}) only now the distribution does not necessarily remain microcanonical.
In particular, a snapshot at $t=t_{\fin}$ would reveal a collection of final conditions that are distributed non-uniformly with respect to $w$ -- see Fig. 3c of Ref.~\cite{2017Jarzynski} for an illustration.

The result that GCD driving preserves the microcanonical distribution while LCD driving (in general) does not, can be traced back to the fact that in the former case the auxiliary Hamiltonian is the same for all $I_{\ini}$, whereas in the latter case $H_{\rm aux}$ generally depends on $I_{\ini}$,
as discussed in Ref.~\cite{2017Jarzynski}, Appendix D.

For FF driving, the trajectories depart from the adiabatic energy shell at intermediate times but return to that shell at the final time.
From Eq.~\eref{eq:FFtraj} it follows that the final conditions are generally distributed non-uniformly (i.e.\ non-microcanonically), as they coincide with the final conditions achieved under LCD driving.

At intermediate times the LCD and FF phase space distributions differ.
However, the projections of these distributions onto the $x$-axis are identical, by Eq.~\eref{eq:FFboost}.
An analogous situation holds in the quantum case: the LCD and FF wavefunctions differ for $t \in (0,t_{\fin})$, but their $x$-space probability distributions coincide, $\vert\psi_{FF}(x,t)\vert^2 = \vert\psi_{LCD}(x,t)\vert^2 = \vert\langle x\vert n(t) \rangle\vert^2$.\footnote{
This conclusion follows by comparing Eqs.(6) and (22) of Ref.~\cite{2017Patra}, which give the exact time-dependent wavefunctions for LCD and FF dynamics, respectively.
}

Under scale-invariant driving, GCD and LCD trajectories are identical, as mentioned earlier.
In this special situation, all three flavors of shortcuts (GCD, LCD, FF) map an initial microcanonical distribution to a final microcanonical distribution.

\subsection{Shortcuts in classical kinetic theory}
\label{subsec:boltzmann}

We now discuss a situation in which classical shortcuts arise in the context of kinetic theory.

The discussion so far has focused on a single particle in a one-dimensional potential, Eq.~\eref{eq:Hzt}.
The situation becomes more complicated when dealing with $N>1$ mutually interacting particles.
Under appropriate conditions, however, \STA-inspired tools have found interesting applications to many-body, interacting systems.

Ref.~\cite{2014Guery-Odelin} considers a dilute gas of identical particles of mass $m$, governed by the Boltzmann equation
\begin{equation}
\label{eq:Boltzmann}
\frac{\partial f}{\partial t} + \mathbf{v}\cdot\nabla_{\mathbf{r}} f + \frac{1}{m}\mathbf{F}(\mathbf{r},t)\cdot\nabla_{\mathbf{v}} f = I_{\rm coll}[\mathbf{v}\vert f,f]
\end{equation}
where $f(\mathbf{r},\mathbf{v},t)$ is the single-particle density, $\mathbf{r}$ and $\mathbf{v}$ denote three-dimensional position and velocity, $\mathbf{F}=-\nabla_{\mathbf{r}}U(\mathbf{r},t)$ is a conservative force, and $I_{\rm coll}$ is a term that models the effect of two-body collisions.
When $I_{\rm coll}=0$ we get the collisionless equation
\begin{equation}
\label{eq:collisionless}
\partial_t f + \mathbf{v}\cdot\nabla_{\mathbf{r}} f + \frac{\mathbf{F}}{m}\cdot\nabla_{\mathbf{v}} f = 0 \, ,
\end{equation}
describing a gas of non-interacting particles.
If the potential $U$ is time-independent, then under Eq.~\eref{eq:Boltzmann} the gas generically evolves to a canonical distribution, by Boltmann's H theorem~\cite{1988Cercignani,2010Kremer}.
However, Boltzmann realized already in the 1870's that when $U\propto r^2$ (with $r=|\mathbf{r}|$), 
oscillatory ``breathing mode'' solutions of Eq.~\eref{eq:Boltzmann} also exist.

It is known that $I_{\rm coll}=0$ when $f$ has the form
\begin{equation}
\label{eq:nullifyCollisions}
f(\mathbf{r},\mathbf{v},t) = \exp\left( -\alpha-\eta v^2 - \boldsymbol{\gamma}\cdot\mathbf{v}\right)
\end{equation}
where $\alpha$, $\eta$ and $\boldsymbol{\gamma}$ are arbitrary functions of position and time~\cite{2014Guery-Odelin}.
In other words the distribution $f$ given by Eq.~\eref{eq:nullifyCollisions} belongs to the kernel of $I$.
Thus any solution of the collisionless equation (Eq.~\eref{eq:collisionless}) that has the form given by Eq.~\eref{eq:nullifyCollisions}, is also a solution of the Boltzmann equation (Eq.~\eref{eq:Boltzmann}).
Building on this insight, the authors of Ref.~\cite{2014Guery-Odelin} discovered novel exact solutions of the Boltzmann equation.
In particular, they constructed time-dependent potentials of the form
\begin{equation}
\label{eq:2014G-Opotential}
    U(\mathbf{r},t) = \frac{m}{2}\omega^2(t) r^2 + \frac{b}{r^2} \, ,
\end{equation}
with appropriately engineered $\omega(t)$, that rapidly steer the dilute gas from an initial canonical distribution at temperature $T_{\ini}$, to a final canonical distribution at temperature $T_{\fin}$, under the assumption that the evolution of the gas is accurately described by the Boltzmann equation.
Here, $b\ge 0$ should be time independent.

We now extend these results to include scale-invariant potentials as well as time-periodic driving.
Consider the time-dependent, spherically symmetric potential
\begin{equation}
\label{eq:tdssp}
\overline{U}(\mathbf{r},t) = \frac{1}{\sigma^2} U_0\left( \frac{r}{\sigma} \right) + \frac{b}{r^2} \, ,    
\end{equation}
where $\sigma(t) > 0$ is a twice-differentiable function of time, and $b\ge 0$ is a constant.
Next, let $\beta(0)>0$ denote an inverse temperature, and define
\begin{equation}
\label{eq:betadef}
    \beta(t) = \frac{\sigma^2(t)}{\sigma^2(0)} \beta(0)
    \quad,\quad
    \bar{\mathbf{v}}(\mathbf{r},\mathbf{v},t) = \mathbf{v} - \frac{\dot\sigma(t)}{\sigma(t)} \mathbf{r}
\end{equation}
and
\begin{equation}
\label{eq:BoltzPot}
    U(\mathbf{r},t) = \overline{U}(\mathbf{r},t) - \frac{m}{2} \frac{\ddot\sigma(t)}{\sigma(t)}r^2
\end{equation}
(compare with Eqs.~\eref{eq:vsi}, \eref{eq:siuff}).
We claim that the time-dependent distribution
\begin{equation}
\label{eq:Boltzmann_exactf}
    f(\mathbf{r},\mathbf{v},t) = \frac{1}{Z_0} e^{-\beta(t) [ m\bar v^2/2 + \overline{U}(\mathbf{r},t) ]}
\end{equation}
is an exact solution of the Boltzmann equation (Eq.~\eref{eq:Boltzmann}), when the force $\mathbf{F}(\mathbf{r},t)$ is obtained from the potential given by Eq.~\eref{eq:BoltzPot}.\footnote{The time-{\it in}-dependence of the normalization constant $Z_0$ follows from the form of $\overline{U}(\mathbf{r},t)$, as can be shown by inspection.}

To establish this claim, note that Eq.~\eref{eq:Boltzmann_exactf} has the form given by Eq.~\eref{eq:nullifyCollisions}, therefore $I_{\rm coll}=0$.
Also, it follows from direct substitution that Eq.~\eref{eq:Boltzmann_exactf} satisfies the collisionless equation, Eq.~\eref{eq:collisionless}.
Thus Eq.~\eref{eq:Boltzmann_exactf} solves Eq.~\eref{eq:Boltzmann} exactly.

We can extend these results further by replacing Eq.~\eref{eq:tdssp} with
\begin{equation}
\label{eq:BoltzPot2}
\overline{U}(\mathbf{r},t) = \frac{1}{\sigma^2} \sum_{j=1}^3 U_{0j}\left( \frac{x_j}{\sigma} \right) +  \,   U_h(x_1,x_2,x_3)
\end{equation}
where the $x_j$'s are the Cartesian components of $\mathbf{r}$, the $U_{0j}$'s are three (generally unrelated) potential functions, and $U_h$ is a homogeneous function of degree -2:
\begin{equation}
    U_h(x_1,x_2,x_3) \,=\, \frac{1}{(x_1x_2x_3)^{2/3}} \, \psi\left(\frac{x_1}{x_2},\frac{x_1}{x_3} \right),
\end{equation}
where $\psi$ is an arbitrary function.
This form for $U_h$ includes a wide range of potentials, including as special cases $b/r^2$ (see Eq.~\eref{eq:tdssp}) and $a_1/x_1^{2} + a_2/x_2^{2} + a_3/x_3^{2}$. The potential $U_h$ should bear no explicit time dependence, unlike $\overline{U}$, which depends explicitly on time through $\sigma$.
If we again define $U(\mathbf{r},t) = \overline{U} - m\ddot\sigma r^2/2\sigma$ (see Eq.~\eref{eq:BoltzPot}), then $f(\mathbf{r},\mathbf{v},t)$ given by Eq.~\eref{eq:Boltzmann_exactf} remains an exact solution of the Boltzmann equation \eref{eq:Boltzmann}.
This result can be established as in the previous paragraph.

If we set $U_0(x)=U_{0j}(x)=\kappa x^2/2$, for some fixed $\kappa >0$ and for $j=1,2,3$, and $U_h = b/r^2$ for some $b\ge 0$, then Eqs.~\eref{eq:tdssp} and \eref{eq:BoltzPot2} describe the same potential, and $U(\mathbf{r},t)$ has the form given by Eq.~\eref{eq:2014G-Opotential}.
If $U_{0j}(x)=\kappa_jx^2/2$ for different fixed values $\kappa_1, \kappa_2, \kappa_3>0$, and $U_h=0$, then $\overline{U}(\mathbf{r},t)$ is an anisotropic harmonic potential of the form considered in Ref.~\cite{2015Papoular}, where exact solutions of the Boltzmann equation were derived for such potentials.
Note that, in general, even if the three $U_{0j}$'s are identical functions, the resulting potential $\overline{U}(\mathbf{r},t)$ (Eq.~\eref{eq:BoltzPot2}) is not spherically symmetric.

We can use these results to design a protocol for driving the gas from a canonical distribution at initial inverse temperature $\beta(0)^{-1}$ to a canonical distribution at final inverse temperature $\beta(t_{\fin})^{-1}$, in an arbitrary, finite time $t_{\fin}$.
To do this, we simply choose $\sigma(t)$ to satisfy the boundary conditions $\dot\sigma=\ddot\sigma=0$ at $t=0$ and $t=t_{\fin}$, and
\begin{equation}
    \sigma(t_{\fin})/\sigma(0)= \sqrt{\beta(t_{\fin})/\beta(0)} \, .
\end{equation}
This protocol can be used with either of the potential forms given by Eqs.~\eref{eq:BoltzPot} or \eref{eq:BoltzPot2}.
In the isotropic and anisotropic harmonic cases discussed in the previous paragraph, these protocols reduce to the ones obtained in Refs.~\cite{2014Guery-Odelin} and \cite{2015Papoular}, respectively.

Alternatively, we can design a driven breathing mode of the Boltzmann equation by choosing the protocol
\begin{equation}
    \sigma(t) = \bar\sigma + \Delta\sigma \cos^2(\omega t) \quad, \quad 0 < \Delta\sigma < \bar\sigma
\end{equation}
again using either Eq.~\eref{eq:BoltzPot} or Eq.~\eref{eq:BoltzPot2}.
The exact solution given by Eq.~\eref{eq:Boltzmann_exactf} then oscillates periodically in time, with frequency $2\omega$.

Finally, let us consider how the above results relate to the {\it undriven} breathing modes discovered by 
Boltzmann \cite{2014Guery-Odelin}.
In Eq.~\eref{eq:tdssp}, take $U_0(x) = mx^2/2$ and $b=0$ so that $\overline{U}(\mathbf{r},t)$ becomes a spherically symmetric harmonic oscillator with a time-dependent stiffness $\sigma^{-4}(t)$.
By Eq.~\eref{eq:BoltzPot} this choice leads to the total potential
\begin{equation}
    U(\mathbf{r},t) = \frac{m}{2} \left( \frac{1}{\sigma^4} - \frac{\ddot\sigma}{\sigma} \right) r^2 \equiv \frac{m}{2}\Omega^2(t) r^2 \, .
\end{equation}
Note that the relationship between $\sigma$ and $\Omega$ here is described by the Ermakov equation \eref{eq:ermakov}.
If we now take $\Omega(t)$ to be constant rather than time-dependent, and we solve for $\sigma(t)$ (see Eq.~\eref{eq:ermakovSolution} below), then Eq.~\eref{eq:Boltzmann_exactf} becomes a time-periodic solution of the Boltzmann equation for a dilute gas in a fixed harmonic potential $U(\mathbf{r},t) = m\Omega^2 r^2/2$.
This class of solutions coincides with Boltzmann's breathing modes.
A related question pertains to the type of static confining potential for which a breathing solution can exist. 
It is addressed in~\ref{app:breathers}.

When $\Omega$ is constant, the general solution of the Ermakov equation
\eref{eq:ermakov} is given by
\begin{equation}
\label{eq:ermakovSolution}
    \sigma(t) = \left[ \tau - \sqrt{\tau^2-\frac{1}{\Omega^2}} \cos(2\Omega t + \phi) \right]^{1/2}
\end{equation}
where $\phi$ and $\tau\ge 1/\Omega$ are constants.
Substituting this result into Eq.~\eref{eq:betadef} for $\beta(t)$ reveals that the effective inverse temperature $\beta(t)$ oscillates harmonically with frequency $2\Omega$.
Moreover, taking the first three derivatives of $\beta(t)$ with respect to time, we straightforwardly obtain
\begin{equation}
    \frac{d^3}{dt^3}{\beta}(t) + 4\Omega^2 \dot\beta(t) = 0
    \label{eq:PRL2014_9b}
\end{equation}
which is equivalent (for static $\Omega$) to Eq. (9b) of Ref.~\cite{2014Guery-Odelin}. 

\subsection{Relations to quantum shortcuts to adiabaticity}
\label{subsec:classical-quantum}

The classical shortcuts described above have quantum counterparts as we now briefly discuss.

We have already seen this correspondence in the global counterdiabatic case: just as the classical term $H_{\rm GCD}(z,t)$ defined by Eqs.~\eref{eq:HCD-c2a} and \eref{eq:HCD-c2b} generates the desired shortcut for any choice of initial energy shell (Eq.~\eref{eq:GCDtraj}), so too its quantum counterpart $\widehat H_{\rm GCD}(t)$ given by Eqs.~\eref{eq:HCD-q2a} and \eref{eq:HCD-q2b} generates the desired shortcut for any energy eigenstate~\cite{2003Demirplak,2009Berry}.

For local counterdiabatic driving, the auxiliary term $H_{\rm LCD}=pv(x,t)$ is designed for a specific choice of initial energy shell, as discussed in Sec.~\ref{subsubsec:lcd}.
In the quantum case, let us instead choose an initial energy eigenstate $\vert n(0)\rangle$ and construct the cumulative distribution
\begin{equation}
{\cal F}(x,t) = \int_{-\infty}^x dx^\prime \, \left\vert \phi_n(x^\prime,t) \right\vert^2
\quad,\quad
\phi_n(x,t) \equiv \langle x \vert n(t) \rangle.
\end{equation}
Now define a velocity field $v(x,t)$ in a manner analogous to Eq.~\eref{eq:vdef}, but with ${\cal F}(x,t)$ playing the role of ${\cal S}(x,t)$:
\begin{equation}
\label{eq:qvdef}
v(x,t) = -\frac{\partial_t{\cal F}}{\partial_x{\cal F}} \, .
\end{equation}
Using this field we construct a quantum auxiliary term
\begin{equation}
\label{eq:HLCD-q}
\widehat H_{\rm LCD}(t) = \frac{\widehat p\widehat v + \widehat v\widehat p}{2}
\quad , \quad
\widehat v \equiv v(\widehat x,t)
\end{equation}
(compare with Eq.~\eref{eq:HLCD}).
If the system begins in the state $\vert n(0)\rangle$ and then evolves under $\widehat H(t) + \widehat H_{\rm LCD}(t)$, it will remain in the instantaneous eigenstate $\vert n(t)\rangle$ (up to an overall time-dependent phase) for all $t\in [0,t_{\fin}]$~\cite{2017Patra}.

Using the velocity field $v(x,t)$ given by Eq.~\eref{eq:qvdef}, we next construct an acceleration field $a(x,t) = v\partial_x v + \partial_t v$ and a corresponding fast-forward potential $U_{\rm FF}(x,t)$ via $-\partial_x U_{\rm FF} = ma$ (see Eqs.~\eref{eq:adef} and \eref{eq:uffdef}).
If the system evolves under $\widehat H(t) + \widehat U_{\rm FF}(x,t)$, from an initial state $\vert n(0)\rangle$, then at the final time $t=t_{\fin}$ it will arrive at the eigenstate $\vert n(t_{\fin})\rangle$ (up to a phase), though at intermediate times it generally will not be in the state $\vert n(t)\rangle$~\cite{2017Patra}.
The potential $U_{\rm FF}(x,t)$ obtained in this manner -- using ${\cal F}$ rather than ${\cal S}$ to construct $v$ -- is equivalent to the fast-forward potential originally derived by Masuda and Nakamura~\cite{2010Masuda} and further studied in Refs.~\cite{2012Torrontegui,2016Martinez-Garaot,2015Takahashi}.

It is interesting to note that both  quantum and classical auxiliary terms are constructed using a velocity field determined from a cumulative function, either ${\cal F}(x,t)$ or ${\cal S}(x,t)$.
As we shall see later, this pattern applies as well to stochastic shortcuts to adiabaticity, where the cumulative function ${\cal F}(x,t)$ is given in terms of a canonical probability distribution.

The discussion in the previous three paragraphs requires a rather strong caveat.
For excited states $n>0$, the field $v(x,t)$ given by Eq.~\eref{eq:qvdef} generically diverges at the nodes of the eigenstate, i.e.\ where $\phi_n(x,t) = 0$, leading to ill-behaved auxiliary terms $\widehat H_{\rm LCD}$ and $\widehat U_{\rm FF}$.
Thus in general the quantum LCD and FF approaches described above are limited to ground states $n=0$, which have no nodes;
scale-invariant systems are an exception to this statement~\cite{2014Deffner}.

Ref.~\cite{2021Patra} develops a semiclassical fast-forward approach that avoids the problem posed by the nodes of $\phi_n(x)$.
In this approach, the fields $v$ and $a$ are constructed using ${\cal S}$ (Eq.~\eref{eq:vdef}) rather than ${\cal F}$ (Eq.~\eref{eq:qvdef}).
The resulting auxiliary term $\widehat U_{\rm FF}$ is well-behaved, but no longer guides the wavefunction {\it exactly} to the desired final state $\vert n(t_{\fin})\rangle$.
Instead it provides an approximate shortcut that is expected to work well in the semiclassical regime of large $n$.
Numerical simulations support this expectation~\cite{2021Patra}.


\section{Shortcuts for classical systems in contact with a thermal bath}
\label{sec:systems-with-bath}

In the previous section, we have obtained finite-time protocols
for driving an isolated system, with the same final state 
as with an infinitely long driving. In the remainder, 
our interest goes to systems that are (strongly) coupled to an environment; the systems we have in mind are epitomized by a Brownian object such as a colloid in a
fluid. The latter plays the role of a thermal bath \cite{2017Ciliberto}, which introduces fluctuations. Therefore, the system necessitates a stochastic/probabilistic description,
a new feature compared to the treatment in Sec.~\ref{sec:isolated-systems}. 
This goes with a point-of-view-change, 
from single trajectories to probability distributions.
From an experimental perspective, two routes can be proposed,
where the question, formulated in a probabilistic fashion, becomes meaningful.
a) One may be interested in repetitions of an experiment
involving a Brownian object, such as a macromolecule,
or a nano device; statistics is then gathered by automatizing
the protocol, as e.g. in~\cite{2016Martinez_b} or~\cite{2020Besga};
b) A single experiment can allow for measuring a distribution function, provided it involves a collection of Brownian objects,
manipulated simultaneously. When they do not interact, 
or when interactions are weak such as in a 
low density colloidal system, routes a) and b) are equivalent. 

\setcounter{footnote}{0} 
In a nutshell, what we now aim at suitably driving with an external force
is a probability density, which shall be denoted $\rho(x,t)$:
What should the drive be to evolve the system from an initial density 
$\rho(x,t_{\ini})$ at time $t_{\ini}$ to a target distribution $\rho(x,t_{\fin})$ at time $t_{\fin}$? 
Such a question of fast driving a system from a given equilibrium state to another appeared under different names in the literature:
Engineered Swift Equilibration \cite{2016Martinez},\footnote{Also variants such as Engineered Swift Relaxation for the connection of non-equilibrium states~\cite{2021Prados}, see Sec.~\ref{sec:beyond-equilibrium}.} or Shortcuts to Isothermality \cite{2017Li}.\footnote{The latter work is akin to the counterdiabatic method presented in Sec.~\ref{sec:counterdiabatic}.} 
It has been investigated experimentally for small systems (e.g. colloids \cite{2016Martinez} and AFM tips \cite{2016Le-Cunuder}) in contact with a thermostat. There, one searches for the proper variation of the control parameters to ensure a transfer from the initial state at time $t_{\ini}$ to the desired final state at time $t_{\fin}$. 

\subsection{Inverse Engineering}\label{sec:inverse-engineering}

A method that often proves efficient in practice is the commonly-called inverse engineering technique. This approach has been successfully applied to ordinary differential equations in classical and quantum physics \cite{2019Guery-Odelin}. If we denote by $\mathbf{X}$ the dynamical variables (state vector
with $n$ components), the control problem is encapsulated in a set of coupled differential equations
\begin{equation}
\dot{ \mathbf{ X}} =  {\mathbf{f}} (  {\mathbf{ X}},  {\mathbf{ \lambda}}(t) ).
\label{controlequation}
\end{equation}
There is no general statement for the reachability of the desired state under the driving provided by  ${\mathbf \lambda}(t)$ 
(control vector with $r$ components) except when ${\mathbf f}$ is a linear function with time-invariant coefficients, for which the Kalman rank condition applies \cite{1963Kalman,1987Pontryagin}.\footnote{For a linear system of the type ${\mathbf f} ({\mathbf X},{\mathbf \lambda(t)})=A{\mathbf X}(t) + B{\mathbf \lambda}(t)$, where $A$ and $B$ are time-independent matrices of size $n\times n$ and $n\times r$, respectively, we define the $n\times nr$ controllability matrix $C=[B\;AB\;...\;A^{n-1}B]$. The system is controllable if $C$ has $n$ linearly independent columns, i.e. rank$(C)=n$ \cite{1963Kalman,1987Pontryagin}.} When the desired state belongs to the reachable set of solutions, there exists a wide variety of possible protocols to reach the final state. 

The inverse engineering method provides a convenient way to find out a proper driving. It amounts to impose the evolution of the dynamical variable ${\mathbf X}$ and to infer from Eq.~\eref{controlequation}, the expression for $ {\mathbf \lambda}(t)$. However, this inverse use of the differential equation that governs the dynamics is not always easy to handle. The mathematical property which allows for such an inverse use of the dynamical (including nonlinear) equations is known as the flatness property, and can be considered as an extension of the Kalman's controllability criterion \cite{1995FLIESS}. This strategy has been successfully used to transport a particle in a moving harmonic potential, both in classical and quantum physics \cite{2011Torrontegui,2014Guery-Odelin_b,2021Zhang}, or to shuttle the particle, i.e. to set a given velocity to the particle---see \cite{2021Qi} and references therein.

For a system in contact with a thermal bath, the dynamical variables obey stochastic differential equations (Langevin-like) that cannot be directly written as a set of continuous equations such as that of Eqs.~(\ref{controlequation}). For Brownian motion, this difficulty has been circumvented in two steps. First, as emphasized above, 
the discussion is made not on the individual trajectories but on the density distribution that obeys the Smoluchowski, or Fokker-Planck (FP),  equation. (See~\ref{app:stoch-frame} for a crash recapitulation of the essential aspects of the Langevin and  frameworks
for stochastic processes.)
Second, an ansatz depending on a set of a few effective dynamical variables, such as the moments of the PDF, is usually proposed to get a finite set of equations in the form of Eqs.~(\ref{controlequation}). 
The coupled equations on the moments can be alternatively derived from the Langevin equation.

Consider the overdamped motion of a bead of micron size immersed in a fluid and trapped by a general confining potential $U(x,t)$. The density $\rho(x,t)$ obeys the overdamped ---or Smoluchowski---equation
\begin{equation}
\label{eq:FP}
    \gamma \partial_t\rho(x,t) = \partial_x \left[ \partial_xU(x,t)  \rho(x,t) \right] + \beta^{-1} \partial^2_{x}\rho (x,t),
\end{equation}
where $\gamma$ is the friction coefficient and $\beta=(k_BT)^{-1}$ refers to the inverse of the temperature, $k_B$ being Boltzmann's constant---see~\ref{app:stoch-frame} for a detailed discussion of the stochastic dynamics framework.
Particular attention has been paid to the harmonic potential, which models paradigmatic systems such as optical tweezers. Equation \eref{eq:FP} then becomes
\begin{equation}
\label{eq:FP-harm}
  \gamma  \partial_t \rho(x,t) = \partial_x \left[ \kappa x \rho(x,t) \right] + \beta^{-1} \partial^2_{x}\rho(x,t).
\end{equation}
In the harmonic problem, the control (time-dependent)
parameters are {\it a priori} the temperature $T$ (see 
Sec.~\ref{sec:engineering-thermal-env}) and the stiffness $\kappa$ of the potential---typically, the friction coefficient is assumed to be constant, see \ref{app:stoch-frame} for details. Here, the statistics of the bead's position remains Gaussian
if it is initially, with a standard deviation $\sigma(t)$ that 
plays the role of an effective dynamical variable and obeys
\begin{equation}
\label{eq:standard-dev-harmonic}
\dot \sigma = - \frac{\kappa}{\gamma}\sigma + \frac{k_BT}{\gamma}\frac{1}{\sigma} .
\end{equation}
For both time-independent temperature and stiffness, the system approaches the canonical equilibrium distribution, for which the variance of the position is
\begin{equation}
\label{eq:sigma-eq}
   \sigma_{eq}^2=\frac{k_B T}{\kappa},
\end{equation}
i.e. the equilibrium equation of state. 

Let us assume for instance that the temperature is constant and we vary the stiffness of the trap. The initial state corresponds to thermal equilibrium state with the initial stiffness $\kappa_{\ini}=\kappa(t_{\ini})$. The objective is to reach the equilibrium state with the desired final value of $\kappa_{\fin}=\kappa(t_{\fin})$ in a chosen amount of time $t_{\fin}$. Those boundaries conditions define the values $\sigma_{\ini}=(k_BT/\kappa_{\ini})^{1/2}$ and $\sigma_{\fin}=(k_BT/\kappa_{\fin})^{1/2}$ at initial and final time. The inverse engineering technique involves the choice of an interpolation function $\sigma(t)$ between those two values, and to subsequently infer the time-dependent stiffness $\kappa(t)$ to be applied, directly from Eq.~\eref{eq:standard-dev-harmonic},
since both $\sigma$ and $\dot\sigma$ are then known. For the very same problem in the underdamped regime, one shall solve the Kramers equation for the phase space distribution. In this latter case, the effective dynamical system boils down to a set of 3 coupled linear equations for the time evolution of the 3 moments $\langle x^2 \rangle$, $\langle v^2 \rangle$, and $\langle x v \rangle$. As a result, the strategy to extract $\kappa(t)$ is slightly more involved \cite{2018Chupeau_b}.

The inverse engineering method applied here 
is specific to the manipulation of Gaussian states. As already commented, a remarkable property is that such an initial condition, under time-dependent harmonic forcing, remains Gaussian
at all times, and thus preserves its shape.\footnote{This can be viewed as a consequence of the conservation of the Gaussian character when summing Gaussian variables, and dwells on the fact that the solution to Langevin equation is then linear in the noise history.}
Yet, while it is difficult to compute analytically the potential required
to connect two arbitrary non-Gaussian states, it turns out that if both have the same shape, a simple solution  can be found, therefore generalizing the aforementioned
Gaussian result. We here impose the following  shape for the PDF $\rho(x,t)$:
\begin{equation}
\label{eq:sh-pres}
\rho(x,t) = \frac{1}{\sigma(t)Z_{\ini}} \exp \left[ -\beta U_{\ini}\left( \frac{x-\mu(t)}{\sigma(t)}\right)\right] ,  \end{equation}
with $\mu(t)$ and $\sigma(t)$ continuous real functions, such that $\mu(t_{\ini})=0$ and $\sigma(t_{\ini})=1$, which accounts for shifting and re-scaling the space dependence. Note that, although we used similar ideas in the scale-invariant protocols introduced in Eq.~\eref{eq:scaleInvart}---justifying the same notation for the parameters, processes preserving the shape are different from those scale-invariant protocols.\footnote{Specifically, one can be mapped onto the other if $U_{\ini}(x)$ is proportional to a power of $x$. For example, if we consider the harmonic potential $U_{\ini} (x)=k_{\ini} x^2/2$, the potential $U$ remains harmonic for both cases, shape-preserving and scale-invariant, but with a different stiffness, $k_{\ini}/\sigma^2$ and $k_{\ini}/\sigma^4$ respectively. For general $U_{\ini}$, the map is not guaranteed.}  The quantity  $Z_{\ini}=\int_{\cal D} dx\, e^{-\beta U_{\ini}(x)}$ is the partition function that guarantees the correct normalization of the distribution. Such distributions only connect states that belong to the same family of potentials, whatever this family is. The final potential then reads \cite{2021Plata}
\begin{equation}
    U_{\fin} (x)=U_{\ini}\left[ \frac{x-\mu(t_{\fin})}{\sigma(t_{\fin})}\right].
\end{equation} 
The time-dependent driving potential required to ensure  shape preservation is found by introducing the ansatz \eref{eq:sh-pres} into the  equation. One finds \cite{2021Plata}
\begin{eqnarray}
U(x,t)  =&      U_{\ini} \left[ \frac{x - \mu(t)}{\sigma(t)}\right] \nonumber \\    
& -\gamma  \frac{2\dot{\mu}(t)\sigma(t)[x-\mu(t)]+[x-\mu(t)]^2\dot{\sigma}(t)}{2\sigma(t)}.
\label{eq:driv-sh-pres-U}
\end{eqnarray}
Such a driving potential involves two different contributions: the shape-bearing potential itself and an additional harmonic potential, whose stiffness and center are determined by  certain combinations of the shift and scaling functions $\mu(t)$ and $\sigma(t)$. The extra time-dependent harmonic potential is nothing but the counterdiabatic term, to be discussed further below. To enforce the smoothness of the potential, one can add the following extra conditions on the parameters at initial and final time: $\dot{\mu}(t_{\ini})=\dot{\mu}(t_{\fin})=0$ and $\dot{\sigma}(t_{\ini})=\dot{\sigma}(t_{\fin})=0$.

Inverse engineering techniques have also been employed to address the underdamped situation~\cite{2018Chupeau_b}. 
The authors worked out eligible conservative, velocity-independent, drivings $U(x,t)$. However, the problem becomes involved and limitations appear, presumably inherent to the functional forms chosen for constructing explicit solutions.

Related in spirit to inverse engineering are stochastic methods
that generate Brownian paths conditioned to start and end at prescribed (ensemble of) points
~\cite{2015Majumdar}.
The conditioning, that endows these paths with precise statistical properties, can be of various types: a bridge, meaning a path that starts at some $x_{\ini}$ at $t=t_{\ini}$, and ends at a given $x_{\fin}$ at $t=t_{\fin}$;
an excursion, meaning a bridge with $x_{\fin}=x_{\ini}$ that is furthermore constrained to lie 
at all times to the right of $x_{\ini}$, etc.
These methods can be used to generate computationally 
the constrained paths in an efficient manner; a naive variant would amount to pruning an ensemble of unconstrained paths, keeping only those
trajectories that fulfil the imposed constraints, a highly inefficient way of proceeding. For overdamped Langevin dynamics,
it was shown on general grounds that irrespective of $x_{\ini}$,
addition of an external harmonic force centered at $x_{\fin}$ and with 
stiffness $\gamma /(t_{\fin}-t) $, which thus diverges for $t\to t_{\fin}$,
generates a {\it bona fide} bridge \cite{2015Majumdar}. More generally, as might be anticipated from the previous example,
the gist of the approach is to add a time-dependent entropic potential of the form 
$-2 k_BT \log Q(x,t)$; $Q$, encoding the constraints considered (such as remaining in the
allowed half-space for an excursion), is the probability density to be at $x_{\fin}$ at time $t_{\fin}$,
having started at point $x$ at time $t<t_{\fin}$
\cite{2015Majumdar,1957Doob}:
\begin{equation}
    Q(x,t) \equiv P(x_{\fin},t_{\fin}| x,t) .
\end{equation}
While this result can be obtained by inverse engineering,
it should be stressed that it does feature an important 
difference with short-cutting ideas discussed here, in the sense
that there does not exist an infinitely slow process that 
is being accelerated in some form.
Note that $Q$ introduced above fulfills the backwards  equation, and is as such intimately related to first passage problems~\cite{2001Redner}.

\subsection{Counterdiabatic method}
\label{sec:counterdiabatic}


In this section, we extend the counterdiabatic method (see Secs.~\ref{sec:intro} and \ref{sec:isolated-systems}) to systems in contact with a thermal bath.
As above, we model the evolution of such systems with overdamped Langevin dynamics at the single-trajectory level, and with the FP equation at the ensemble level, and we restrict ourselves to systems with a single degree of freedom, $x$.
Because momentum is ignored in the overdamped limit, we will use the generic notation $U(x)$ rather than $H(x,p)$ to denote the system's energy function.
This potential, used to drive the system, can be viewed as a Hamiltonian.

In the isolated quantum and classical cases described in Secs.~\ref{sec:intro} and \ref{sec:isolated-systems}, counterdiabatic driving aims to preserve an adiabatic invariant under rapid driving.
In the present context the role of the adiabatic invariant is played by the functional form of the probability distribution function (PDF), $\rho(x,t)$.
Specifically, when the potential $U$ is driven very slowly, its PDF evolves through a continuous sequence of equilibrium states $\rho_{eq}(x,t)$, 
\begin{eqnarray}
\rho_{eq}(x,t)= e^{\beta[F(t)-U(x,t)]}, \nonumber \\  F(t)=-\beta^{-1} \ln \left[ \int dx \,e^{-\beta U(x,t)}  \right].
\label{eq:rho_eq_preserved}
\end{eqnarray}
In the counterdiabatic method we seek to construct a term $U_{CD}(x,t)$ such that under the full driving potential $U_{\rm \STA}=U+U_{CD}$, the system evolves through the same equilibrium states $\rho_{eq}(x,t)$, even when the time dependence of $U(x,t)$ is not slow.

Note that, although the fine details may be different, the idea behind the counterdiabatic method could be cast under the umbrella of inverse engineering techniques. That is, we want to find an auxiliary $H_{CD}$ that enforces the conservation of the adiabatic invariant. The separation into categories of the techniques analyzed in this review has to be understood as a choice made more by pedagogical purposes than by a rigorous categorization. 


Consider an overdamped Brownian particle whose statistical state $\rho(x,t)$ obeys the FP equation~\eref{eq:FP}---see~\ref{app:stoch-frame} for details---with the potential $U_{\rm \STA}(x,t)=U(x,t)+U_{CD}(x,t)$. The instantaneous equilibrium distribution associated with $U$,
$\rho_{eq}$ given by Eq.~\eref{eq:rho_eq_preserved},
plays the role of adiabatic invariant.  In other words, we aim to steer the system so that the equilibrium PDF of the unperturbed potential $U(x,t)$ is maintained throughout the process.

When $U(x,t)$ varies at finite rate, $U_{CD}\neq 0$ is needed to preserve the prescribed evolution $\rho_{eq}$. Specifically, the counterdiabatic term   escorting the adiabatic evolution is given by~\cite{2017Li}
\begin{equation}
- \partial_x U_{CD}(x,t) = -\gamma \frac{\int^x dx' \partial_t \rho_{eq}(x',t)}{\rho_{eq}(x,t)}.
\label{eq:UCD_FP}
\end{equation}  
This expression can be derived from the cumulative distribution
\begin{equation}\label{eq:cumulative-distr}
\mathcal{F}(x,t)= \int_{-\infty}^{x} dx' \rho(x',t)
\end{equation}
and its velocity field
\begin{equation}\label{eq:veloc-field-CD}
v(x,t)=\partial_t x(\mathcal {F},t) = - \frac{\partial_t \mathcal{F}(x,t)}{\partial_x \mathcal{F}(x,t)},
\end{equation} 
where $x(\mathcal {F},t)$ is the function obtained by inverting $\mathcal{F}(x,t)$ for fixed $t$. Specifically, using the velocity field, we obtain
\begin{equation}
\label{eq:UCD-stoch}
- \partial_x U_{CD}(x,t) = \gamma v(x,t).
\end{equation}
This approach is particularly appealing since it has been shown to be useful as a unified procedure to derive counterdiabatic terms in  quantum, classical and thermal systems \cite{2017Patra}---as in Eqs.~\eref{eq:vdef} and \eref{eq:HLCD}.

The counterdiabatic force given in Eq. \eref{eq:UCD_FP} allows to
recover the inverse engineering results derived for shape preserving
potentials in Sec.~\ref{sec:inverse-engineering}. Indeed,
injecting relation \eref{eq:sh-pres}
into \eref{eq:UCD_FP} (with $\rho$ playing the role of $\rho_{eq}$) yields, rather unexpectedly,
a counterdiabatic force that does not depend on $U$ and thus holds irrespective of the functional form chosen for the PDF \cite{2021Plata}, provided this form is conserved
(and is thus a preserved ``adiabatic invariant''):
\begin{equation}
\label{eq:driv-sh-pres}
\partial_x U_{CD}(x,t) \,=\,   -\gamma \, \frac{\dot{\mu}(t)\sigma(t)+[x-\mu(t)]\dot{\sigma}(t)}{\sigma(t)},
\end{equation}
which is consistent with the potential given in Eq. 
\eref{eq:driv-sh-pres-U}, remembering
$U_{\rm \STA}=U+U_{CD}$.

The counterdiabatic method has been also employed in the underdamped situation~\cite{2014Tu,2017Li,2021Li_b}. At variance with the conservative drivings $U(x,t)$ derived with the inverse engineering technique~\cite{2018Chupeau_b}, the counterdiabatic potential contains a term depending on the momentum of the particle. This entails serious difficulties for the experimental implementation of such a counterdiabatic driving. 
Counterdiabatic protocols have also been devised for systems described by master equations, i.e. Markov processes with discrete states---see Sec.~\ref{ssec:otherApplications}. Finally, we present in~\ref{app:fluctuation-relation} a derivation of the work fluctuation relation~\cite{1997Jarzynski} from counterdiabatic type of arguments.

\subsection{Fast-forward}\label{sec:fast-forward}

Now, let us consider another strategy for the swift connection of equilibrium states: the so-called fast-forward procedure. In the quantum case~\cite{2008Masuda,2010Masuda}, fast-forward refers to a protocol that makes it possible to reach a desired final state, independently of the path swept to do so. For the isolated classical systems analyzed in Sec.~\ref{sec:isolated-systems} of this review, fast-forward has been employed in a similar sense. Therein, the fast-forward protocol led the system to the target state, with the same value of the adiabatic invariant as the initial state, by adding a velocity-independent potential $U_{FF}(x,t)$. The price of such a procedure was the adiabatic invariant no longer preserved at intermediate times. Derivation of fast-forward protocols are unified in both quantum and classical mechanics by using the acceleration flow field~\cite{2017Patra}, as discussed in Sec.~\ref{subsubsec:ff}.

Here, we consider the extension of the fast-forward idea to the context of systems with stochastic dynamics, again described by the overdamped FP equation~\eref{eq:FP}. In contrast to the counterdiabatic driving just described, there is no underlying adiabatic transformation over which the shortcut is built. The idea is the following: one considers a certain reference process---not necessarily slow---that connects two given states and then searches for a tailor-made external potential that accelerates this reference process~\cite{2021Plata}. In this way, the ``frames" of the ``movie" are fixed, given by the reference process, but are played at a higher rate in the fast-forward protocol. As explained below, the same idea  allows for reproducing the frames at a lower rate (slow-forward) or even play the movie backwards, at a higher or a lower rate (fast-backward or slow-backward, respectively)---somehow generalizing the shortcuts described in this review.



Going into specifics, and following Ref.~\cite{2021Plata}, we consider a reference solution $\rho_r(x,t)$ of Eq.~\eref{eq:FP} under a reference potential $U_r(x,t)$. Then, we introduce a time distortion $\Lambda(t)$ of the reference, i.e., $\rho(x,t)=\rho_r(x,\Lambda(t))$, and look for the potential $U(x,t)$ required to drive the evolution following the given prescription $\rho_r(x,\Lambda(t))$. It is important to remark that, at variance with the classical case, the connecting path is fixed and given by $\rho_r(x,\Lambda(t))$.

The FP equation~\eref{eq:FP} is rewritten as a continuity equation
\numparts
\ba
\label{eq:FP-continuity}
\partial_t \rho=-\partial_x \left(\rho v\right), \\
\label{eq:veloc-field-FP}
    v(x,t)= -\gamma^{-1} \left[ \partial_x U(x,t) +\beta^{-1} \partial_x \ln\rho(x,t) \right]
\ea
\endnumparts
where $v(x,t)$ is a velocity field.\footnote{This velocity field is exactly the same defined in the counterdiabatic driving, since spatial integration of the FP equation, written in the form \eref{eq:FP-continuity},  from $-\infty$ to $x$ leads precisely to Eq.~\eref{eq:veloc-field-CD}.}  The derivation of the fast-forward protocol is based on the relation between the velocity fields 
in the reference and the manipulated process,
\begin{equation}
    v(x,t)=\dot{\Lambda}(t) v_r(x,\Lambda(t)).
\end{equation}
Hence, the driving potential can be solved. For the sake of clarity, we display the solution for the auxiliary potential,
\begin{eqnarray}
-\partial_x U_{FF}(x,t) = &[1-\dot{\Lambda}(t)] \nonumber \\ 
&  \times [\partial_x U_r(x,\Lambda(t))+ \beta^{-1} \partial_x \ln \rho_r(x,\Lambda(t))  ] .
\end{eqnarray}
The total driving potential is $U(x,t)=U_r(x,\Lambda(t))+U_{FF}(x,t)$. The solution is given in terms of the reference process and the time map $\Lambda(t)$. As expected, when the reference and manipulated dynamics coincide, i.e.,
$\Lambda(t)=t$, one simply has $\partial_x U(x,t) =\partial_x U_r(x,t)$ or, equivalently, $\partial_x U_{FF}(x,t) = 0$.

\begin{figure}
    \centering
    \includegraphics[width=0.5\textwidth]{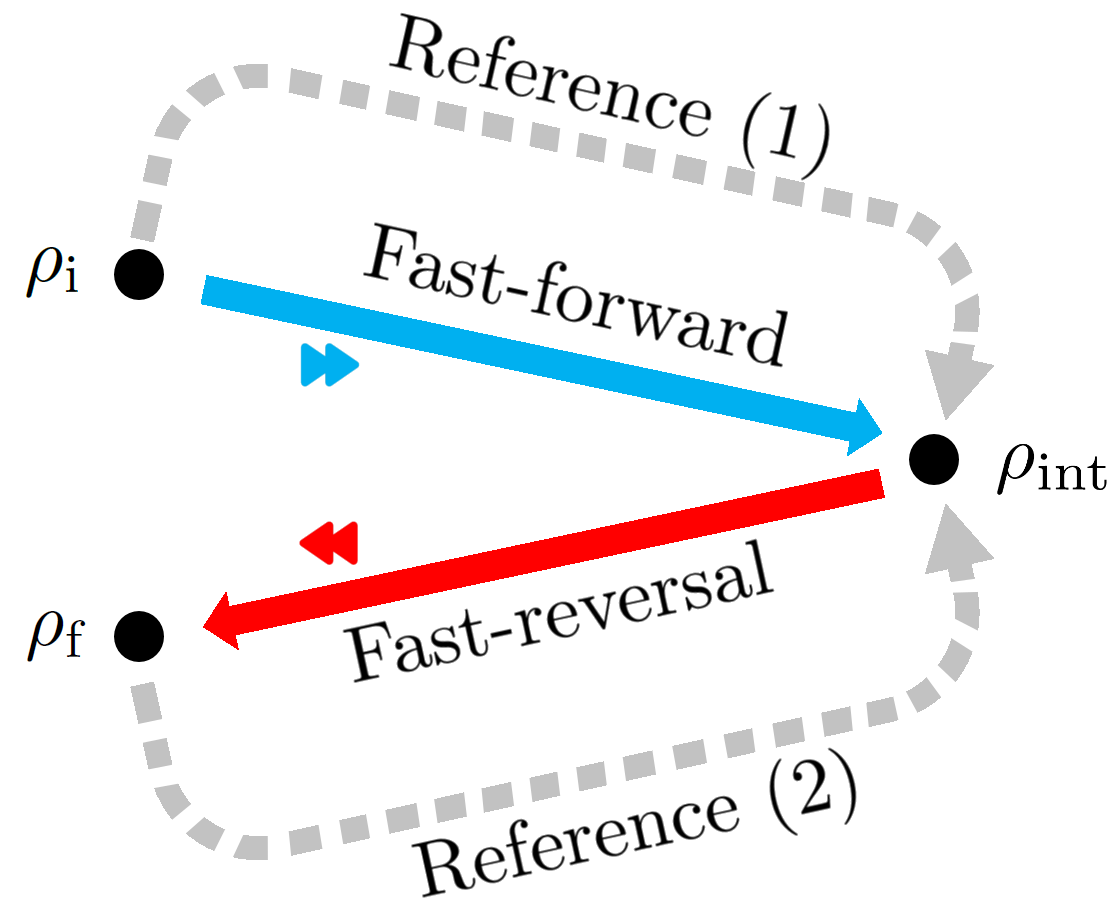}
    \caption{Sketch of the welding strategy to connect two arbitrary distributions $\rho_{\ini}$ and $\rho_{\fin}$. The connection is made in two steps. In the first step, a reference process from $\rho_{\ini}$ to $\rho_{\inter}$ is submitted to a fast-forward evolution. In the second step, a reference process starting at $\rho_{\fin}$ and finishing at $\rho_{\inter}$ is time-reversed and accelerated (fast backward).
    In doing so, one achieves the desired connection in a chosen time.}
    \label{fig:welding}
\end{figure}

This type of driving not only allows for acceleration
($\dot\Lambda>1$) but also for deceleration ($0<\dot\Lambda<1$) and even for the inversion of time's arrow ($\dot\Lambda<0$, meaning that the reference dynamics can be ``played backwards''). Combining simple reference processes, it is possible to build up an operational welding protocol that connects arbitrary states \cite{2021Plata}. Specifically, one can produce a welding connection between an initial state $\rho_{\ini}(x)$ and the target state $\rho_{\fin}(x)$ through an intermediate state $\rho_{\inter}(x)$, distorting two consecutive reference processes. This construction relies on the acceleration of a first reference relaxation process from  $\rho_{\ini}(x)$ to $\rho_{\inter}(x)$; and accelerating and reversing a second  relaxation process from  $\rho_{\fin}(x)$ to $\rho_{\inter}(x)$---see Fig.~\ref{fig:welding}. If one seeks an operating time $t_{\fin}$
(with a starting time $t_{\ini}=0$), one can assign a time duration 
$t_{\fin}/2$ for each of the two steps, but other choices are
possible.

Note that the counterdiabatic method could be understood as a limit of the fast-forward protocol presented above. Let us consider a reference process of duration $t_r$ and take the limit $t_r\to\infty$, so that the reference process becomes quasistatic. Hence, the limit reference process would sweep equilibrium states. 
In this way, the resulting limit of the fast-forward protocol would converge to the counteradiabatic method.


\subsection{Engineering the thermal environment}\label{sec:engineering-thermal-env}

In previous developments, we addressed Brownian objects in some environment at thermal equilibrium, meaning the temperature
entering the Langevin and FP equations 
is fixed. For colloidal systems, the environment is usually
water, and it may be difficult to impose a chosen
time dependence for its temperature, 
especially if a massive heating is sought \cite{2013Martinez}. 
Yet, it is possible 
to ``fool'' the colloidal beads, subjecting them to a random 
forcing that will emulate an effective 
temperature exceeding several thousand kelvins \cite{2013Martinez,2017Ciliberto}.
The method is quite robust, and essentially requires that the 
forcing frequency be large compared to the bead inverse relaxation time.
It is then possible to finely control the time dependence
of the effective temperature, 
by playing on the amplitude of the forcing,
which opens new means for driving the system
\cite{2018Chupeau_b}, and paves the way towards the more general goal
of reservoir engineering. 

Following this idea, a micrometric silica 
sphere has been driven in \cite{2018Chupeau_b} by the joint monitoring of a
harmonic trap stiffness ($\kappa$), and the point of zero force
($x_0$).
In other words, the confining potential is of the form
\begin{equation}
    U(x,t) \,=\, \frac{1}{2} \kappa(t) \left( x-x_0(t)
\right)^2 ,
\label{eq:randomx0}
\end{equation}
and the idea is to impose the proper time dependence
jointly on both $\kappa$ and $x_0$.
Compared to the more usual situation where $x_0$ is fixed,
a new contribution $\kappa(t) x_0(t)$ arises in the force balance.
It is important that a) $x_0$ remain small compared to the 
bead size, in order not to affect the effective stiffness
and b) that the correlation time of the signal $x_0(t)$ be small compared to the protocol 
duration (itself by construction smaller than the intrinsic relaxation time). Then, the forcing of $x_0$ results in an effective
heat bath for the colloidal degree of freedom: this forcing
has a time-dependent
amplitude and, for practical purposes, can be viewed as 
delta-correlated in time. In Ref. \cite{2018Chupeau_b}, the control of $x_0$ was achieved with an acousto-optic deflector, 
and the bath engineering made it possible to quickly deconfine a 
colloidal state. Should one be able to play only on the
stiffness $\kappa(t)$, equivalent transformations would require 
transiently expulsive forces with $\kappa<0$, which 
represent an experimental challenge \cite{2020Albay_b,2021Bayati}. 
A limitation of the approach is that the
extra stochastic forcing applied results in enhanced 
Brownian fluctuations, and in an effective temperature increase. Other techniques would have to be applied
when it comes to cooling the center-of-mass motion of trapped
beads, such as feedback-based approaches \cite{2011Li}.

Recently, the effective heating of optically trapped object 
allowed to devise finite-time adiabatic processes \cite{2020Plata_b}. 
Here, we stress that ``adiabatic'' is understood in its 
usual thermodynamics meaning, of heat-exchange free~\cite{Adiabatic}. A Brownian object is inherently fluctuating, and strongly coupled
to its environment. Consider the compression at fixed temperature $T$
of a colloidal bead,
where for instance the stiffness $\kappa$ of a harmonic potential 
is increased. When the bead has relaxed, the internal energy difference
$\Delta U$ vanishes on average between the initial and final states,
which have the same temperature. 
The first principle of thermodynamics~\cite{2021Peliti}
states that, since the bead received work from the confining
force, heat flew towards the bath, on average~\cite{Sekimoto}. 
If one seeks a vanishing heat exchange on average, 
it is mandatory that the environment temperature increase.
In the quasi-static limit, this increase has to be proportional
to $\sqrt{\kappa}$. This can be seen as a consequence of Laplace relation 
between temperature $T$ and volume $V$ for the adiabatic reversible transformation of an ideal gas:  $T V^{2/3}= \textnormal{const}$ for a monoatomic gas. Here, the confinement length is
$\sigma_{eq}$, so that the role 
of the volume is played by $\sigma_{eq}^3$. Since $\sigma^2_{eq}=k_BT/\kappa$, this yields 
a Laplace condition $T\sigma_{eq}^2=\textnormal{const}$ or, equivalently, $T^2/\kappa = \textnormal{const}$. This can be viewed as the statement that $n\Lambda^3 =\textnormal{const}$, where $n$ is the typical density and $\Lambda$ is De Broglie wavelength, which guarantees that volume in phase space is conserved. This volume is computed from  the (cubed) product of the typical length in real space,  $\sigma$, times the typical velocity, scaling like $\sqrt{T}$. This discussion also illustrates that the quasi-static criterion $T^2/\kappa = \textnormal{const}$ is space-dimension independent~\cite{2013Bo,2015Martinez_b}.

\begin{figure}
\begin{center}
         \includegraphics[width=0.45\textwidth]{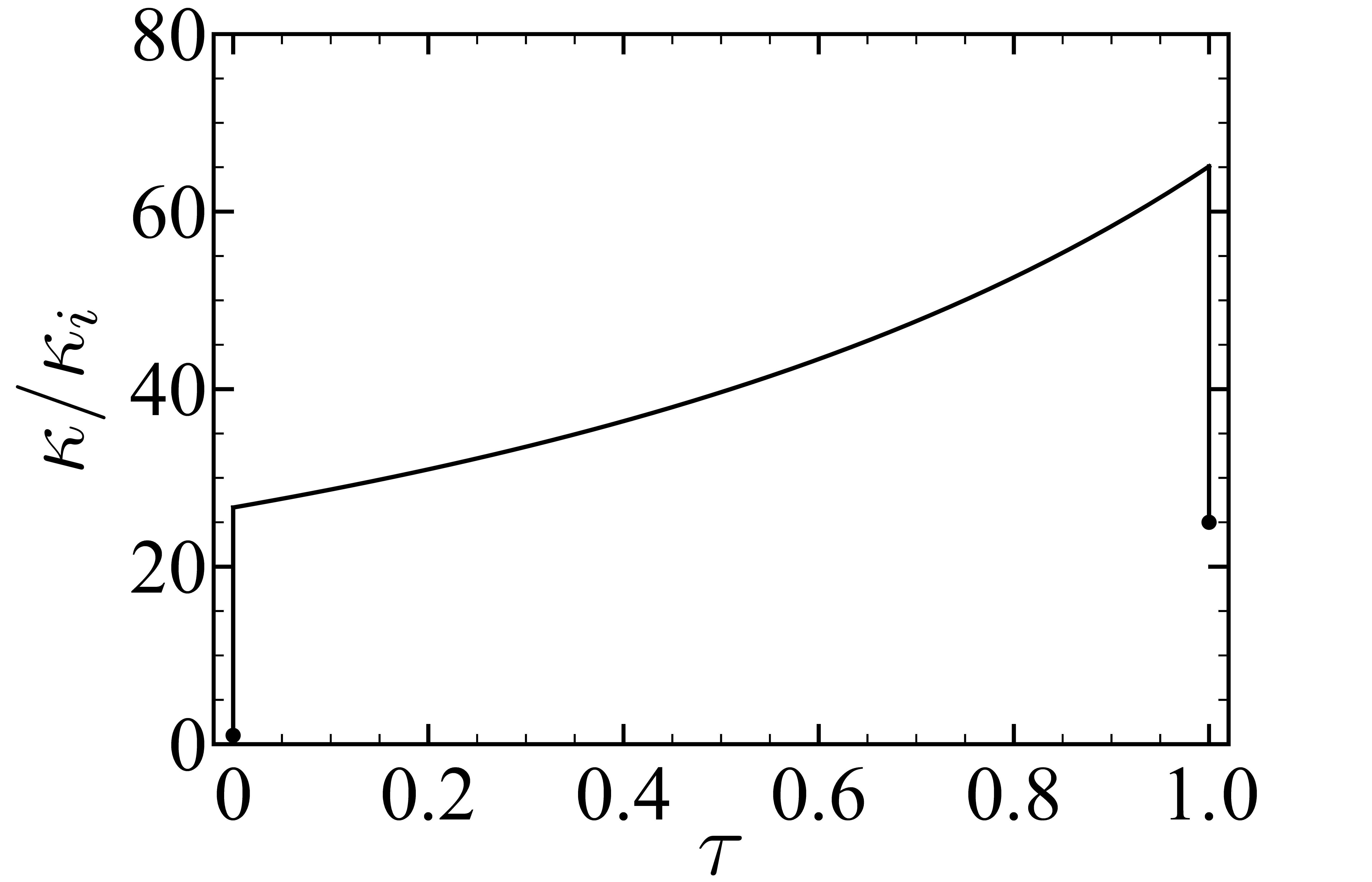}
        \includegraphics[width=0.45\textwidth]{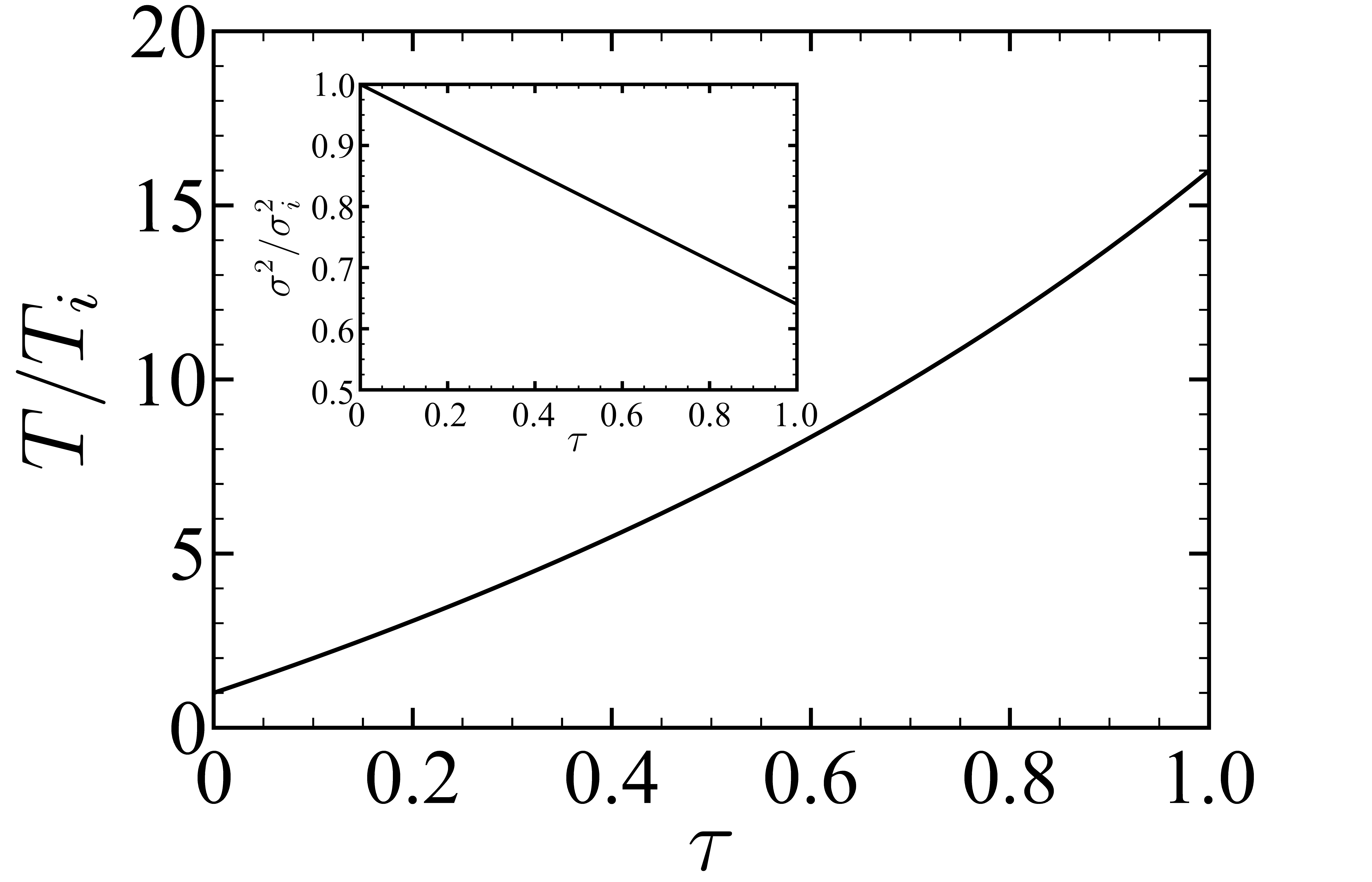}
       \includegraphics[width=0.45\textwidth]{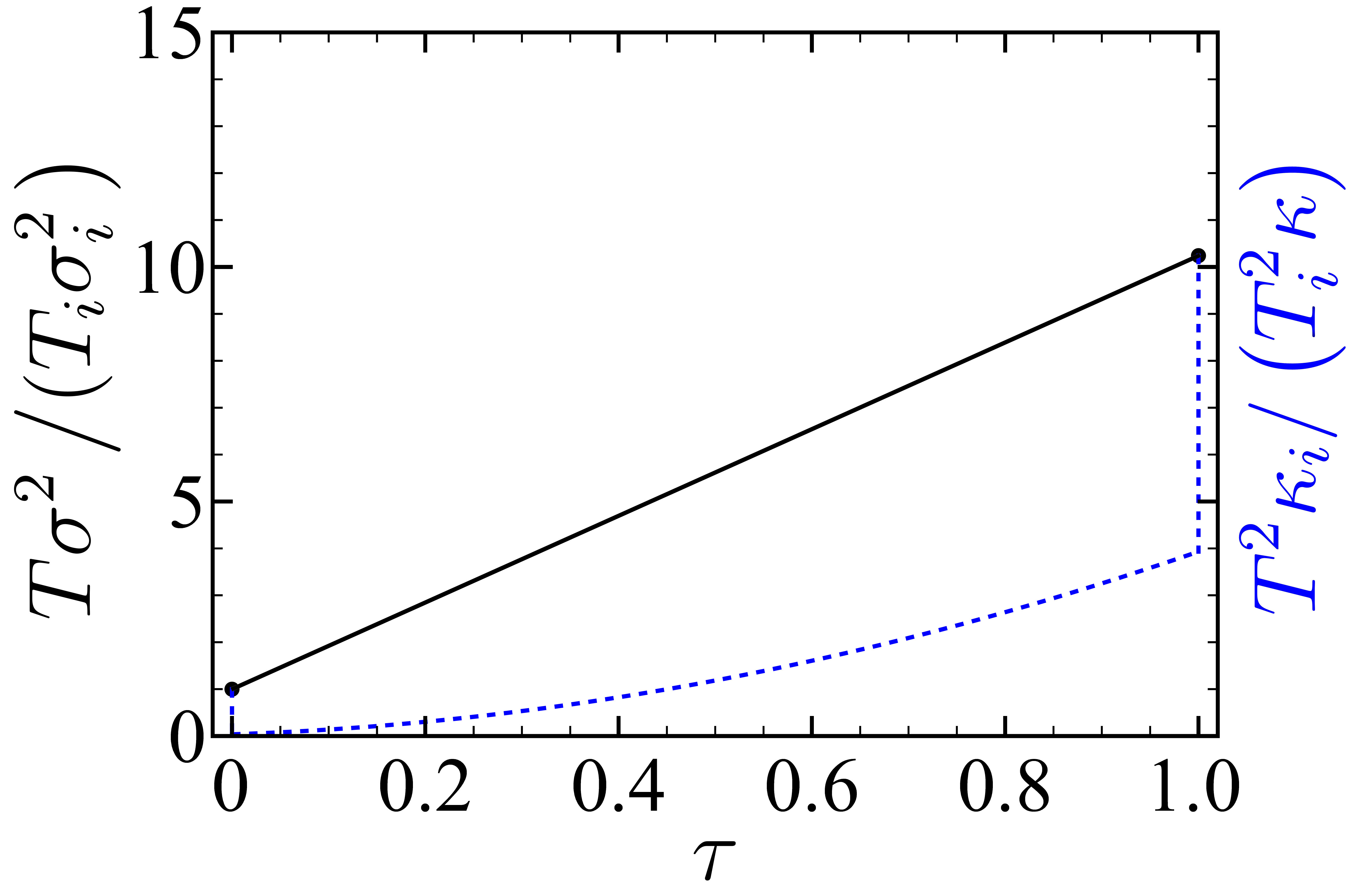}
\end{center}
   \caption{Fastest control and evolution for an adiabatic (in the sense of zero average heat) $20\%$ compression of a harmonically trapped particle. Time $\tau=t/t_{\fin}^*$ has been made dimensionless with the shortest possible duration of the process $t_{\fin}^*$, which is reached for a linear evolution of $\sigma_x^2$. The fastest connection requires the stiffness to be discontinuous at the initial and final times.  The example shown corresponds to $T_{\fin}/T_{\ini}=16$, $\kappa_{\fin}/\kappa_{\ini}=25$ and 
   entails an acceleration
   of a factor around $5.7$ with respect to the relaxation time scale  $t_{\scriptsize\textnormal{rel}}=\gamma / \kappa_{\fin}$.
   See Sec.~\ref{subsec:OCT-other-fig-merit} and Ref.~\cite{2020Plata_b} for further details.
\label{fig:adia-cexample}
}
\end{figure}

For finite-time adiabatic processes, not only does the mean heat released to  the thermal environment vanish between the initial and final states of the transformation, but it also does at any time in between.  The operating time of these irreversible adiabats can be optimized by jointly controlling the potential and the temperature. The condition of zero heat involves the kinetic contribution to the energy: the only assumption being, consistently with the overdamped description, that the velocity degree of freedom is always at equilibrium with the time-dependent value of the temperature. Some general results emerge~\cite{2020Plata_b}, like (i) forbidden regions, i.e., final states that cannot be reached adiabatically, and (ii)  a speed limit, the existence of a minimum, in general nonvanishing, time 
$t_{\fin}^*$
for the adiabatic connection.
For the specific case of a harmonic confining potential, it was shown that $T\sigma^2$ is nondecreasing over the adiabats. This implies that $T_{\fin}/T_{\ini} \geq  \sqrt{\kappa_{\fin}/\kappa_{\ini}}$, taking into account that the system is at equilibrium at the initial and final times---see Eq.~\eref{eq:sigma-eq}. It is only in the quasistatic limit that $T\sigma^2$ remains constant and, moreover, the equilibrium equation of state \eref{eq:sigma-eq} holds for all times, which leads to recover Bo and Celani's result of constant $T^2/\kappa$~\cite{2013Bo}.
Advantage was taken of these finite-time adiabatic transformations to construct an irreversible Carnot engine featuring interesting efficiency properties~\cite{2020Plata}, see also Sec.~\ref{sec:heat-engines}. In Fig.~\ref{fig:adia-cexample}, the fastest possible adiabatic connection is illustrated for a $20\%$ compression, $\sigma_{\fin}=0.8\sigma_{i}$. 


\section{Optimal control theory}\label{sec:OCT}

Hitherto, this review has focused on techniques that allow for connecting two given states. Naturally, having established that going from A to B in a finite time is feasible raises the question of what the best path is. Answering this kind of question is the main goal of optimal control theory~\cite{1987Pontryagin,2012Liberzon}, which combines well with inverse engineering problems. However, the choice of the best pathway depends 
on the quantity to be optimized
(time, some kind of cost function\ldots) 
that should thus be defined in the first place.
(See for instance Refs.~\cite{2021Zhang,2021Prados} for specific examples, ranging from ultracold atoms to granular systems.)

\subsection{Minimization of the mean work}\label{sec:irr-work-opt}

\subsubsection{Harmonic connections}\label{sec:irr-work-opt-harmonic}

Although optimal connection problems in the context of finite-time thermodynamics date back to the 1970's and 1980's~\cite{1975Curzon,1982Band,1984Andresen}, the first solution of an optimal connection problem in the ``modern'' context  of stochastic thermodynamics and \STA\ is due to  Schmiedl and Seifert~\cite{2007Schmiedl}. They considered an overdamped Brownian particle submitted to harmonic trapping, where either the position or the stiffness of the trap is controlled. We recall that experiments with colloidal particles are usually well described by the overdamped FP, or Smoluchowski, equation~\eref{eq:FP}~\cite{2017Ciliberto}---see also~\ref{app:stoch-frame}. 

In Ref.~\cite{2007Schmiedl}, the optimal control needed to minimize the mean work for an isothermal process
\begin{eqnarray}
\mean{W}&= \int_0^{t_{\fin}} \!\!dt \int_{-\infty}^{+\infty} \!\!\! dx \, \partial_t U(x,t)  \,\rho(x,t) \nonumber \\
& =\int_0^{t_{\fin}} \!\!dt \int_{-\infty}^{+\infty} \!\!\! dx\, \dot{\boldsymbol{\lambda}}(t) \cdot \partial_{\boldsymbol{\lambda}} U(x,\boldsymbol{\lambda}(t)) \,\rho(x,t) .
\label{eq:mean-work}
\end{eqnarray}
was derived. Note that we are assuming that the potential $U$ depends on time through some externally controlled parameters $\boldsymbol{\lambda}$.\footnote{See~\ref{app:stoch-frame} for a brief account of the definitions of work and heat in the context of stochastic thermodynamics.} Dealing with the variation of the work, an Euler-Lagrange equation for the control parameter was obtained and analytically solved. The assumption of harmonic potential, and therefore Gaussian states, has much to do with the fact that the problem is analytically solvable. The dynamics of the system, in principle codified in the FP equation, can be simplified to an ordinary differential equation for the only relevant moment of the distribution---e.g. its standard deviation, see Eq.~\eref{eq:standard-dev-harmonic}.

It is natural to try to transpose optimal protocols to actual experiments, for example the compression or decompression of a harmonically trapped Brownian particle---by controlling the stiffness of the trap. Still, the experimental implementation of the optimal protocols presents some difficulties. Specifically, the negative values of the stiffness needed for decompression---for short enough connecting times \footnote{From Eq.~\eref{eq:standard-dev-harmonic}, it is seen that for a fast decompression, where $\dot\sigma$ is strongly positive
 in some time window, then $\kappa \simeq -\gamma \dot\sigma /\sigma$,
 and is thus negative in the same time window.}---are experimentally challenging. A step forward to solve this issue has been made by employing an optical feedback trap~\cite{2020Albay_b}. Although the connection considered therein is not optimal, it is neatly shown that it is possible to decompress the Brownian particle in a finite-time with a potential that becomes repulsive---i.e., with negative stiffness---inside a certain time window. Also, the optimal control---the stiffness of the trap---possesses finite discontinuities at the initial and final times in the overdamped limit. These discontinuities have the same formal origin as in the classic problems of finite-time thermodynamics~\cite{1982Band}, the linearity of the Lagrangian in its highest derivative~\cite{1955Newman}. 

Discontinuities in the ``control functions'' are better rationalized in the context of Pontryagin's maximum principle of optimal control theory than within the framework of variational calculus~\cite{1987Pontryagin,2012Liberzon}. In optimal control theory, the control function only has to be piecewise continuous, and thus discontinuities in the control like those appearing in the stiffness of the harmonic trap are treated in a natural and mathematical rigorous way, see e.g. Ref.~\cite{2019Plata}. More general potentials, beyond the harmonic case, represent a challenge because solving the FP equation cannot be mapped onto solving an ordinary differential equation.  However, numerical minimization of the mean work has been carried out, which shows that the predicted discontinuities are robust features of the optimal control~\cite{2008Then}. In the underdamped case, the discontinuities of the control become harsher, they do not involve finite jumps but delta peaks~\cite{2008Gomez-Marin}.

Optimal harmonic connections considering the limitation stemming from bounded stiffness have also been investigated in the overdamped case \cite{2019Plata}. Specifically, the stiffness has been assumed to be bounded between 0 and a maximum value, $0<\kappa(t)<\kappa_{\max}$. Pontryagin's principle provides the adequate framework to solve such constrained optimal problem. The time evolution of the stiffness turns out to be built by two pieces. In the first one, the equations for the protocol are similar to those coming out from the unconstrained problem~\cite{2007Schmiedl} whereas, in the second piece, the control is kept fixed and equal to one of its limiting values. These two pieces smoothly match, in the sense that the dynamical variable---the variance of the position of the Brownian particle---is continuous and has continuous time derivative. As a consequence of the bounds, and depending on the target value of the stiffness and the desired connection time, the target state may become inaccessible. When the connection is possible, the minimum work is greater than the minimum one found for the unconstrained case, the difference between them becomes large in some situations. Similarly to the unconstrained case, the control develops finite jumps at the initial and final times.

\subsubsection{Beyond the harmonic case}\label{sec:irr-work-opt-nonharmonic}

\setcounter{footnote}{0}
After the problem of optimally connecting Gaussian states---in the sense of minimizing the mean work---our interest goes to the analytical derivation of the optimal connection for an arbitrary nonlinear potential---still in the isothermal case. This challenging problem has been first addressed in a series of related papers~\cite{2011Aurell,2012Aurell,2012Aurell_b,2013Muratore,2013Muratore_b,2014Muratore}. 
Starting from the Langevin description, instead of the equivalent FP equation for the PDF, they show that the above question  has quite a general answer in the overdamped limit.  Building on long-established relations between dynamical systems and stochastic control theory~\cite{1981Accardi,1983Guerra}, the minimization of the average heat released to the reservoir can be mapped onto an optimal mass transport problem, ruled by the Burgers equation, and explicitly solved for several physical situations~\cite{2011Aurell,2012Aurell}. Since optimizing average heat and work is the same problem,\footnote{Note that the first principle states that $\mean{Q}+\mean{W}=\Delta U$. As the initial and final states of the \STA\ are fixed, $\Delta U$ is also fixed and minimizing $\mean{W}$ thus entails minimizing $-\mean{Q}$, i.e. the average heat released to the thermal bath.} for the particular case of a harmonic trap the optimal protocol that minimizes the mean work in Refs.~\cite{2007Schmiedl,2008Schmiedl} is recovered. Once more, the optimal control presents discontinuities at the initial and final times.\footnote{The authors addressed the regularization of these discontinuities by introducing a penalty for the current acceleration~\cite{2012Aurell_b}. A different ``surgery'' procedure, which also introduces boundary layers, has been considered to avoid these jumps~\cite{2019Plata}.} It is worth stressing that the minimization of heat is shown to be equivalent to the minimization of entropy production, providing a refined version of the second law of thermodynamics~\cite{2012Aurell}. In general, this line of research evidences that  differential geometry concepts can also be useful to investigate optimization problems in stochastic thermodynamics~\cite{2013Muratore_b,2012Sivak}.

The generalization of the above results to Markov jump processes, governed by a master rather than a FP equation, has also been carried out~\cite{2013Muratore}. In the continuum limit, the results converge to those  previously described. It has been shown that optimal protocols---in the same sense of minimizing entropy production---for systems with discrete states may involve non-conservative forces~\cite{ref2_01}. Here, non-conservative means that the logarithm of the ratio of forward and backward rates cannot be written as a difference of state functions. Also, the relevance of information geometry concepts like the Wasserstein distance has been unveiled in this context~\cite{ref2_02}. Interestingly, the Wasserstein distance is closely related to entropy production on quite general grounds, also for continuous states~\cite{2021Nakazato,2022Dechant}.

Furthermore, the possibility of extending the above results to the underdamped, Langevin-Kramers, case has been investigated~\cite{2014Muratore}. Therein, the emergence of singularities and also of momentum dependence in the optimal driving potential makes the situation less clear-cut than that found in the overdamped situation. Recently, the minimization of the work for the particular case of a counterdiabatic connection in the underdamped regime has been investigated~\cite{2021Li_b}, but the driving is once more velocity-dependent. In fact, the derivation of the optimal conservative driving potential  $U(x,t)$ that provides the minimum irreversible work in the underdamped situation is, to the best of our knowledge, an open question. An advance in this direction has been accomplished by working out a momentum-independent protocol that approximates the momentum-dependent optimal protocol for counterdiabatic driving~\cite{2022Li}. Therein, the momentum-dependent terms of the auxiliary Hamiltonian are removed by combining a variational method with a gauge transformation.

In the overdamped limit, the general problem for the optimization of the mean work performed during the connection between arbitrary states has also been worked in the FP framework~\cite{2019Zhang,2020Zhang}. Specifically, the starting point is the FP equation for the probability density $\rho(x,t)$, written as a continuity equation, Eq.~\eref{eq:FP-continuity}. Still, the main role is played by the cumulative distribution introduced in Eq.~\eref{eq:cumulative-distr}.
Making use of the method of characteristics, the general results for the optimal connecting potential obtained from the Langevin equation~\cite{2011Aurell} are recovered. Here, we give the main results for deriving the minimum work and the associated optimal protocol---for a more detailed derivation thereof, see~\ref{app:Aurell-Zhang-opt-work}.

For a quasi-static process in which the system remains at equilibrium for all times, the average work $\mean{W}$ equals the free energy difference $\Delta F$ between the final and initial states. For a finite-time process, the second principle implies $\mean{W}>\Delta F$ and the irreversible (or excess) work is defined as
\begin{equation}
    W_{\irr}\equiv \mean{W}-\Delta F \geq 0.
\end{equation}
Starting from Eq.~\eref{eq:mean-work} for the mean work, repeated use of integration by parts and the FP equation leads to
\numparts
\ba 
 W_{\irr}=\int_0^{t_{\fin}} dt\, P_{\irr}(t), \\
  P_{\irr}(t)=\gamma \int_{-\infty}^{+\infty} \!\!\! dx \, v^2(x,t) \rho(x,t) \geq 0, 
\ea
\endnumparts
i.e. $P_{\irr}$ stands for the irreversible power in the considered finite-time process and $v$ is given in \eref{eq:veloc-field-FP}. Note that, consistently with our discussion above, $W_{\irr}$ (or $P_{\irr}$) vanishes for a reversible process only: $v(x,t)$ then identically equals zero---and thus the PDF has the equilibrium shape $\rho(x,t)\propto e^{-\beta U(x,t)}$---for all $(x,t)$. 

The problem of minimizing the average work $\mean{W}$ is then equivalent to finding  the fields $\rho(x,t)$ and $v(x,t)$ such that $W_{\irr}$ becomes minimum, while verifying the FP equation~\eref{eq:FP-continuity}. This can be done by employing the method of Lagrange multipliers, introducing an auxiliary field $\psi(x,t)$ and seeking the unconstrained minimum of
\begin{equation}
    J[\rho,v,\psi]\equiv \int_0^{t_{\fin}}\! dt \!\int_{-\infty}^{+\infty} \!\!\! dx\, L, \; L=v^2\rho+\psi\left[ \partial_t\rho + \partial_x (\rho v )\right].
\end{equation}
The Euler-Lagrange equations for this problem read
\begin{equation}
    \partial_t\psi + v\partial_x \psi=v^2, \quad \partial_x \psi=2 v,
\end{equation}
plus Eq.~\eref{eq:FP-continuity}. Combining them, one gets the closed equation $\partial_t\psi+\frac{1}{4}\left( \partial_x\psi\right)^2=0$, which is nothing but the Burgers equation for the auxiliary field $\psi(x,t)$~\cite{2011Aurell}. Equivalently, one may write the Burgers equation in terms of the velocity field $v(x,t)$,
\begin{equation}\label{eq:Burgers}
    \partial_t v+ v \partial_x v=0.
\end{equation}

In order to find the optimal profiles for $\rho(x,t)$ and $v(x,t)$---note that the latter provides us with the driving potential $U(x,t)$, the FP equation \eref{eq:FP-continuity} and the Burgers equation \eref{eq:Burgers} must be simultaneously solved. The solution is
\numparts
\ba
\label{eq:v-rho-sol-main}
    v=  \varphi(x-vt), 
    \label{eq:v-sol-main}\\
    \rho(x,t)=  \frac{\rho_{\ini}(x-vt)}{1+t\varphi'(x-vt)},
    \label{eq:rho-sol-main}
\ea
\endnumparts
where 
\begin{equation}
    \rho_{\ini}(x)\equiv \rho(x,t=0)
\end{equation}
is the initial distribution and $\varphi$ is---for the time being---an arbitrary function. 

The function $\varphi(x)$ is determined by the system reaching the target distribution at the final time $t=t_{\fin}$, i.e. $\rho(x,t_{\fin})=\rho_{\fin}(x)$. This condition is easier to implement by employing the cumulative distribution introduced in Eq.~\eref{eq:cumulative-distr}. Specifically, the initial and final cumulative distributions $\mathcal{F}_{\ini,\fin}$, corresponding to the initial and final probability distributions $\rho_{\ini,\fin}$, are needed. After defining
\begin{equation}
    \label{eq:Xi-def}
    \Xi(x)\equiv \mathcal{F}_{\ini}^{-1}\left(\mathcal{F}_{\fin}(x)\right),
\end{equation}
one gets
\begin{equation}\label{eq:phi-Xi}
    \varphi(x)=\frac{\Xi^{-1}(x)-x}{t_{\fin}},
\end{equation}
where $A^{-1}(x)$ stands for the inverse function of $A$, i.e. $A^{-1}(A(x))=x$. Note that finding $\varphi$ makes it possible to obtain the driving potential, making use of Eq.~\eref{eq:v-rho-sol-main} and the definition of  the velocity field $v(x,t)$, Eq.~\eref{eq:veloc-field-FP},
\begin{equation}
    \partial_x U=-\gamma v -k_B T\partial_x\ln\rho.
\end{equation}

The irreversible power over the optimal protocol is shown to be
\begin{eqnarray}
\label{eq:Pirr-opt}
    P_{\irr}^* \,= \,
    \frac{\gamma}{t_{\fin}^2}\int_{-\infty}^{+\infty} \!\!\! dx\, \rho_{\ini}(x) \left[\Xi^{-1}(x)-x\right]^2.
\end{eqnarray}
Note that, as emphasized by our notation, the optimal irreversible power does not depend on time; it is a constant proportional to $t_{\fin}^{-2}$. The irreversible work immediately follows,
\begin{eqnarray}
\label{eq:Wirr*}
    W_{\irr}^*=t_{\fin} P_{\irr}^*=\frac{\gamma}{t_{\fin}}\int_{-\infty}^{+\infty} \!\!\! dx\, \rho_{\ini}(x) \left[\Xi^{-1}(x)-x\right]^2,
\end{eqnarray}
which is then proportional to $t_{\fin}^{-1}$. Quite expectedly, it vanishes in the limit $t_{\fin}\to\infty$, where the optimal process tends to be quasi-static.

In general, the optimal potential stemming from the optimal velocity field, by employing Eq.~\eref{eq:veloc-field-FP}, has discontinuities at both the initial and final times. For the harmonic case, this approach leads to the optimal connections that can be worked out by using simpler methods, as described in Sec.~\ref{sec:irr-work-opt-nonharmonic}. For the general non-harmonic case, finding the optimal driving potential in an explicit closed form can  only be done in a few examples~\cite{2019Zhang,2020Zhang}.  The main difficulty stems from the expression for the mapping $\Xi(x)$ defined in Eq.~\eref{eq:Xi-def}: only in simple cases is it possible to calculate explicitly the cumulative distributions $\mathcal{F}_{\ini}$ and $\mathcal{F}_{\fin}$, and to invert $\mathcal{F}_{\ini}$ and compose it with $\mathcal{F}_{\fin}$. 

The above difficulties limit the usefulness of the exact optimal protocol for practical implementations. Non-optimal driving potentials but with values of $W_{\irr}$ close to the optimal one, which can be expressed in closed form and do not present discontinuities, can be derived with the welding procedure introduced in Ref.~\cite{2021Plata}---see also Sec.~\ref{sec:fast-forward}. Therein, the authors look for the (sub)optimal connection belonging to fast-forward protocols that leads to minimal work. Remarkably, such a protocol conserves the property of the global optimum of delivering work at constant power.

Linear response theory~\cite{2008Marconi} is a standard framework for understanding  nonequilibrium fluctuations. In fact, the problem above---i.e. optimization of the mean work performed during the connection between arbitrary states---has also been addressed using linear response theory~\cite{2012Sivak,2014Bonanca}.  Sivak and Crooks obtained the optimal protocol minimizing the irreversible work in the linear regime, making use of information geometry concepts---see Sec.~\ref{sec:speed-limits} for further details. Once more, this approach leads to an optimal work with constant power~\cite{2007Crooks}. However, the assumption of linear response prevents this analysis from capturing the discontinuities  of the optimal potential at  the initial and final times. Bonan\ifmmode \mbox{\c{c}}\else \c{c}\fi{}a and Deffner deepened in the linear response approach, thoroughly discussing its range of validity and comparing its predictions with exact results~\cite{2014Bonanca}.

\subsection{Optimization of other figures of merit}
\label{subsec:OCT-other-fig-merit}

In the previous subsection, we have gone over the literature related to the optimization of average work. Nonetheless, the optimal approach in the context of \STA\ is not limited to the minimization of the average work. Herein, we present different studies where the optimization of other relevant quantities has been examined.

The minimization of the statistical error of the free energy---an estimator of the difference between the free energy obtained after averaging over a certain number of individual measurements and the real free energy change---has been addressed \cite{2010Geiger}. Studying this quantity makes it possible to estimate the number of experiments needed to attain a certain accuracy when performing free energy measurements. Analytical results are not available even for simple (harmonic) cases, but numerical optimization provides step-like protocols with a significant reduction of the statistical error. Interestingly, this problem is intimately related to the reverse process~\cite{2007Lechner,2020Albay}. 

Minimization of the average work carries no insight on work fluctuations. Hence, interest aroused in the analysis of alternative figure of merits, combining information of both average and fluctuations. Solon and Horowitz studied the minimization  of an objective function, which is a linear combination of the work average $\langle W \rangle$ and the work standard deviation $\sigma_W$, specifically $\mathcal{J}_\alpha= \alpha \langle W \rangle + (1-\alpha) \sigma_W$,  \cite{2018Solon}. By varying the coefficient $\alpha$ from $0$ to $1$, the weight of work fluctuations (mean work) is reduced (increased) in the search of the optimum protocol. The notion of Pareto-optimal solutions is applied to classify all possible optimal protocols.  Due to the mathematical complexity of the optimization problem, this  is carried out numerically by tuning the value of the control in a finite set of times.  A first-order phase transition is found when illustrating the optimization procedure above in a quantum dot. 

The physical approach to information and memory storage~\cite{1991Landauer,2012Berut,2015Parrondo} may also give rise to optimization problems in the general framework described in this review.  Landauer's principle states the minimum heat cost exhausted to erase one random bit stored in a memory device~\cite{1961Landauer,2021Dago}. Such a bound refers to a quasistatic process. Finite-time processes have been considered as well, posing new relevant problems, not only from the theoretical point of view but also from the technological one, where fast computational operations with memory devices are necessary~\cite{2018Chiuchiu,2020Proesmans,2020Touzo,2021Zhen,2021Lucero,2021Boyd, ref1_04}.

All previous instances belong in the optimization of energetic observables. Remarkably, optimal problems involving time
have also been investigated. The minimization of time related observables is closely linked to information geometry concepts and the so-called speed limits, which are discussed in Sec.~\ref{sec:speed-limits}. Below, we briefly report some results that have been obtained outside the general framework of information geometry ideas.

The optimal static external potential required to minimize the escape time $\tau$, of a Brownian particle confined in a box has been investigated~\cite{2012Palyulin,2020Chupeau}. The escape time is defined here as the mean first passage time to the end of the box, and it is considered that the external potential does not introduce any bias between the starting and final points. Rather surprisingly, in the overdamped regime, the escape time
can approach zero arbitrarily close \cite{2012Palyulin}, which requires
divergent and strongly ``squeezing'' potentials \cite{2012Palyulin,2020Chupeau}.
However, when some constraints are considered (e.g., the maximal potential difference is below  a certain threshold $\Delta U$) an expression, which reminds Heisenberg's time-energy uncertainty principle, is obtained $\tau_{opt} \Delta U=\textnormal{const}$. Related to this, the optimization problem for static external potential minimizing the first passage time to a certain target, distributed according to a certain symmetric probability distribution  with respect the initial position, has been also worked out~\cite{2017Kusmierz}. On a different note, in the context of stochastic resetting, optimal first passage time is a hot topic~\cite{2011Evans,2020Evans,2014Kusmierz,2020Besga,2021Stanislavsky}. Usually, in those studies, the external potential is fixed and the resetting rate is the object that plays the role of the external control.

Another relevant time optimization problem is the minimization of the connection time between the initial and target states. For the \STA\ between equilibrium states, the minimum connection time is zero for the unconstrained problem---similarly to the situation described above for the escape time of the Brownian particle. A different situation arises when the connection problem has additional restrictions, for example when the protocol has to be adiabatic in the thermodynamic sense of zero average heat. Therein, the adiabatic constraint together with the second principle gives rise to the emergence of a speed limit, i.e. the emergence of a minimum time for the adiabatic connection~\cite{2020Plata_b}. In general, the instantaneous adiabat does not exist, and moreover there appear forbidden regions beyond the quasi-static curve $T^2/\kappa=\textnormal{const}$ at which the minimum connection time diverges---as discussed in Sec.~\ref{sec:engineering-thermal-env}. Another problem in which a non-vanishing minimum connection time emerges is the thermal bath engineering of harmonically confined Brownian particles~\cite{2022Patron}. Therein, the restriction comes from the bath temperature being bounded---in a certain interval, from a practical point of view, and to non-negative values, from a fundamental point of view. The optimal protocol, i.e. the brachistochrone, is of bang-bang type and comprises as many bangs as the dimension of the system, with the temperature alternating between its maximum and minimum value. As a consequence, the minimum connection time increases with the dimension of the system  $d$, even for spherically symmetric confinement.

\subsection{Information thermodynamics and speed limits}\label{sec:speed-limits}

In the previous sections, we have discussed the optimization of the work and other figures of merit over \STA\ protocols---including the minimization of the connection time for some specific situations. Here, we focus on the emergence of the so-called classical speed limits, which are closely related to information thermodynamics concepts, and their relevance in the context of \STA. 

The acceleration of the connection entailed by \STA\ protocols comes at a price: for example, we have already discussed that there appears a non-vanishing irreversible contribution to the average work, which only vanishes for infinite connection time, for the \STA\ connecting equilibrium states at constant temperature. In fact, the minimum irreversible work, as given by Eq.~\eref{eq:Wirr*}, is proportional to $t_{\fin}^{-1}$ and then blows up for very short connecting times. Therefore, a natural question arises, whether or not there exists a speed limit for \STA, i.e. a minimum value for the connection time $t_{\fin}$.

The existence of a speed limit in the context of quantum mechanics dates back to the 1940s~\cite{1991Tamm}. It is related to the time-energy uncertainty relation; for review of the subject, see Ref.~\cite{2017Deffner}. The simplest situation is that of the time evolution of a pure state in a conservative system. In this case, there appear two inequalities for the time $t_{\fin}$ necessary to evolve from an initial state $\ket{\psi(0)}$ 
to an orthogonal target state $\ket{\psi(t_{\fin})}$, $\braket{\psi(0)|\psi(t_{\fin})}=0$. 
First, the Mandelstamm-Tamm bound~\cite{1991Tamm}, $t_{\fin}\Delta H\geq \pi \hbar/2$, and, second, the Margolus-Levitin bound~\cite{1998Margolus}, $t_{\fin}\braket{H}\geq \pi \hbar/2$. In these expressions, $\braket{H}$ and $\Delta H$ are the (time-independent) expectation value and standard deviation of the energy, respectively. In fact, it has been proven that the combination of the Mandelstamm-Tamm and Margolus-Levitin inequalities gives the tightest bound for $t_{\fin}$~\cite{2009Levitin}, i.e.
\begin{equation}
    t_{\fin}\geq\max\left(\frac{\pi\hbar}{2\Delta H},\frac{\pi\hbar}{2\braket{H}}\right).
\end{equation}
This inequality expresses the existence of a \textit{natural} time scale, which cannot be beaten for the evolution of a quantum system with time-independent Hamiltonian $H$.

In statistical mechanics, only more recently the possible existence of inequalities resembling the quantum time-energy uncertainty relation---and the possible associated emergence of a speed limit---has been investigated~\cite{2012Sivak,2018Okuyama,2018Shanahan,2018Ito,2018Nicholson,2018Shiraishi,2020Ito,2020Nicholson,2020Falasco,2020Plata_b,2020Rosales-Cabara,2021Prados,2021Nakazato,2022Yan}. One of the first derivations of an inequality for the connection time is probably that of Ref.~\cite{2012Sivak}, within the linear response regime. For a review of these ideas, see Ref.~\cite{2020Deffner}. Let us consider a general system with control functions $\boldsymbol{\lambda(t)}$, and a protocol that drives the system from an initial state with $\boldsymbol{\lambda}(t=0)=\boldsymbol{\lambda}_{\ini}$ to the target state with $\boldsymbol{\lambda}(t_{\fin})=\boldsymbol{\lambda}_{\fin}$; the initial and final states must be close enough in order to use linear response theory. The irreversible power can be written as a bilinear function of the time derivative of the control functions $\boldsymbol{\lambda}(t)$, with coefficients provided by the---positive definite---friction tensor $\boldsymbol{\zeta}(\boldsymbol{\lambda})$,
\begin{equation}\label{eq:Pirr-zeta-rel}
    P_{\irr}(t)=\frac{d\boldsymbol{\lambda^T(t)}}{dt} \cdot \boldsymbol{\zeta}(\boldsymbol{\lambda(t)}) \cdot
     \frac{d\boldsymbol{\lambda(t)}}{dt},
\end{equation}
where the superindex $T$ stands for transpose. This expression naturally introduces a metric in the problem and, in fact, a \textit{statistical length} $\calL_{\lin}$~\footnote{The subindex ``$\lin$'' stresses that this definition of statistical length is valid in linear response, see below for a more general definition based on the Fisher information.} is defined as 
\begin{equation}
    \calL_{\lin}=\int_{0}^{t_{\fin}} dt \sqrt{P_{\irr}(t)}.
\end{equation}
This length somehow measures the distance swept by the system in parameter space along the \STA\ path from the initial to the target state. The connection time $t_{\fin}$ and the irreversible work $W_{\irr}$ are shown to verify the inequality
\begin{equation}\label{eq:sl-Sivak-Crooks}
    t_{\fin}\, W_{\irr} \geq \calL_{\lin}^2.
\end{equation}
The equality only holds when $P_{\irr}$ is constant, as it was for the protocol that minimizes the irreversible work---but note that Eq.~\eref{eq:Pirr-opt} holds for arbitrary initial and final equilibrium states, not necessarily close, and thus it is not restricted to linear response.\footnote{A perturbative approach to solve the FP equation has been introduced with the aim of extending the range of validity of Sivak and Crook's results~\cite{2021Wadia}.}

Speed limits have been discussed in a broader context, beyond the linear response regime~\cite{2018Ito,2018Shiraishi,2018Nicholson,2019Hasegawa,2019Dechant,2020Ito,2020Falasco,2020Plata_b,2020Rosales-Cabara,2020Nicholson,2020Ito,2021Prados,2021Nakazato,2022Yan}. Here, we focus on processes that involve a net transformation of states; for time-periodic or stationary processes, there are specific inequalities showing that the entropy production rate bounds the rate at which physical processes can be carried out~\cite{2020Falasco}---which has been experimentally checked~\cite{2022Yan}. On the one hand, the first results were derived under the assumption of Markovian dynamics~\cite{2018Ito,2018Shiraishi} and are thus restricted to system with dynamics described by master, for discrete variables, and FP (or Langevin) equations, for continuous variables. On the other hand,  approaches based on information geometry concepts hold for general dynamics~\cite{2020Nicholson,2020Ito}, not necessarily Markovian. Key to the latter results is the Fisher information $I(t)$,
\begin{equation}
    I(t)\equiv \mean{\left(\partial_t \ln P\right)^2}=\int d\mathbf{x} \frac{\left(\partial_t \rho(\mathbf{x},t)\right)^2}{\rho(\mathbf{x},t)},
\end{equation}
which is the curvature of the Kullback-Leibler divergence,
\begin{equation}
    \int\! d\mathbf{x} \rho(\mathbf{x},t+dt) \ln\! \left[\! \frac{\rho(\mathbf{x},t+dt)}{\rho(\mathbf{x},t)}\right]\!=\frac{1}{2}(dt)^2 I(t)+O(dt)^3.
\end{equation}
The Fisher information can be connected with entropy production and the so-called thermodynamic uncertainty relations~\cite{2018Nicholson,2018Ito,2019Hasegawa,2020Ito}, and it is directly linked with the thermodynamic length $\calL$, first introduced in the 1980s~\cite{1983Salamon,1985Salamon,1985Feldmann}. For equilibrium systems, this relation was addressed in a pioneering work by Crooks~\cite{2007Crooks}, which showed that
\begin{equation}
    \calL=\int_0^{t_{\fin}} dt\, \sqrt{I(t)}.
\end{equation}
Note that, in general, $\calL\ne\calL_{\lin}$. 

Also relevant to our discussion is the divergence---also called the thermodynamic cost~\cite{2020Ito}---of the path,
\begin{equation}
    \calC=\frac{1}{2}\int_0^{t_{\fin}} dt\, I(t).
\end{equation}
The Cauchy-Schwarz inequality makes it possible to establish the following lower bound for the connection time,
\begin{equation}\label{eq:sl-Ito-Dechant}
  2 t_{\fin} \,\calC \geq \calL^2.
\end{equation}
Interestingly, Eq.~\eref{eq:sl-Ito-Dechant} is implicitly written in Ref.~\cite{2007Crooks} [see Eq.~(9) therein], although it has not been explicitly stated as establishing a speed limit for finite-time \STA\  until recently---see Ref.~\cite{2018Ito} for the case of Markovian dynamics and Ref.~\cite{2020Ito} for arbitrary dynamics. At variance with the quantum case, it must be remarked that the tightness of the bound in Eq.~\eref{eq:sl-Ito-Dechant} has not been proven, to the best of our knowledge. Indeed, Eq.~\eref{eq:sl-Ito-Dechant} is valid for an arbitrary dynamics---see Ref.~\cite{2021Prados} for an analysis thereof for granular fluids described at the kinetic level, where the PDF 
obeys the non-linear (inelastic) Boltzmann equation---but there may exist a larger bound for the connection time.

Equations~\eref{eq:sl-Sivak-Crooks} and \eref{eq:sl-Ito-Dechant} provide us with two inequalities that must hold in \STA. In principle, one could argue that Eq.~\eref{eq:sl-Sivak-Crooks} was derived under the framework of linear response but it is a direct consequence of the definition of irreversible work and the Cauchy-Schwarz inequality---it is linking $P_{\irr}(t)$ with a Riemannian metric that linear response ensures~\footnote{Very recently, the geometric ideas introduced in Ref.~\cite{2012Sivak} have been extended beyond linear response~\cite{2022Li}. Therein, it has been shown that, for connecting equilibrium states, the protocol that minimizes the energetic cost corresponds to the geodesic path.}. A comparison between the predictions of both inequalities for the harmonic case is given in~\ref{app:inequal-comparison}.

\section{Beyond equilibrium: connecting NESS or arbitrary states}\label{sec:beyond-equilibrium}

In the statistical mechanics context, \STA\ have been mainly employed to connect equilibrium states~\cite{2007Schmiedl,2011Aurell,2012Aurell,2012Aurell_b,2013Muratore,2013Muratore_b,2014Muratore,2013Martinez,2016Martinez,2016Le-Cunuder,2017Ciliberto,2017Li,2018Chupeau_b,2018Chupeau,2020Albay_b,2020Plata_b,2021Bayati,2021Plata,2019Plata,2008Then,2008Gomez-Marin}. The analysis of accelerated connection between NESSs has been initiated in \cite{2020Baldassarri,2021Prados}. The delay in the engineering of \STA\ connections between NESSs stems from some difficulties that are inherent to the initial and final states being non-equilibrium, as explained below.

For the sake of concreteness, we revisit the already addressed paradigmatic example of a colloidal Brownian particle trapped in a one-dimensional potential $U(x,t)$ and immersed in a fluid at equilibrium with temperature $T$. In the overdamped regime, the distribution function $\rho(x,t)$ obeys the FP equation \eref{eq:FP} and the typical \STA\ problem is the connection between two equilibrium states, those corresponding to the initial potential $U_{\ini}(x)$ and the target one $U_{\fin}(x)$. An advantage of the equilibrium situation is our perfect knowledge of the initial and target distributions, $\rho_{\ini}(x)\propto\exp[-\beta U_{\ini}(x)]$ and $\rho_{\fin}(x)\propto\exp[-\beta U_{\fin}(x)]$. 

An especially simple case is that of harmonic confinement, in which the potential $U(x,t)$ is harmonic for all times. As described in Sec.~\ref{sec:inverse-engineering}, the PDF is Gaussian for all times therein, and thus completely determined by its average and variance. Moreover, the evolution equations of the average and variance are analytically solvable in closed form. Therefore, although the intermediate states between the initial and target PDFs are indeed non-equilibrium ones, \STA\ connections can be exactly worked out, both non-optimal~\cite{2016Martinez} and optimal in some sense~\cite{2007Schmiedl,2008Then,2008Gomez-Marin,2019Plata}---as already analyzed in Sec.~\ref{sec:irr-work-opt-harmonic}.

In principle, \STA\ methods are transposable to situations in which the initial and final states are NESSs, instead of equilibrium states. Still, some  problems emerge because there is not a general form, playing the role of the canonical distribution, for the PDF corresponding to any NESS. As a consequence, the initial and target states are not perfectly known in general; this constitutes a first limitation for the catalogue of NESS that can be considered as candidates to be \STA -connected.

The \STA\ connection between two NESS of the Brownian gyrator is the subject of study in Ref.~\cite{2020Baldassarri}.  The Brownian gyrator is an overdamped particle moving in a two-dimensional potential
\begin{equation}
    U(\mathbf{x},t)=\frac{1}{2}\kappa_x(t) x^2+\frac{1}{2}\kappa_y(t) y^2+ u(t) x y, \quad \mathbf{x}\equiv (x,y),
\end{equation}
where $\kappa_x$ and $\kappa_y$ are both positive and $\kappa_x \kappa_y-u^2>0$, in order to have a confining potential. The gyrator is coupled to two heat baths with temperatures $T_x$ and $T_y$ in the $x$ and $y$ directions, respectively. The two-dimensional position $\mathbf{x}$ is a Markov process and the FP equation for its PDF $\rho(\mathbf{x},t)$ reads 
\begin{eqnarray}
    \gamma \partial_t \rho=\partial_x \left(\rho\, \partial_x U\right)+\partial_y \left(\rho\,\partial_y U\right)+k_B \left(T_x \partial_x^2 \rho+T_y \partial_y^2 \rho\right).
\end{eqnarray}
If $T_x=T_y$, and $(\kappa_x,\kappa_y,u)$ are time-independent parameters, the stationary solution of this FP equation is the canonical PDF at temperature $T=T_x=T_y$ corresponding to the static potential $U(\mathbf{x})$. If $T_x\ne T_y$, the stationary solution of this FP equation can be exactly computed and is Gaussian, although it does not have the canonical shape~\cite{2020Baldassarri} and induces a current which is rotational. 

The approach of Ref.~\cite{2020Baldassarri} is quite similar to that of Ref.~\cite{2016Martinez} for the engineered swift equilibration of a Brownian particle moving in a one-dimensional harmonic potential. A non-optimal \STA\ connection of two NESS of the Brownian gyrator, corresponding to different values of the triplet $(\kappa_x,\kappa_y,u)$, can be built in a simple way, because the PDF is Gaussian, not only in the initial and target states, but for all times. Therefore, one can write
\begin{equation}
    \rho(\mathbf{x},t)=\frac{e^{-\frac{1}{2}\nu_1(t) x^2-\frac{1}{2}\nu_2(t) y^2-\nu_3(t) x y}}{2\pi \left[\nu_1(t) \nu_2(t)-\nu_3^2(t)\right]^{-1/2}}.
\end{equation}
The functions $\boldsymbol{\nu}(t)\equiv(\nu_1(t),\nu_2(t),\nu_3(t))$ obey a closed system of first-order ODEs, in which $\boldsymbol{\lambda}\equiv(\kappa_x,\kappa_y,u)$ play the role of control functions and appear linearly. Thus, the controls $\boldsymbol{\lambda}$ can be explicitly written as $\mathbf{A}(\boldsymbol{\nu})+\mathbf{B}(\boldsymbol{\nu})\dot{\boldsymbol{\nu}}$, where $\mathbf{A}$ and $\mathbf{B}$ are certain functions of $\boldsymbol{\nu}$---the exact expression for which is not relevant here and can be found in Ref.~\cite{2020Baldassarri}. In this way, the ``equilibrium swift equilibration'' technique introduced in Ref.~\cite{2016Martinez} is generalized to the connection of two NESS, taking advantage of the simplicity of the mathematical problem for the Brownian gyrator. Although the problem is two-dimensional and the initial and target states are NESS, the main characteristic features that facilitates the \STA\ connection of a harmonically trapped particle still hold: perfect knowledge of the initial and target distributions, Gaussianity of the PDF for all times, and simple enough evolution equations for the relevant variables (making it possible to obtain exact analytical solutions thereof). Interestingly, the Brownian gyrator has been used to build a thermodynamic cycle, which is studied through the lens of thermodynamic geometry~\cite{ref1_03}. This represents an alternative proposal to those described in Sec.~\ref{sec:heat-engines} for the construction of a mesoscopic heat engine.

Another physical situation in which the \STA\ connection between two NESS has been considered is the uniformly heated granular fluid~\cite{2021Prados}. This case is more involved than the Brownian gyrator, due to the intrinsically dissipative character of the dynamics. The evolution equation for the PDF $P(\mathbf{v},t)$ has two terms: a diffusive, FP, term stemming from the stochastic forcing and an inelastic Boltzmann collision term, which makes the evolution equation non-linear in the PDF.

In the long-time limit, the granular fluid reaches a homogeneous NESS due to the balance---in average---of the energy loss in collisions and the energy input by the applied stochastic forcing. Even for this homogeneous NESS, the velocity PDF is non-Gaussian and thus is not completely characterized by the variance---the granular (kinetic) temperature. In addition, nor is the velocity PDF perfectly known: it is necessary to resort to  approximate schemes, 
keeping track of non-Gaussianities that are essential. 
The simplest way of doing so is through the so-called first Sonine approximation, in which the granular fluid is assumed to be described by its granular (kinetic) temperature $T$ and the excess kurtosis $a_2$ (the fourth cumulant of the velocity), 
\begin{equation}\label{eq:P1-Sonine}
  P(\mathbf{v};t)=\frac{e^{-w^{2}}}{\left[\pi v_{T}^2(t)\right]^{d/2}}\left[1+a_{2}(t) 
    S_{2}\left(w^{2}\right)\right], \quad \mathbf{w}\equiv\frac{\mathbf{v}}{v_{T}(t)}.
\end{equation}
The thermal velocity $v_{T}$ is defined by $v_{T}^{2}\equiv 2T/m$ in terms of the granular temperature, and  $S_{2}(x)=x^{2}/2-(d+2)x/2+d(d+2)/8$ is the second Sonine polynomial where $d$ is space dimension.
Within this approximation, the evolution equations of $T$ and $a_2$ constitute a system of two coupled ODES, in which the control function is the intensity of the thermostat $\chi(t)$. The problem of finding the protocol that minimizes the connection time between two NESSs corresponding to different values of the driving intensity, $\chi_{\ini}$ and $\chi_{\fin}$, has been analyzed in Ref.~\cite{2021Prados}. This control problem is, at first, non-trivial, since the evolution equations in the Sonine approximation  are non-linear in $(T,a_2)$ and then not analytically solvable. Still, the control function $\chi(t)$ enters linearly in the evolution equations, and this simplifies the mathematical problem: the optimal controls are of bang-bang type. That is, the optimal control comprises two time-windows, inside each of them $\chi(t)$ is constant and equal to either its maximum possible value $\chi_{\max}$ or its minimum possible value $\chi_{\min}$, with one switching between these extreme values at a certain intermediate time.\footnote{For evolution equations that are linear both in the controls and in the variables, there exist rigorous theorems that ensure that the optimal control is of bang-bang type, with $n-1$ switchings for a system with $n$ variables. For evolution equations that are non-linear in the variables but linear in the controls, Pontryagin's maximum principle implies that a similar situation is expected and, since here $n=2$, there is one switching~\cite{2021Prados}.} For a full-power thermostat, $\chi_{\min}=0$ and $\chi_{\max}=\infty$, the optimal control problem can be exactly solved---the bangs with $\chi_{\max}=\infty$ are instantaneous whereas the bangs with $\chi_{\min}=0$ correspond to time windows where the granular fluid freely cools. Both the thermodynamic length and the information geometry cost can be evaluated over the optimal protocols, being consistent with the recently derived general inequalities that impose the existence of a speed limit for arbitrary dynamics, not necessarily Markovian~\cite{2020Ito,2020Nicholson}---see also Sec.~\ref{sec:speed-limits}. The case of a more realistic thermostat, where $\chi_{\min}>0$ and $\chi_{\max}<\infty$, has been  addressed in the linear response regime~\cite{2022Ruiz-Pino}.

\section{Applications: heat engines and beyond}
\label{sec:heat_engine_and_beyond}

\subsection{Heat engines}\label{sec:heat-engines}

\STA\ have been employed in the last decade to design irreversible heat-engines~\cite{2008Schmiedl,2010Esposito,2012Blickle,2013Bo,2013Deng,2014Holubec,2014Sheng,2014Tu,2015Muratore,2016Holubec,2016Martinez_b,2017Polettini,2017Lee,2018Abah,2019Abah,2019Albay,2020Albay_b,2020Nakamura,2020Plata,2021Li,2021Tu}. Loosely speaking, these heat engines can be considered as the mesoscopic counterparts of the classical, macroscopic, heat engines, in which the branches of the corresponding cycle last for a finite time. Note that we focus on classical heat engines, described at the mesoscopic level by nonequilibrium classical statistical mechanics---their quantum counterparts are thus not addressed in this review. 

Let us go back to the colloidal particle trapped in a harmonic potential of stiffness $\kappa$, and immersed in a fluid at equilibrium with temperature $T$, thus described by the Smoluchowski equation~\eref{eq:FP-harm}. Recall that both the stiffness of the trap and the bath temperature can be time-dependent, with their time-dependence being externally controlled. This system is experimentally realizable, see for example Refs.~\cite{2012Blickle} and \cite{2016Martinez_b} for experimental implementations of Stirling-like and Carnot-like cycles, respectively. Moreover, since it has been shown that active baths can in some situations be mapped into regular ``passive'' baths with an effective temperature, the approach below may be also relevant within the context of active Brownian heat engines~\cite{ref1_05,ref1_06}. 

A difficulty arises in the designing of these mesoscopic heat engines when they involve, as is the case of the Otto or the Carnot cycles, adiabatic---in the sense of zero heat, not in the sense of infinitely slow employed before \cite{Adiabatic}---branches. Complete decoupling of the system from the heat bath is not possible, since the interaction between the mesoscopic object---i.e., the Brownian particle---and the heat bath cannot be switched off. In addition, zero-heat (in average, fluctuations are unavoidable) and isoentropic processes are not equivalent for finite-time protocols like those in \STA. This makes the definition of adiabatic processes a subtle issue in the mesoscopic world~\cite{2008Schmiedl,2013Bo,2016Martinez_b,2019Plata,2020Nakamura,2020Plata}. 

In a pioneering work~\cite{2008Schmiedl}, Schmiedl and Seifert analyzed an irreversible Carnot-like heat engine, built on an overdamped Brownian particle confined in a harmonic trap. This heat engine works cyclically, with two isothermal and two pseudoadiabatic branches~\cite{2008Schmiedl}---like the original Carnot engine---that connects equilibrium states. However, unlike with the original Carnot engine, all the branches are irreversible. More specifically, during the hot (cold) isotherm, the system is in contact with a heat bath at temperature $T_h$ ($T_c$) for a certain finite time $t_h$ ($t_c$). In the \textit{instantaneous pseudoadiabats}, the temperature of the system suddenly jumps from $T_h$ to $T_c$ (or vice versa) while keeping the probability distribution for the position of the Brownian particle unchanged. Therefore, what remains constant over the adiabats is the configurational contribution to the entropy or, equivalenty, the ratio $T/\kappa$.  As already noted in Ref.~\cite{2008Schmiedl}, this means that there appears a nonzero heat---and a nonzero entropy increment---along the pseudoadiabatic branches associated with the instantaneous change of the kinetic contribution to the energy. This is the reason why we term these branches pseudoadiabatic, to differentiate them from the actual adiabats, over which the heat vanishes in average, as described below.

This heat engine is optimized in the following way: keeping the isotherm times $t_h$ and $t_c$ fixed, one can minimize the irreversible work, i.e., maximize the output work $-W$ (which is a function of $t_h$ and $t_c$). Afterwards, one can maximize the output power $P=-W/(t_h+t_c)$ and obtain the corresponding efficiency at maximum power $\teta$, which reads
\begin{equation}\label{eq:etaSS}
    \etaSS=\frac{2\etaC}{4-\etaC},
\end{equation}
where $\etaC$ is Carnot's efficiency,
\begin{equation}\label{eq:etaC}
    \eta_C=1-\frac{T_c}{T_h}.
\end{equation}

The study of efficiency at maximum power is a classic problem in the field of finite-time thermodynamics~\cite{1975Curzon,1977Andresen,1985DeVos,1985Hoffmann,1989Chen,1994Chen}. The so-called Curzon-Ahlborn efficiency at maximum power is
\begin{equation}\label{eq:etaCA}
    \etaCA=1-\sqrt{\frac{T_c}{T_h}},
\end{equation}
which was obtained for an endoreversible heat engine~\cite{1975Curzon}---an engine that operates reversibly but for the irreversible coupling to the heat-baths.\footnote{A microscopic theory for the Curzon-Ahlborn cycle has been proposed~\cite{2021Chen}.} It must be noted that efficiency at maximum power is a rather delicate concept, which in general depends on the parameters with respect to which the maximisation is carried out. Still, even if not completely general, the Curzon-Ahlborn value is a quite frequent bound for the efficiency at maximum power of a wide range of heat engines. In fact, $\etaSS<\etaCA$, although in the limit of small temperature difference, for which $\eta_C\ll 1$, both efficiencies coincide up to second order in $\eta_C$,
\begin{equation}\label{eq:etaCA-SS-comparison}
    \etaCA=\frac{\eta}{2}+\frac{\eta^2}{8}+O(\eta^3), \quad \etaSS-\etaCA=O(\eta^3).
\end{equation}
Nevertheless,  the universality of the efficiency at maximum power up to quadratic order in $\etaC$ is quite a general feature~\cite{2009Esposito_b,2010Esposito,2014Sheng,2014Holubec,2015Sheng,2016Holubec,2020Nakamura,2020Plata}, which can be shown to hold for heat engines with left-right symmetry (switching of the baths entail an inversion of the fluxes)~\cite{2009Esposito}. Less strict conditions for this universality have been derived in Refs.~\cite{2014Sheng,2015Sheng}, which also apply to heat engines without left-right symmetry such as the Feynman ratchet~\cite{2008Tu} or the Curzon-Ahlborn endoreversible heat engine~\cite{1975Curzon}. Besides, this kind of universality has been extended to other figures of merit, beyond the maximum power regime~\cite{2010Salas}.

Different variants of this Carnot-like heat engine have been investigated in the literature. Holubec considered an overdamped Brownian particle moving in a log-harmonic potential~\cite{2014Holubec}. The cycle is exactly the same as in Ref.~\cite{2008Schmiedl}, with two optimal isotherms and two instantaneous pseudoadiabats, adapted to the different binding potential. The efficiency at maximum power also agrees with $\etaCA$ up to second order in $\etaC$. Bo and Celani analyzed a Brownian particle immersed in a fluid with inhomogeneous temperature~\cite{2013Bo}. The linear temperature profile makes it necessary to consider the underdamped description, because of the so-called \textit{entropic anomaly} stemming from the temperature gradient~\cite{2012Celani}. In this work, the authors also derive the condition to obtain quasistatic (reversible) adiabatic branches: $T^2/\kappa$ 
is the quantity that must be kept constant, see Sec.~\ref{sec:engineering-thermal-env}. Martinez et al. employed these quasistatic adiabatic branches to realize experimentally an irreversible Carnot-like engine~\cite{2016Martinez_b}, with a charged colloidal particle immersed in water and optically trapped. The charge of the particle allows for adding a noisy electrostatic force: in this way, the effective temperature felt by the particle can be varied from room temperature to thousands of kelvins. Over the ``adiabats'', there is a nonzero fluctuating heat  that must be taken into account when defining the efficiency of such an engine. The authors analyzed the probability distribution function of the efficiency as a function of the number of cycles and observed the appearance of super-Carnot efficiencies, even far from the quasistatic regime. Tu also investigated a Carnot-like engine built with a Brownian particle in the underdamped description, but with the temperature of the thermal bath being homogeneous~\cite{2014Tu}. Suitable shortcuts are introduced along the isothermal branches, whereas the Brownian particle is decoupled from the thermal bath and submitted to a velocity-dependent force during the ``adiabatic'' branches. Both the decoupling from the thermal bath and the velocity-dependent force limit the experimental feasibility of such an engine, but makes it possible to exactly derive the Curzon-Ahlborn value for the efficiency at maximum power. 

\begin{figure}
    \centering
    \includegraphics[width=0.49\textwidth]{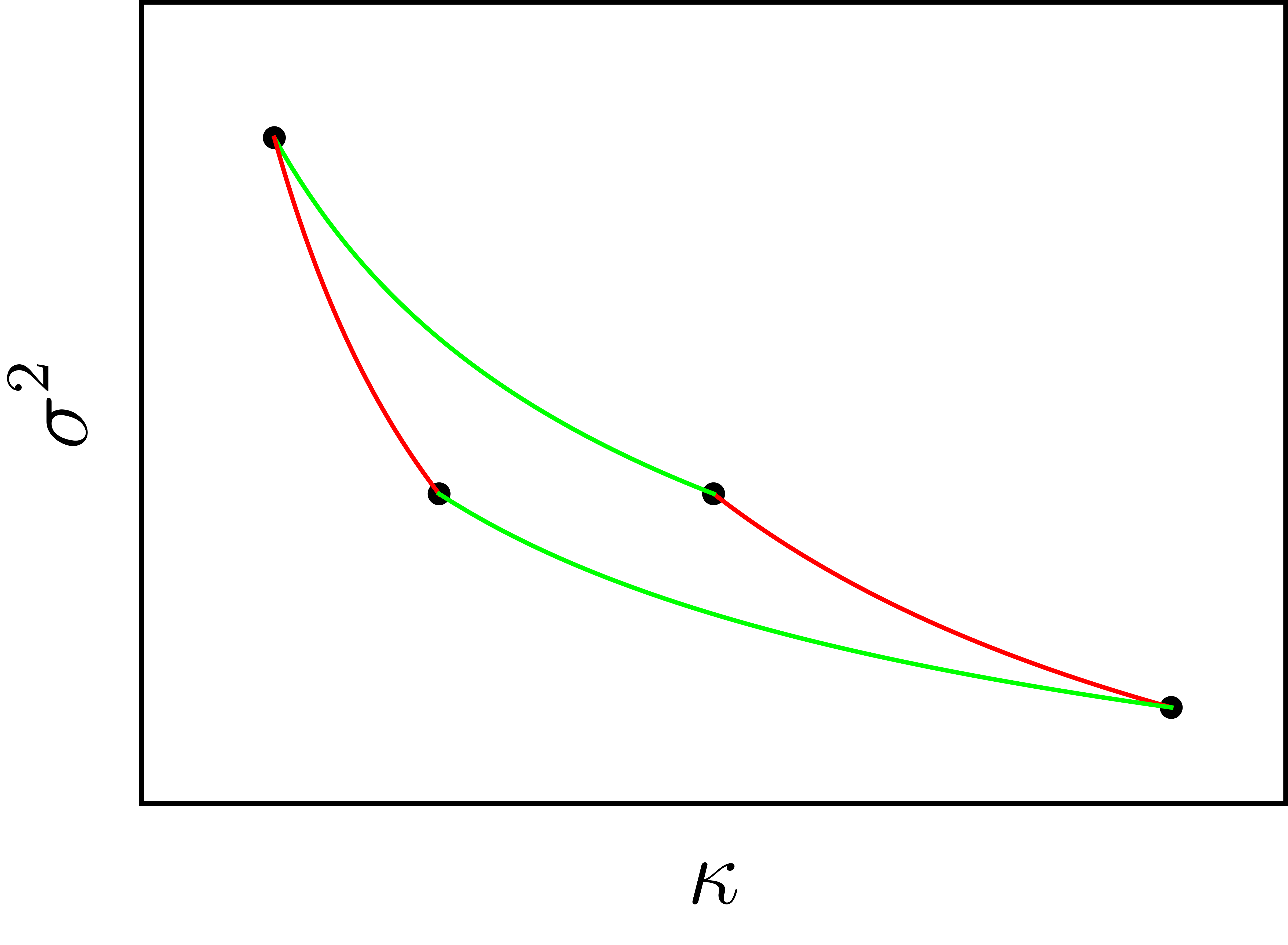}
    \includegraphics[width=0.49\textwidth]{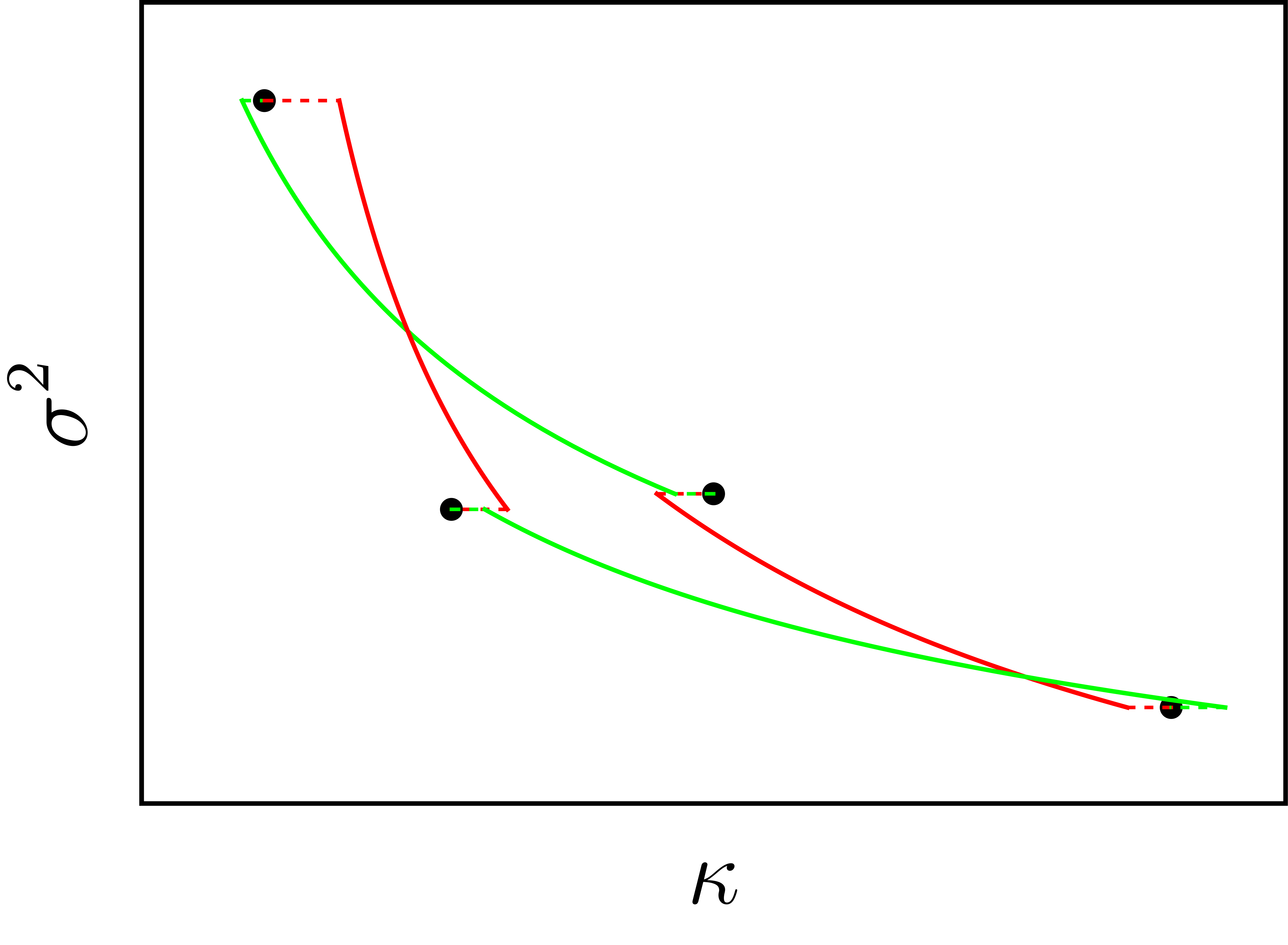}
    \caption{Sketch of a Carnot cycle with a overdamped harmonic oscillator. The cycle is made by two isothermal branches (red lines) and two adiabatic ones (green lines). On the one hand, in the traditional Carnot cycle (top panel), the states are swept quasistatically, and the system keeps at equilibrium at all times. On the other hand, the optimal finite-time version (bottom panel) involves non-equilibrium evolution between operation equilibrium points. See Ref.~\cite{2020Plata_b} for further details.}
    \label{fig:cycles}
\end{figure}

As the discussion above suggests, the construction of the adiabatic branches of the Carnot-like engine is a subtle issue. Plata et al. showed how to build finite-time actually adiabatic---in the sense of zero average heat---branches for an overdamped Brownian particle trapped in an arbitrary nonlinear potential~\cite{2020Plata_b}---see also Secs.~\ref{sec:engineering-thermal-env} and \ref{subsec:OCT-other-fig-merit}. Figure \ref{fig:cycles} illustrates the difference between the classical quasistatic Carnot cycle and its optimal finite-time counterpart.  Nakamura et al. also thoroughly studied the adiabatic connection---again in the sense of zero average heat---for the specific case of harmonic confinement  within the underdamped description of the dynamics~\cite{2020Nakamura}. Extending ideas of the fast-forward protocol---already discussed in Sec.~\ref{sec:fast-forward}, they built a Carnot-like heat engine with fast-forwarded isothermal and adiabatic branches. On the one hand, similarly to the situation in Ref.~\cite{2014Tu}, the extra potential contains terms involving the velocity; the efficiency at maximum power then equals the Curzon-Ahlborn bound for ``weak dissipation" (in the ``strongly'' underdamped regime). On the other hand, for strong dissipation (in the overdamped limit), the efficiency at maximum power is smaller than $\etaSS$, as given by \eref{eq:etaSS}. Nakamura et al. claimed that the overdamped approximation cannot be used to describe the full cycle of a Carnot-like stochastic heat engine. Still, Plata et al. built an irreversible Carnot-like engine with an overdamped colloidal particle trapped in a harmonic potential~\cite{2020Plata}. The main novelty with regard to Ref.~\cite{2008Schmiedl} is the authors' employing of the finite-time adiabatic branches derived in Ref.~\cite{2020Plata_b}, instead of the instantaneous pseudoadiabats that involve a nonvanishing heat. In addition to minimum work (maximum work output) isotherms, minimum time adiabats are employed to construct the Carnot-like cycle. Further optimization of the cycle shows that the efficiency at maximum power of this overdamped Carnot-like heat engine is very close to the Curzon-Ahlborn value throughout  the whole range of temperature ratios $T_c/T_h$. 

It is worth mentioning that several works have addressed the problem of attaining the Carnot efficiency at finite power~\cite{2013Allahverdyan,2015Shiraishi,2016Campisi,2016Koning,2017Lee,2017Polettini,2017Holubec,2018Holubec,2021Cangemi,2021Miura}. Mostly, attaining---or even exceeding~\cite{2021Cangemi}---Carnot's efficiency is connected with the antiadiabatic limit of infinitely fast transformations~\cite{2016Campisi,2017Lee,2017Polettini,2017Holubec,2021Miura}. In Ref.~\cite{2021Miura}, working in the small temperature difference regime, this has been shown to be consistent with the trade-off relation between efficiency and power recently derived on quite general grounds~\cite{2016Shiraishi}.

Other cycles, of the Stirling or Otto types, have been studied in the literature. In fact, the first experimental realization of a mesoscopic heat engine was achieved with the Stirling cycle~\cite{2012Blickle}, which comprises two isothermal and two isochoric branches. A Brownian particle (a single melamine bead of diameter close to $3\mu$m) suspended in water was trapped in a two-dimensional harmonic potential created with optical tweezers. The stiffness of the trap linearly varies with time along the isothermal  branches, whereas it is kept constant in the almost instantaneous processes that connect the isotherms---which are thus isochoric, the average work vanishes. The determined efficiency at maximum power is, within experimental error, in agreement with the Curzon-Ahlborn value. Optimal protocols for the overdamped Stirling engine have been derived in Ref.~\cite{2015Muratore} by a mapping to an optimal mass transport problem, with an efficiency at maximum power given by Eq.~\eref{eq:etaSS}. Also, Stirling engines with active baths have been investigated~\cite{2016Krishnamurthy,2017Zakine}. Non-Gaussianities seem to be responsible for the improved efficiency---as compared to usual engine with non-active baths---of the active engine analyzed in Ref.~\cite{2016Krishnamurthy}. Still, these results are changed if a different definition of the temperature, based on the diffusion constant of the particle without any external potential, is employed~\cite{2017Zakine}.

\STA\ for the Otto cycle have been considered in Ref.~\cite{2013Deng}. In this paper, it is shown that the accelerated connection intrinsic to \STA\ does not only increase power output of the engine but also may increase its efficiency, both in the quantum and the classical case. Abah et al. further investigated the quantum Otto engine, establishing the energetic cost of driving with \STA\ and finding that the efficiency at maximum power is very close to the Curzon-Ahlborn bound~\cite{2019Abah}. 

Recently, the optimization of quite general heat engines has also been considered~\cite{2020Zhang,2020Brandner,2021Movilla,2021Izumida,2022Frim,2022Watanabe}. In Ref.~\cite{2020Zhang}, the cycle comprises two isothermal branches connected by the instantaneous \textit{pseudoadiabats} described above. The driving potential that minimizes irreversible work or heat dissipation is obtained, which in relevant cases leads to results previously derived in the framework of optimal mass transport~\cite{2011Aurell,2012Aurell,2013Bo_b}---see also Sec.~\ref{sec:OCT} of this review. In Ref.~\cite{2020Brandner}, a very general class of microscopic heat engines driven by arbitrary periodic protocols of temperature and mechanical control parameters is considered. For slow driving, a universal trade-off relation between efficiency and power has been derived by using thermodynamic geometry arguments. In Ref.~\cite{2021Movilla}, a system in contact with a heat bath with an arbitrary time-varying peri- odic temperature profile is studied. Optimal control protocols that maximize power or efficiency are derived. In Ref.~\cite{2021Izumida},  a linear response theory for generic Gaussian heat engines obeying Fokker-Planck dynamics is developed. In this context, the structure of the Onsager coefficients is investigated in detail. In Ref.~\cite{2022Frim}, the full space---within the linear response regime---of non-equilibrium thermodynamic cycles is exploredI . Therein, the authors also employ information geometry ideas for deriving an upper bound for the efficiency and building finite-time heat engines that outperform---in terms of efficiency---other recent proposals. In Ref.~\cite{2022Watanabe}, the simultaneous optimization of the average and the fluctuation of the efficiency in a finite-time Carnot cycle has been investigated. The approach is also based on information geometry ideas and holds for slow driving and over a coarse-grained timescale.

\subsection{Other Applications}\label{ssec:otherApplications}

The idea to control the time required to reach a given transformation,
or perform a certain task has a respectable history in the field of
engineering \cite{1958Smith}, where a common situation is that pertaining to 
crane driving (trolley displacement and rope length). Yet, the similarities
between such a macroscopic system and the dynamical equations ruling
the transport of ultracold
ions or neutral atoms in effectively one-dimensional traps
with time-dependent controllable parameters (as used in
 advances towards scalable quantum-information
processing), opens the way for a transfer of method between the two 
fields \cite{2017Gonzalez}.

Besides, \STA\  allow to circumvent some shortcomings of
existing manipulation techniques. In particular, optical manipulations
of colloids are plagued by the  impracticality of creating time-dependent
expulsive confinements, that are required when seeking to deconfine 
a colloidal state \cite{2016Martinez,2018Chupeau}. While feedback protocols offer a way out \cite{2020Albay_b}, we seek a purely feedforward technique, without applying
any retroaction on the system.
Taking advantage of diffusiophoresis, {i.e.} the migration of colloids induced by solute gradients, one can drive the system by
 a proper time-dependent control of the salt concentration
in a buffer in contact with the solution. 
A fast decompression can thereby be achieved
\cite{2021Bayati}, from a joint optical and diffusiophoretic driving.
To remain in the field of soft matter, a related 
proposal of driving a bulk system from its boundaries was put
forward in \cite{2019PalaiaPhD}: by monitoring the potential
(or the charge) difference between the two plates of a nanocapacitor, 
one can accelerate significantly electrolyte relaxation
following a charging of the system.
We also report applications of \STA\ to systems driven by L\'evy noise \cite{Baldovin2022}, as well as
in the field of active matter for active Brownian particles,
to control the system's state by a joint monitoring of activity and the strength of trap confinement.

A rather unexpected application of \STA\ has emerged in the context of biological evolution~\cite{2021Iram}. The setting is provided by the Wright-Fisher model~\cite{1932Wright}, describing the evolution of a population of organisms through a space of $M$ possible genotypes. The state of the population is characterized by a time-dependent probability distribution $\rho(\mathbf{x},t)$, where $\mathbf{x}\equiv (x_1,\cdots x_M)$ is a frequency vector whose component $x_m$ gives the fraction of the population found in genotype $m\in[1,M]$ at time $t$. The  Wright-Fisher model describes the diffusive evolution of $\rho(\mathbf{x},t)$ due to mutations that induce a random walk among genotypes. It is additionally assumed that there exists an externally controlled environmental parameter $\lambda$, corresponding for example to a drug concentration that affects the fitness of a pathogen.
If $\lambda$ is held fixed, then under the  Wright-Fisher model the population evolves to an evolutionary equilibrium state $\rho^{\rm eq}(\mathbf{x},\lambda)$. If the parameter is varied with time, then the actual state of the population $\rho(\mathbf{x},t)$ lags behind the instantaneous equilibrium state $\rho^{\rm eq}(\mathbf{x},\lambda(t))$. Applying the tools of \STA, Ref.~\cite{2021Iram} shows how to construct a counterdiabatic control protocol that eliminates this lag. This opens the possibility of controlling biological evolution to drive a population from an initial equilibrium state to a desired final equilibrium state in finite time. This kind of counterdiabatic driving has been generalized to discrete-state continuous-time Markov networks and applied to control chaperon-assisted protein folding. Interestingly, the obtained control protocols resemble natural strategies observed in actual experiments with \textit{E. coli} and yeast~\cite{2022Ilker}.

\STA\ have also been used to develop strategies for implementing computational tasks rapidly and accurately, and to study the inherent thermodynamic costs associated with these tasks~\cite{2021Boyd,ref1_04}.
Specifically, protocols were designed to create, erase and transfer a single bit of information stored in a double-well potential.
These tasks dissipate energy into the bit's thermal surroundings.
Ref.~\cite{2021Boyd} explored how this dissipated work scales with the rate of computation, the characteristic length scale of the system, and the robustness of the (inherently metastable) information storage.
A key finding of Ref.~\cite{2021Boyd} was that although Landauer's principle~\cite{1961Landauer,2015Parrondo} establishes the minimum dissipated energy needed to erase a bit of information, there is no similar bound on the amount of time needed to erase the bit. 
Using \STA, a bit can be erased arbitrarily rapidly and with arbitrarily high fidelity, provided one is willing to pay the requisite thermodynamic cost in dissipated energy. A tight bound on this irreversible cost has been derived~\cite{ref1_04}.
The results of Refs.~\cite{2021Boyd,ref1_04} illustrate, in a computational thermodynamic setting, the refined second law of thermodynamics obtained in Ref.~\cite{2012Aurell}. Generalizing these
approaches to underdamped systems is challenging 
\cite{2021Dago}.

To conclude, we emphasize that the key ideas behind \STA,
and inverse engineering, can be 
illustrated in an early undergraduate course, with RC, LC or RLC circuits. An experimental demonstration was proposed in 
\cite{2019Faure}, that furthermore provides an original 
approach to the venerable teaching of differential
equations.


\section{Perspectives and conclusions}
\label{sec:conclusions}

Shortcuts to adiabaticity (STA) are finite-time protocols that produce the same state as
an infinitely slow (so-called adiabatic) driving. We have here generalized this 
approach to finite-time protocols connecting arbitrary states, referred to under the terminology of 
\STA\ (Swift State-to-state Transformations). For both Hamiltonian and stochastic dynamics, we have classified \STA\ into different types, inverse-engineering, counterdiabatic (CD), and fast-forward (FF), mirroring the usual categorization of quantum STA. This categorization is indeed useful for presentation purposes but not so clear-cut as it may seem: in the stochastic case, CD protocols can be thought both as a special case of inverse-engineering ones and as a certain limit of FF protocols---which somehow smears the boundaries between the different categories. One of the goals of our review is to highlight the common fundamentals behind all of them. To conclude the paper, we offer below some perspectives that emerge.

For isolated Hamiltonian systems (Sec.~\ref{sec:isolated-systems}) we foresee three topics of potential future investigation. In systems with one degree of freedom, one could drop the assumption that the steering potential $U(x,t)$ is a confining potential with a single minimum, and instead consider a confining potential with a double-well structure, with a local maximum separating the two wells.  If $U_{\rm max}$ gives the value of the potential at this local maximum, then each energy shell with $E>U_{\rm max}$ forms a single closed loop in phase space, while each energy shell $E<U_{\rm max}$ forms a pair of non-intersecting closed loops.  The shell $E=U_{\rm max}$ is a {\it separatrix}, with the characteristic shape of a ``figure-eight'' closed loop.  The adiabatic invariance of the action $I(E,t)$ breaks down in the vicinity of the separatrix~\cite{1986Tennyson} and this in turn leads to subtle topological effects~\cite{2015Lu}, which may have interesting implications for the design of shortcuts for such systems. Next, as already noted in Sec.~\ref{subsubsec:beyond1dof}, it is plausible that Hamiltonian shortcuts can be generalized to integrable systems with $N>1$ degrees of freedom, as such systems effectively decompose into $N$ independent one-degree-of-freedom systems. Finally, as also suggested in Sec.~\ref{subsubsec:beyond1dof}, it would be worthwhile to investigate the design of \STA\ for ergodic and chaotic Hamiltonian systems that constitute {\it canonical families}.  These are systems with the (non-generic) property that $H(\mathbf{x},\mathbf{p},t)$ and $H(\mathbf{x},\mathbf{p},t^\prime)$ are related by a canonical transformation, for any $t,t^\prime \in [0,t_{\fin}]$. Progress on any of these three questions may in turn provide insight into quantum shortcuts for corresponding systems.

Most \STA\ have been developed for one-dimensional systems. As a rule, the generalization of these protocols---either inverse engineering, counterdiabatic, fast-forward, or thermal engineering---to higher-dimensional systems is not straightforward. Still, inverse engineering protocols, akin to those originally devised for a colloidal particle in a one-dimensional harmonic trap~\cite{2016Martinez}, have been worked out for an  arbitrary curved configuration space~\cite{2021Frim}. This shows the feasibility of extending \STA\ to arbitrary geometries, but engineering optimal protocols---in the sense of minimizing work, heat, entropy production, connection time or any other relevant physical property over them---remains a challenge in this case. On a related note, the protocols detailed in our review are mostly designed in the overdamped regime. Future theoretical developments in \STA\ are expected in the underdamped regime to accompany the development of new experimental setups having a tunable damping rate \cite{2017rondin}.

The irreversible work unavoidably appearing in \STA\ has been linked to information geometry ideas in linear response~\cite{2012Sivak}. In the information geometry approach, a key role is played by the Fisher information, which at equilibrium is related to derivatives of the distribution function and, in certain specific situations, to fluctuations of relevant quantities~\cite{1983Salamon,1984Andresen,2007Crooks,2012Crooks}.  This theoretical framework also makes it possible to establish a speed-limit inequality for \STA, Eq.~\eref{eq:sl-Sivak-Crooks}, which resembles the quantum speed limit (the time-energy uncertainty relation~\cite{1991Tamm,1998Margolus}). Recently, the information geometry approach has been employed beyond linear response. This has led to the emergence of several versions of classical speed-limits inequalities---also resembling the quantum time-energy uncertainty relation---for systems described by stochastic dynamics~\cite{2018Shanahan,2018Okuyama,2018Ito,2018Shiraishi,2020Nicholson,2020Ito,2021Nakazato}. Nevertheless, the tightness of the obtained classical bounds, for instance that given by Eq.~\eref{eq:sl-Ito-Dechant}, is not well established---unlike in the quantum case~\cite{2009Levitin,2017Deffner}. Clarifying this specific issue and, on a more general note, linking information geometry concepts and physical quantities beyond linear response are open questions. Answering them constitutes an enticing perspective for future work, and may provide a novel way to tackle and move forward non-equilibrium statistical mechanics and, more specifically, the field of \STA. 

We have exposed here several strategies for the control of systems in contact with a thermostat. Closed-loop control protocols, i.e. those with feedback, have not been discussed since they are intrinsically subject to the system's
own time scales. However, such approaches have the advantage of not requiring a perfect modeling of the system under study. This is to be contrasted with open-loop control protocols detailed in this
review article, that are of the feed-forward type. 
They provide a wide range of strategies to accelerate the transition from a given state to another over an {\it a priori} arbitrarily short amount of time. The methods presented here
provide such solutions, despite the intrinsic randomness associated to the coupling with a reservoir, and even propose to engineer the randomness.

\appendix

\section{Summary of acronyms}\label{app:acronyms}

\begin{tabular}{ll}
CD & Counterdiabatic \\
FF & Fast-forward \\
FP &  Fokker-Planck \\
GCD & Global Counterdiabatic \\
LCD & Lobal Counterdiabatic \\
NESS & Non-equilibrium Steady State \\
PDF & Probability density function \\
STA & Shortcut(s) to adiabaticity \\
\STA  & Swift State-to-state Transformation(s)
\end{tabular}

\section{Stochastic processes framework}\label{app:stoch-frame}

\subsection{Langevin and Fokker-Planck descriptions}\label{appsub:Langevin-FP}

For the sake of concreteness, let us consider a colloidal particle immersed in a fluid in equilibrium at temperature $T$ in the overdamped regime. For one-dimensional motion, the evolution equation for the position $x$ of the Brownian particle is
\begin{equation}
    \label{eq:app-Langevin}
    \gamma \dot x =-\partial_x U(x,t)+ \sqrt{2\gamma k_B T(t)}\, \eta(t),
\end{equation}
where $\gamma$ is the friction coefficient, $U(x,t)$ is the driving potential, $T(t)$ is the temperature, $k_B$ is the Boltzmann constant, and $\eta(t)$ is a Gaussian white noise of unit variance,
\begin{equation}
    \langle \eta(t)\rangle=0, \quad \langle \eta(t)\eta(t')\rangle=\delta(t-t').
\end{equation}
Note that, in general, both the potential and the temperature may depend on time---as explicitly stated in our notation. The diffusion coefficient and the temperature verify Einstein's relation $D=k_B T/\gamma$, so the diffusion coefficient is in general also time-dependent. 
Usually, the friction coefficient $\gamma$ is assumed to be constant, including in the situation 
where colloid temperature is artificially controlled, as discussed below Eq. (\ref{eq:randomx0}),
by randomly shaking the center-of-force of the trapping
\footnote{Temperature control is done by adding an independent stochastic white-noise force as detailed in Refs.~\cite{2013Martinez,2017Ciliberto}, which increases the variance of the noise acting on the Brownian particle and effectively heats it, with the temperature of the background fluid remaining unchanged.}. Different realizations of the noise $\eta(t)$ give rise to different trajectories for the colloidal particle. In the Langevin description, $\langle\cdots\rangle$ thus mean averaging over the different trajectories of the stochastic process.

It is also possible to analyze the stochastic motion of the colloid by using the ensemble picture. Therein, we introduce the PDF $\rho(x,t)$, which obeys the FP equation
\begin{equation}
    \label{eq:app-FP}
    \gamma \partial_t\rho(x,t) =\partial_x \left\{[\partial_x U(x,t)]\rho(x,t)+k_B T(t)\partial_x \rho(x,t)\right\}.
\end{equation}
For time-independent potential and temperature, the FP equation has only one stationary solution that is the equilibrium one, $\rho_{eq}(x)\propto \exp[-\beta U(x)]$, with $\beta=(k_B T)^{-1}$. 

It is interesting to note that the FP equation can be cast in the form of a continuity equation,
\begin{equation}
\label{eq:app-FP-continuity}
    \partial_t \rho(x,t)=-\partial_x J(x,t),
\end{equation}
where the probability current $J(x,t)$ is defined as
\begin{equation}
\label{eq:app-J-FP}
    J(x,t)\equiv -\gamma^{-1}\left\{ \left[ \partial_x U(x,t)\right] \rho(x,t)+k_B T(t)\partial_x\rho(x,t) \right\}.
\end{equation}
Writing the probability current as $J(x,t)=\rho(x,t)v(x,t)$, a velocity field $v(x,t)$ is  introduced as
\begin{equation}
\label{eq:app-v-FP}
    v(x,t)=-\gamma^{-1}\left[ \partial_x U(x,t) +k_B T(t)\partial_x\ln\rho(x,t) \right].
\end{equation}
Note that, despite our assumption of constant $\gamma$, Eqs.~\eref{eq:app-FP}-\eref{eq:app-v-FP} remain valid for time-dependent friction.

\subsection{Stochastic thermodynamics}\label{appsub:stoch-thermod}

Following Sekimoto~\cite{Sekimoto}, we can write down the first principle at the mesoscale by defining stochastic work $W$ and heat $Q$.
As usual, work is the potential energy change associated with the variation of the external parameters $\boldsymbol{\lambda}(t)$ in the potential  (volume, stiffness of the trap, etc.). In other words, we consider 
\begin{eqnarray}
    U(x,t)&=U(x,\boldsymbol{\lambda}(t)), \\ 
    \partial_t U(x,t)&= \boldsymbol{\dot{\lambda}}(t)\cdot  \left[\partial_{\boldsymbol{\lambda}}U(x,\boldsymbol{\lambda})\right]_{\boldsymbol{\lambda}=\boldsymbol{\lambda}(t)}.
\end{eqnarray}
For each trajectory of the stochastic process, i.e. for each realization of the white noise in the Langevin equation, work is thus calculated as
\begin{eqnarray}
    W&=\int_0^{t_{\fin}} \!\! dt\, \left[\partial_t U(x,t)\right]_{x=x(t)} \nonumber \\
    &=\int_0^{t_{\fin}} \!\! dt\, \boldsymbol{\dot{\lambda}}(t)\cdot  \left[\partial_{\boldsymbol{\lambda}}U(x,\boldsymbol{\lambda}(t))\right]_{x=x(t)}.
    \label{eq:app-stoch-work}
\end{eqnarray}
Note that $W$ changes from trajectory to trajectory, i.e. it is stochastic. We can also introduce the stochastic instantaneous power, 
\begin{equation}
    \label{eq:app-stoch-power}
    \dot{W}=\boldsymbol{\dot{\lambda}}(t)\cdot  \left[\partial_{\boldsymbol{\lambda}}U(x,\boldsymbol{\lambda}(t))\right]_{x=x(t)},
\end{equation}
such that $W=\int_0^{t_{\fin}}dt\, \dot{W}$. For each trajectory of the stochastic process, the configurational contribution to  heat is thus the potential energy change due to the time evolution of the particle's position $x(t)$, i.e.
\begin{equation}
\label{eq:app-stoch-conf-heat}
    Q_{\conf}=\int_0^{t_{\fin}} \!\! dt\, \dot{x}(t) \left[\partial_x U(x,t)\right]_{x=x(t)}.
\end{equation}
For mathematical consistency, the above integral must be understood in the Stratonovich sense~\cite{2011Aurell}. Physically, this entails 
\begin{equation}\label{eq:app-Qconf+W}
    Q_{\conf}+W=\int_0^{t_{\fin}} \!\! dt\, \frac{d}{dt} U(x(t),t)=\Delta U,
\end{equation}
where $\Delta U$ is the potential energy increment over the considered trajectory, i.e. $\Delta U$ is also stochastic. 

In the Langevin picture, mean work and heat are obtained by averaging over noise realizations. For example, the average work reads
\begin{equation}\label{eq:app-av-work-Langevin}
\mean{W}=\mean{\int_0^{t_{\fin}} \!\! dt\, \boldsymbol{\dot{\lambda}}(t)\cdot \left[\partial_{\boldsymbol{\lambda}}U(x,\boldsymbol{\lambda}(t))\right]_{x=x(t)}},
\end{equation}
where $\mean{\cdots}$ on the rhs means average over the noise $\eta(t)$. In the equivalent ensemble picture, the PDF $\rho(x,t)$ obeys the FP equation~\eref{eq:app-FP} and one has
\begin{equation}\label{eq:app-av-work-FP}
\mean{W}=\int_{-\infty}^{+\infty} \!\!dx\int_0^{t_{\fin}} \!\! dt\, \boldsymbol{\dot{\lambda}}(t)\cdot \partial_{\boldsymbol{\lambda}}U(x,\boldsymbol{\lambda}(t))\, \rho(x,t).
\end{equation}
Consistently, the average configurational heat ensures that Eq.~\eref{eq:app-Qconf+W} holds in average,
\begin{eqnarray}
    \label{eq:app-av-conf-heat}
    \mean{Q_{\conf}}&=\mean{\Delta U}-\mean{W}=\int_{-\infty}^{+\infty} \!\!dx \int_0^{t_{\fin}} \!\! dt\,U(x,t) \partial_t\rho(x,t) \nonumber \\
    &= \int_{-\infty}^{+\infty} \!\!dx \int_0^{t_{\fin}} \!\! dt\,[\partial_x U(x,t)] v(x,t) \rho(x,t).
\end{eqnarray}

In order to have energy conservation---i.e. the first principle---we must incorporate the kinetic contribution to the energy. In the overdamped regime, the velocity degree of freedom instantaneously relaxes to equilibrium, so that the average kinetic energy $\mean{K}=k_B T(t)/2$ and therefore average heat has a kinetic contribution,
\begin{equation}
    \label{eq:app-av-heat}
    \mean{Q}=\underbrace{\frac{1}{2}k_B \Delta T}_{\mean{Q_{\kin}}}+\mean{Q_{\conf}}.
\end{equation}
The first principle states
\begin{equation}
    \mean{Q}+\mean{W}=\mean{\Delta E},
\end{equation}
where $E$ is the total energy, $E=K+U$. For isothermal processes, $T$ is constant and then there is no kinetic contribution to the heat. Conversely, $\mean{Q_{\kin}}\ne 0$ for non-isothermal processes: this is important, for example, to correctly define finite-time adiabatic (in the sense of zero average heat) processes~\cite{2020Plata_b}---see also Sec.~\ref{sec:engineering-thermal-env}.

\section{Conditions for the existence of Boltzmann's breathers under static confinement}
\label{app:breathers}

We work here in the framework of the Boltzmann equation 
for dilute gases, as described in Sec.~\ref{subsec:boltzmann}.
We have reported that when the external confining potential $U$ is time-independent and harmonic,
permanent (undamped) undriven breathing modes do exist, with a time-dependent and periodic temperature. We address here the reciprocal question: under which conditions on $U$ do such
solutions exist? We restrict to a one-dimensional dynamics 
for simplicity.

We start from the potential given by \eref{eq:BoltzPot}, where 
the contribution $\bar U$ stems from Eq. 
\eref{eq:tdssp}. The confining potential then reads
\begin{equation}
    U(x,t) \,=\, \frac{1}{\sigma^2} U_0\left(\frac{x}{\sigma} \right) - \frac{m}{2} \frac{\ddot\sigma}{\sigma} x^2 \, + \,
    \frac{b}{x^2},
    \label{eq:app_U_U0}
\end{equation}
in which we require that $\sigma$ be time dependent,
with nevertheless a time independent $U$ on the left hand side.
Taking the third derivative with respect to $x$, we arrive at 
\begin{equation}
    \partial_x^3 U \,=\, \frac{1}{\sigma^5} U_0^{'''}\left(\frac{x}{\sigma} \right) - \frac{24\, b}{x^5},
\end{equation}
which should be independent of time, and thus independent of $\sigma$. This implies that $\widetilde x\,U_0^{IV}(\widetilde x)=-5\,U_0^{'''}(\widetilde x)=0$, i.e.  $U_0^{'''}(\widetilde x)$ must be a power-law
function in $1/\widetilde x^5$. Subsequent integration yields
\begin{equation}
    U_0(x) \,=\, \frac{A}{x^2} + B x^2 + C x,
\end{equation}
where $A$, $B$ and $C$ are constants.
Plugging this back into Eq. \eref{eq:app_U_U0} 
and requiring once more the time-independence of $U$,
we get $C=0$. This indicates \begin{itemize}
    \item  what differential 
equation $\sigma$ should fulfil:
\begin{equation}
    \frac{B}{\sigma^4} - \frac{m}{2} \frac{\ddot \sigma}{\sigma}
    \,=\, {\cal C}\!\hbox{onstant}.
\end{equation}
This is an Ermakov-like relation for
which $d^3\beta/dt^3 + 4 \Omega^2 \dot\beta = 0$ with $\Omega^2 = 2\, {\cal C}\!\hbox{onstant}/m$,
see Sec.~\ref{subsec:boltzmann} and in particular Eqs. 
\eref{eq:ermakovSolution} and \eref{eq:PRL2014_9b}.
We recover the 
breathing modes found out by Boltzmann himself,
and described in \cite{1988Cercignani,2014Guery-Odelin}.
\item that 
apart from the $1/r^2$ contribution already mentioned in 
Sec.~\ref{subsec:boltzmann} (here $1/x^2$), $U$ 
should be a harmonic potential. 
\end{itemize}
While the present argument transposes directly
to isotropic potentials of the type
\begin{equation}
     U(\mathbf{r},t) \,=\, \frac{1}{\sigma^2} U_0\left(\frac{r}{\sigma} \right) - \frac{m}{2} \frac{\ddot\sigma}{\sigma} r^2 \, + \,
    \frac{b}{r^2},
\end{equation}
generalization to higher dimensions does not seem straightforward.

\section{Derivation of the expression for the minimum irreversible work}\label{app:Aurell-Zhang-opt-work}


In this appendix, we give more details of the derivation of the expression for the minimum irreversible work. The calculation below somehow puts on a common ground the approaches developed in Refs.~\cite{2011Aurell,2019Zhang,2020Zhang}.

Our starting point is the Burgers equation~\eref{eq:Burgers}, which must be  jointly solved with the FP equation~\eref{eq:FP-continuity} to find the optimal profiles for $\rho(x,t)$ and $v(x,t)$. On the one hand, the characteristic system of ODEs (Lagrange-Charpit equations) for the Burgers equation is
\begin{equation}\label{eq-app:Lagrange-Charpit-Burgers}
    \frac{dt}{ds}=1, \quad \frac{dx}{ds}=v, \quad \frac{dv}{ds}=0.
\end{equation}
Therefore, $s=t$, $v=v_{\ini}$, and $x=v_{\ini} s+x_{\ini}$. At the initial time $t=t_{\ini}=0$, we have that $v_{\ini}=\varphi(x_{\ini})$, where $\varphi$ is an an arbitrary smooth function, and thus $v=\varphi(x_{\ini})$, with $x_{\ini}=x-v t$. The solution of the Burgers equation can be given in implicit form,
\begin{equation}\label{eq-app:v-sol}
    v=\varphi(x-vt).
\end{equation}
On the other hand, the characteristic system of ODEs for the FP equation is
\begin{equation}\label{eq-app:Lagrange-Charpit-FP}
    \frac{dt}{ds}=1, \quad \frac{dx}{ds}=v, \quad \frac{d\rho}{ds}=-\rho\,\partial_x v.
\end{equation}
The two first equations are the same as for the Burgers equation, so we focus on the third one,
\begin{equation}
    \frac{d\rho}{ds}=-\frac{\varphi'(x_{\ini})}{1+s\varphi'(x_{\ini})}\rho,
\end{equation}
which can be readily integrated to give
\begin{equation}
    \rho(s)=\frac{\rho_{\ini}(x_{\ini})}{1+s\varphi'(x_{\ini})}.
\end{equation}
We have taken into account that $\rho(s=0)=\rho_{\ini}(x_{\ini})$. Going back to the original variables $(x,t)$,
\begin{equation}\label{eq-app:rho-sol}
    \rho(x,t)=\frac{\rho_{\ini}(x-vt)}{1+t\varphi'(x-vt)},
\end{equation}
in which $v(x,t)$ is in turn implicitly given by Eq.~\eref{eq-app:v-sol}. Equation~\eref{eq-app:rho-sol} can also be written in an equivalent form: making use of Eq.~\eref{eq-app:v-sol}, $v_x=\left(1-t v_x\right)\varphi'(x-vt)$ and
\begin{equation}
    v_x=\frac{\varphi'(x-vt)}{1+t\varphi'(x-vt)} \Rightarrow 1-t v_x=\frac{1}{1+t\varphi'(x-vt)}.
\end{equation}
Therefore, one has that
\begin{equation}\label{eq-app:rho-sol-bis}
    \rho(x,t)=(1-t v_x) \rho_{\ini}(x-vt),
\end{equation}
and
\begin{equation}\label{eq:jacobian-rho}
  dx \rho(x,t)=dx_{\ini} \rho_{\ini}(x_{\ini},t), \quad x_{\ini}\equiv x-vt.
\end{equation}

Now, the function $\varphi$ is determined by imposing that the system must reach the target distribution $\rho_{\fin}(x)$ at the final time $t=t_{\fin}$, i.e.
\numparts
\ba
    \rho_{\fin}(x)&=\frac{\rho_{\ini}(x-v_{\fin}(x)t_{\fin})}{1+t_{\fin}\,\varphi'(x-v_{\fin}(x)t_{\fin})},
\\
    v_{\fin}(x)&\equiv v(x,t_{\fin})=\varphi(x-v_{\fin}(x)t_{\fin}). \label{eq-app:vf-phi}
\ea
\endnumparts
Taking spatial derivative in the last equation and inserting the result into the previous one, we get
\begin{equation}\label{eq-app:rho-f}
    \rho_{\fin}(x)=\rho_{\ini}(x-v_{\fin}(x)t_{\fin})\left[1-v'_{\fin}(x)t_{\fin}\right],
\end{equation}
which determines $v_{\fin}(x)$ or, equivalently, the (up to now) unknown function $\varphi$.

Zhang's solution to the optimization problem~\cite{2019Zhang,2020Zhang} is recovered by going to the cumulative distribution function defined in Eq.~\eref{eq:cumulative-distr}. Integrating Eq.~\eref{eq-app:rho-f} from $-\infty$ to some arbitrary point $x$, one gets
\begin{equation}
    \mathcal{F}_{\fin}(x)=\mathcal{F}_{\ini}(x-v_{\fin}(x)t_{\fin}),
\end{equation}
after assuming that $\lim_{x\to -\infty} \left[x-v_{\fin}(x)t_{\fin}\right]=-\infty$; $\mathcal{F}_{\ini}$ and $\mathcal{F}_{\fin}$ are the cumulative distributions corresponding to $\rho_{\ini}$ and $\rho_{\fin}$, respectively. Since both $\rho_{\ini}$ and $\rho_{\fin}$ are positive definite for equilibrium states, $\mathcal{F}_{\ini}$ and $\mathcal{F}_{\fin}$ are strictly increasing functions and we can explicitly solve the above equation for $v_{\fin}(x)$,
\begin{equation}\label{eq-app:v_f-sol-1}
    x-v_{\fin}(x)t_{\fin}=\mathcal{F}_{\ini}^{-1}\left(\mathcal{F}_{\fin}(x)\right), 
\end{equation}
where $\mathcal{F}_{\ini}^{-1}$ is the inverse function of $\mathcal{F}_{\ini}$.\footnote{Note that Eq.~\eref{eq-app:v_f-sol-1} ensures that $\lim_{x\to -\infty} \left[x-v_{\fin}(x)t_{\fin}\right]=-\infty$, since $\mathcal{F}_{\fin}(-\infty)=0$ and $\mathcal{F}_{\ini}^{-1}(0)=-\infty$.} By defining the function
\begin{equation}
    \label{eq-app:Xi(x)}
    \Xi(x)\equiv \mathcal{F}_{\ini}^{-1}\left(\mathcal{F}_{\fin}(x)\right),
\end{equation}
we get that
\begin{equation}
    v_{\fin}(x)=\frac{x-\Xi(x)}{t_{\fin}}.
\end{equation}
Note that determining $v_{\fin}(x)$ is utterly equivalent to determining $\varphi(x)$, because Eq.~\eref{eq-app:vf-phi} tells us that
\begin{equation}\label{eq-app:phi-Xi}
    v_{\fin}(x)=\varphi(\Xi(x)) \Rightarrow \varphi(x)=v_{\fin}(\Xi^{-1}(x))=\frac{\Xi^{-1}(x)-x}{t_{\fin}}.
\end{equation}
It is useful to recall that, in the connection problem we are addressing, both $\mathcal{F}_{\ini}$ and $\mathcal{F}_{\fin}$ are given, so that the function $\Xi(x)$ in \eref{eq-app:Xi(x)} is perfectly defined. Then, the function $\varphi(x)$ follows from Eq.~\eref{eq-app:phi-Xi} and $v(x,t)$ is thus implicitly provided by \eref{eq-app:v-sol}, which completes the solution to the optimization problem.

It is important to remark that shocks cannot appear in the obtained solution when connecting equilibrium states,
\begin{equation}\label{eq-app:no-shocks}
    1+t\varphi'(x)=1+\frac{t}{t_{\fin}}\left(\frac{d}{dx}\Xi^{-1}(x)-1\right)>0
\end{equation}
for all times $t\in[0,t_{\fin}]$. Proving the above inequality is equivalent to prove that the term inside the parenthesis in Eq.~\eref{eq-app:no-shocks} is always larger than $-1$ or, equivalently, that $\frac{d}{dx}\Xi^{-1}(x)>0$ for all $x$. This is indeed true, since Eq.~\eref{eq-app:Xi(x)} can be rewritten as
\begin{equation}
    \mathcal{F}_{\ini}(x)=\mathcal{F}_{\fin}(\Xi^{-1}(x)) \Rightarrow \frac{d}{dx}\Xi^{-1}(x)=\frac{\rho_{\ini}(x)}{\rho_{\fin}(\Xi^{-1}(x))}>0.
\end{equation}

Bringing to bear Eqs.~\eref{eq-app:v-sol}, \eref{eq-app:rho-sol}, \eref{eq:jacobian-rho}, and \eref{eq-app:phi-Xi}, the irreversible power over the optimal protocol is
\begin{eqnarray}
    P_{\irr}^*&= \gamma \int_{-\infty}^{+\infty} \!\!\! dx\, \rho(x,t) v^2(x,t) \nonumber \\
    &= \gamma \int_{-\infty}^{+\infty} \!\!\! dx\, (1-t v_x) \rho_{\ini}(x-vt) \varphi^2(x-vt)
    \nonumber \\
    &=\gamma \int_{-\infty}^{+\infty} \!\!\! dx_{\ini} \, \rho_{\ini}(x_{\ini}) \left[\frac{\Xi^{-1}(x_{\ini})-x_{\ini}}{t_{\fin}}\right]^2 \nonumber \\
    &=\frac{\gamma}{t_{\fin}^2}\int_{-\infty}^{+\infty} \!\!\! dx_{\ini} \, \rho_{\ini}(x_{\ini}) \left[\Xi^{-1}(x_{\ini})-x_{\ini}\right]^2. \label{eq-app:Pirr-opt}
\end{eqnarray}
where Eq.~\eref{eq-app:rho-sol-bis} has been used and $x_{\ini}$ was defined in Eq.~\eref{eq:jacobian-rho}. The expression~\eref{eq-app:Pirr-opt} for the optimal irreversible power is identical to Eq.~\eref{eq:Pirr-opt} in the main text. Therefrom, the minimum irreversible work is readily obtained,
\begin{eqnarray}
\label{eq-app:Wirr*}
    W_{\irr}^*=t_{\fin} P_{\irr}^*=\frac{\gamma}{t_{\fin}}\int_{-\infty}^{+\infty} \!\!\! dx_{\ini} \, \rho_{\ini}(x_{\ini}) \left[\Xi^{-1}(x_{\ini})-x_{\ini}\right]^2,
\end{eqnarray}
which is again identical to Eq.~\eref{eq:Wirr*}---we repeat it here to keep the appendix self-contained. 

\section{Comparison of speed limit inequalities for the harmonic case}\label{app:inequal-comparison}

It is illuminating to compare inequalities~\eref{eq:sl-Sivak-Crooks} and \eref{eq:sl-Ito-Dechant} for the harmonic case, for which the probability distribution is Gaussian for all times, $\rho(x,t)=(2\pi\sigma^2)^{-1/2} \exp\left(-\frac{x^2}{2\sigma^2}\right)$. A simple calculation leads to
\begin{equation}
    P_{\irr}(t)=\gamma \dot{\sigma}^2, \quad I(t)=2\left(\frac{\dot{\sigma}}{\sigma}\right)^2.
\end{equation}
Making use of these expressions, we have that
\numparts
\ba
    \calL_{\lin}=\sqrt{\gamma}\int_0^{t_{\fin}} dt \left|{\dot{\sigma}} \right| \geq \sqrt{\gamma} \left| {\int_0^{t_{\fin}} dt\, \dot{\sigma}} \right|=\sqrt{\gamma}\left|{\sigma_{\fin}-\sigma_{\ini}}\right|, \\
    \calL=\sqrt{2} \int_0^{t_{\fin}} dt \left|{\frac{d}{dt}\ln\sigma}\right| \geq \sqrt{2}\left|{\int_0^{t_{\fin}} dt \frac{d}{dt}\ln\sigma}\right|=\sqrt{2}\,\ln\!\left|{\frac{\sigma_{\fin}}{\sigma_{\ini}}}\right|,
\ea
\endnumparts
and thus
\begin{equation}\label{eq:inequal-combined}
    t_{\fin}\, W_{\irr}\geq \gamma\left|{\sigma_{\fin}-\sigma_{\ini}}\right|^2, \quad t_{\fin}\, \calC\geq \left(\ln\left|{\frac{\sigma_{\fin}}{\sigma_{\ini}}}\right|\right)^2.
\end{equation}

On the one hand, these inequalities do not immediately translate to bounds for the connecting time, because both $W_{\irr}$ and $\calC$ depend on $t_{\fin}$---and on the path swept by the system to go from the initial to the target state. In fact, in the harmonic case we are analysing
\begin{equation}
    W_{\irr}=\gamma \int_0^{t_{\fin}}dt\, \dot{\sigma}^2, \quad \calC=\int_0^{t_{\fin}}dt\,\left(\frac{d}{dt}\ln\sigma\right)^2.
\end{equation}
If one assumes that all the time dependence occurs through a reduced time $s=t/t_{\fin}$, i.e. $\sigma=\sigma(s)$,
\begin{equation}
    W_{\irr}=\frac{\gamma}{t_{\fin}}  \int_0^{1}ds\, \left(\frac{d\sigma}{ds}\right)^2, \quad \calC=\frac{1}{t_{\fin}} \int_0^{t_{1}}ds\,\left(\frac{d}{ds}\ln\sigma\right)^2,
\end{equation}
and $t_{\fin}$ disappears from Eq.~\eref{eq:inequal-combined}. On the other hand, both inequalities are equivalent in the linear response regime---a property that is somehow expected, since the Fisher information $I(t)$ is directly related to the friction tensor $\boldsymbol{\zeta}(\boldsymbol{\lambda})$ in Eq.~\eref{eq:Pirr-zeta-rel}~\cite{2012Sivak}. Within linear response, $\sigma=\sigma_{\fin}+\delta\sigma$ and non-linearities in $\delta\sigma$ are neglected. Therefore, we have that 
\begin{equation}
    \calL_{\lin}\sim\sqrt{\gamma}\left|{\delta\sigma_{\ini}}\right|, \quad \calL\sim\frac{\sqrt{2}}{\sigma_{\fin}}\left|{\delta\sigma_{\ini}}\right|,
\end{equation}
and
\begin{equation}
    W_{\irr}\sim \gamma \int_0^{t_{\fin}}dt\, \left(\frac{d}{dt}\delta\sigma\right)^2, \quad \calL\sim \frac{1}{\sigma_{\fin}^2}\int_0^{t_{\fin}}dt\, \left(\frac{d}{dt}\delta\sigma\right)^2.
\end{equation}
Finally, both inequalities can be cast in the same form,
\begin{equation}
    t_{\fin} \int_0^{t_{\fin}}dt\, \left(\frac{d}{dt}\delta\sigma\right)^2 \geq \left|{\delta\sigma_{\ini}}\right|^2.
\end{equation}
Consistently with our notation, $\delta\sigma_{\ini}=\sigma_{\ini}-\sigma_{\fin}$.

\section{A shortcut to the fluctuation relation}
\label{app:fluctuation-relation}

The idea of counterdiabatic driving, in the framework of the Langevin/Fokker-Planck equation used in Sec.~\ref{sec:counterdiabatic}, offers an economical derivation of the work fluctuation relation 
\cite{1997Jarzynski}. In a potential $U(x,{\boldsymbol \lambda}(t))$ where $\boldsymbol{\lambda}(t)$ stands for the collection of external parameters controlled by an  operator (say, the experimentalist), the particle density $\rho$ obeys Eq. \eref{eq:FP} with fixed temperature (i.e. fixed $\beta$). 
Here, it is understood that a given protocol is chosen from the outset, through the time
dependence of the steering parameter ${\boldsymbol \lambda}$. 
What would be the counterdiabatic force to be applied, such that the 
particle density would be forced to follow the equilibrium
distribution? In other words, we seek to enforce
\begin{equation}
    \rho(x,t) = \rho_{eq}(x,{\boldsymbol \lambda}(t)) = \frac{1}{Z({\boldsymbol \lambda}(t))} e^{-\beta U(x,{\boldsymbol \lambda}(t))}
    \label{eq:appdefEQ}
\end{equation}
with the partition function 
\begin{equation}
    Z({\boldsymbol \lambda}) = \int_{-\infty}^\infty e^{-\beta U(x,{\boldsymbol \lambda})}\, dx,
\end{equation}
assumed well defined for all ${\boldsymbol \lambda}$ (i.e. $U$ is supposed to be confining 
enough). The required counterdiabatic term is given by Eq. \eref{eq:UCD_FP}.
Introducing the Fokker-Planck operator ${\cal L}_u$ associated to $U(x,{\boldsymbol \lambda}(t))$,
such that
\begin{equation}
    {\cal L}_u \rho \,\equiv\, \partial_x [\rho \, \partial_x U(x,{\boldsymbol \lambda}(t))] + \beta^{-1} \partial^2_x \rho
\end{equation}
the evolution equation for $\rho_{eq}$ can be written
\begin{eqnarray}
    \partial_t \rho_{eq}(x,{\boldsymbol \lambda}(t)) &= {\cal L}_u \rho_{eq} -\beta \left(\dot W - \left\langle \dot W \right\rangle\right) \rho_{eq} 
    \label{eq:appFR_1}\\
    &= -\beta \left(\dot W - \left\langle \dot W \right\rangle\right) \rho_{eq} .
    \label{eq:appFR_2}
\end{eqnarray}
Here, we have introduced the instantaneous power 
\begin{equation}
    \dot W (x,t) \,=\, \dot{\boldsymbol \lambda} \cdot \partial_{\boldsymbol \lambda} U(x,{\boldsymbol \lambda}(t)),
\end{equation}
the dot denotes a time derivative, and the bracket is for an average 
with weight $\rho_{eq}$:
\begin{equation}
    \left\langle \dot W \right\rangle \,=\, \int_{-\infty}^\infty 
    \dot W \,\rho_{eq}(x,{\boldsymbol \lambda})\, dx.
    \end{equation}
Note that integrating $\dot W$
over a trajectory, one obtains the corresponding stochastic work
$W$. 

While relation \eref{eq:appFR_2} is straightforwardly obtained 
from the very definition of the equilibrium density in Eq.
\eref{eq:appdefEQ}, we are here more interested in the seemingly more
complex form in Eq. \eref{eq:appFR_1}. Consider indeed the auxiliary dynamics
defined by the evolution equation 
\begin{equation}
     \partial_t {\cal Q}(x,t) \, = \, {\cal L}_u {\cal Q} -\beta \dot W  {\cal Q}
     \label{eq:appQ}
\end{equation}
and initial condition 
\begin{equation}
    {\cal Q}(x,t=0) \,=\, \rho_{eq}(x,{\boldsymbol \lambda}(0)) 
    = \frac{1}{Z({\boldsymbol \lambda}(0))} e^{-\beta U(x,{\boldsymbol \lambda}(0))} .
\end{equation}
Compared to Eq. \eref{eq:appFR_1}, the term in 
$\beta \langle \dot W \rangle Q$ 
is missing in \eref{eq:appQ}. To proceed, it is instructive to reinterpret the dynamics encoded in \eref{eq:appFR_1} as follows. 
A collection of independent random walkers, diffusing in the force 
field $-\partial_x U$ (hence the presence of the operator ${\cal L}_u$),
is subject to a population/depopulation mechanism, such that walkers 
are added at rate $\beta \dot W$. This leads to an overall population change,
such that when $U$ increases locally, $\dot W>0$ and particles are removed, to match a lesser
probability of presence. Conversely, when $\dot W<0$, particles are added locally.
The counter-term in $\langle \dot W \rangle \rho_{eq}$ has the
effect to correct for this gain/loss, and conserve globally the total number of walkers. The resulting integral $\int_{-\infty}^{+\infty} \rho_{eq} dx$ is thereby conserved.
The counter-term does not change the $x$ dependence; it only affects 
normalization, and can be viewed as the action of a global rescaling 
$\rho_{eq} \to \Lambda \rho_{eq}$.
If this counter-term is absent, as in the case in 
Eq. \eref{eq:appQ}, we obtain the unnormalized solution 
\begin{equation}
    {\cal Q}(x,t) \,=\, \frac{1}{Z({\boldsymbol \lambda}(0))} e^{-\beta U(x,{\boldsymbol \lambda}(t))} .
\end{equation}

The final step in the argument is the realization that 
$\cal Q$ can be reinterpreted as \cite{2001Hummer}
\begin{equation}
    {\cal Q}(x,t) = \left\langle e^{-\beta W}  \delta(x-x(t)) \right\rangle.
\end{equation}
Here, the brackets correspond to averaging over all trajectories
starting from a point $x_0$ drawn according to the distribution $\rho_{eq}(x_0,{\boldsymbol \lambda}(0))$, and that end up at position $x$ at time $t$.
The stochastic work is
\begin{equation}
    W \,=\, \int_0^t \dot{\boldsymbol \lambda}(\tau) \cdot \partial_{\boldsymbol \lambda} U(x(\tau),{\boldsymbol \lambda}(\tau)) \, d\tau .
\end{equation}
This is a facet of the Feynman-Kac correspondence. 
Since
\begin{equation}
    \left\langle e^{-\beta W}  \delta(x-x(t)) \right\rangle \, = \, \frac{1}{Z({\boldsymbol \lambda}(0))} e^{-\beta U(x,{\boldsymbol \lambda}(t))}
\end{equation}
we can integrate over $x$ on both sides to get
\begin{equation}
     \left\langle e^{-\beta W} \right\rangle \, = \, \frac{Z({\boldsymbol \lambda}(t))}{Z({\boldsymbol \lambda}(0))} \equiv
     e^{-\beta \Delta F }
\end{equation}
where the last equality serves to define the free-energy 
difference $\Delta F$. The above relation, valid at all
time $t$ for an arbitrary protocol ${\boldsymbol \lambda}(t)$, is the work fluctuation 
relation \cite{1997Jarzynski}.

\section*{Acknowledgments}
We would like to thank A. Baldassarri, M. Baldovin, L. Bellon, B. Besga, A. Chepelianskii, S. Ciliberto, S. Dago, T. de Guillebon, S.N. Majumdar, I. Mart\'inez, J.G. Muga, I. Palaia, A. Patr\'on, A. Petrossian, A. Puglisi, D. Raynal, L. Rondin, N. Ruiz-Pino. C.A.P. and A.P. acknowledge financial support from Grant PGC2018-093998-B-I00 funded by MCIN/AEI/10.13039/501100011033/ and by ERDF ``A way of making Europe.'' C.A.P.  also acknowledges financial support from Junta de Andaluc\'{\i}a and European Social Fund through the program PAIDI-DOCTOR. This work was supported by the `Agence Nationale de la Recherche' grant No. ANR-18-CE30-0013.  C.J. acknowledges financial support from the Simons Foundation through Award No.\ 681566, and from the U.S. National Science Foundation under Grant No.\ 2127900.

\section*{References}
\bibliographystyle{iopart-num}
\bibliography{bib_shortcut_comp}

\end{document}